\documentclass[11pt]{article}
\usepackage{jheppub}
\usepackage{epsfig}
\usepackage{amssymb}
\usepackage{amsmath}
\usepackage{graphicx, subfigure, array, placeins, float}

\graphicspath{{./fig/}}
\usepackage{makecell}
\usepackage{multirow}
\usepackage{epstopdf}
\usepackage{extarrows}

\usepackage{dsfont,bbm}
\usepackage{shuffle}

\usepackage{booktabs}
\usepackage{longtable}

\usepackage{xtab}
\usepackage{caption}

\title{Analytic two-Loop four-point form factor of the stress-tensor supermultiplet in ${\cal N}=4$ SYM}

\author[a,b]{Yuanhong Guo,}
\emailAdd{guoyuanhong@itp.ac.cn}
\author[c,d]{Lei Wang,}
\emailAdd{wanglei1@bhu.edu.cn}
\author[a,b,e,f]{Gang Yang,}
\emailAdd{yangg@itp.ac.cn}
\author[a,b]{Yixiong Yin}
\emailAdd{yinyixiong@itp.ac.cn}
\affiliation[a]{CAS Key Laboratory of Theoretical Physics, Institute of Theoretical Physics, \\Chinese Academy of Sciences, Beijing 100190, China}
\affiliation[b]{School of Physical Sciences, University of Chinese Academy of Sciences, Beijing 100049, China}
\affiliation[c]{College of Physical Science and Technology, Bohai University, Jinzhou 121013, China}
\affiliation[d]{School of Physics, Peking University, Beijing 100871, China }
\affiliation[e]{School of Fundamental Physics and Mathematical Sciences, Hangzhou Institute for Advanced Study, UCAS, Hangzhou 310024, China}
\affiliation[f]{International Centre for Theoretical Physics Asia-Pacific, Beijing/Hangzhou, China}

\abstract{
We compute the two-loop four-point MHV form factor of the stress-tensor supermultiplet in planar ${\cal N}=4$ super Yang-Mills (SYM).
This form factor is analogous to the Higgs plus four-gluon amplitudes in the heavy-top limit of QCD when translated to the ${\cal N}=4$ SYM context.
We obtain the full $D$-dimensional integrands up to two loops via unitarity-cut methods. 
Subsequently, we utilize IBP reduction to express the result in terms of a set of uniformly transcendental basis integrals, incorporating the two-loop non-planar five-point one-mass integrals recently given by Abreu et al. [PRL 132 (2024) 14]. 
We obtain the two-loop finite remainder in the functional form in terms of the pentagon functions. 
The symbol of our remainder confirms the bootstrap results reported by Dixon et al. [PRL 130 (2023) 11].
We perform various non-trivial checks of our results, including the triple-collinear limit, which recovers the two-loop six-gluon remainder.
We also show that the form factor has a directional dual conformal symmetry at the integrand level.
Our results are expected to shed further light on the study of antipodal dualities and the computation of Higgs plus four-parton amplitudes in QCD.
}

\begin{document}

\maketitle

\setcounter{footnote}{0}

\section{Introduction}
\label{sec:intro}

Scattering amplitudes and correlation functions are central physical quantities in our understanding of quantum field theory (QFT).
Form factors, which represent the matrix elements that connect on-shell states with local operators, can be viewed as an intersection of scattering amplitudes and correlation functions. They offer new insights into the hidden structures of QFT.
An $n$-point form factor is given by the matrix element between a state created by a local operator $\mathcal{O}(x)$ and a multiple-particle outgoing state denoted by $\langle 1 \cdots n|$:
\begin{equation}
	\int \mathrm{d}^{4} x e^{-i q x} \langle 1 \cdots n|\mathcal{O}(x)| 0\rangle = \delta^{(4)} (q-\sum_{i=1}^{n} p_{i} ) \mathcal{F}_{\mathcal{O}, n}(1, \ldots, n) \,,
\end{equation}
where $p_i$ signifies the momentum of the $i$-th external on-shell particle, and $q=\sum_i p_i$ represents the total momentum carried by the operator.

In this paper, we focus on the form factor of the stress-tensor supermultiplet in ${\cal N}=4$ supersymmetric Yang-Mills theory (SYM). 
This form factor has served as a rich playground for extrapolating advancements from amplitude studies and for deepening our understanding of the dynamics of ${\cal N}=4$ SYM.
It also has an intriguing connection to the Higgs-plus-gluons amplitudes in quantum chromodynamics (QCD), as we will see below.
We begin with a review of the studies of this form factor.

The investigation of the form factor of the stress-tensor supermultiplet in ${\cal N}=4$ SYM dates back to 1985, when the two-point form factor, known as the Sudakov form factor, was first calculated by Van Neerven up to two loops \cite{vanNeerven:1985ja}.
Later, the strong coupling picture for the form factor was considered using the AdS/CFT duality \cite{Alday:2007he}, which involves finding the area of the minimal surface bounded by a periodic Wilson loop. 
Interest in higher-point form factors started in 2010, first at strong coupling \cite{Maldacena:2010kp} and then at weak coupling \cite{Brandhuber:2010ad, Bork:2010wf, Brandhuber:2011tv, Bork:2011cj}.
The strong coupling picture \cite{Alday:2007he, Maldacena:2010kp, Gao:2013dza} suggests a duality between the form factor and a periodic Wilson loop, which was studied at weak coupling at one loop \cite{Brandhuber:2010ad}.
The study of the high-point tree-level form factor of the stress-tensor supermultiplet \cite{Brandhuber:2010ad, Brandhuber:2011tv} also revealed that the MHV super form factors follow a Parke-Taylor-like formula.

The two-loop three-point form factor of the stress-tensor supermultiplet was computed using both the unitarity and symbol bootstrap methods 
in \cite{Brandhuber:2012vm}. Intriguingly, the resulting two-loop remainder function was found \cite{Brandhuber:2012vm} to match the maximally transcendental part of the two-loop Higgs-plus-three-gluon amplitudes \cite{Gehrmann:2011aa}.
This finding can be understood as an extension of the maximal transcendental principle \cite{Kotikov:2002ab, Kotikov:2004er}.
Additionally, an interesting coincidence between the symbols of the three-point form factor remainder and the six-gluon MHV amplitude remainder was observed in \cite{Brandhuber:2012vm}. This observation was later generalized in a non-trivial way to higher loops, leading to the antipodal duality between the three-point form factor and the six-gluon MHV amplitude \cite{Dixon:2021tdw}, which will be discussed below.

The Sudakov form factor in ${\cal N}=4$ SYM provides one of the simplest quantities to extract the cusp and the collinear anomalous dimensions, which are crucial to understanding the infrared (IR) divergences in amplitudes. The three-loop Sudakov form factor was obtained using the unitarity method in \cite{Gehrmann:2011xn}. 
The four-loop Sudakov form factor offers the first new insight into the sub-leading $N_c$ corrections to the cusp and collinear anomalous dimensions  \cite{Boels:2012ew, Boels:2015yna, Boels:2017skl, Boels:2017ftb}. These corrections explicitly show that quadratic Casimir scaling breaks down at four loops, a result confirmed by other computations, see \emph{e.g.}~\cite{Moch:2017uml, Grozin:2017css, Moch:2018wjh, Lee:2019zop, Bruser:2019auj, Henn:2019swt, Huber:2019fxe}.

Beyond the planar limit, the form factor of the stress-tensor supermultiplet has also been constructed at high-loop orders. 
The full-color integrand of the five-loop Sudakov form factor was obtained \cite{Yang:2016ear} and the three-point form factor up to four loops \cite{Lin:2021kht, Lin:2021lqo}, thanks to the color-kinematics duality \cite{Bern:2008qj, Bern:2010ue}. 
The integration of the full-color three-loop form factor was also performed in \cite{Lin:2021qol, Guan:2023gsz}.

The planar form factor of the stress-tensor supermultiplet exhibits rich mathematical structures analogous to those found in amplitudes in ${\cal N}=4$ SYM. The Grassmannian and polytope frameworks for the form factor have been explored in \cite{Frassek:2015rka, Bork:2016hst, Bork:2016xfn, Bork:2017qyh, Bork:2014eqa}. The twistor formalism was extended to the form factor in \cite{Koster:2016ebi, Koster:2016loo, Chicherin:2016qsf, Koster:2016fna}. The so-called connected description was developed for the form factor in \cite{He:2016dol, Brandhuber:2016xue, He:2016jdg}. At the integrand level, the recursion relation for the form factor was studied in \cite{Bolshov:2018eos, Bianchi:2018peu}.

The non-perturbative framework known as the form factor operator product expansion (FFOPE) was proposed in \cite{Sever:2020jjx, Sever:2021nsq, Sever:2021xga}.
With the prediction of FFOPE, the three-point form factor of the stress-tensor supermultiplet was bootstrapped to eight loops \cite{Dixon:2020bbt, Dixon:2022rse}.
These results reveal the aforementioned antipodal duality between the three-point form factor and the six-point amplitude \cite{Dixon:2021tdw}.

The two-loop four-point form factor of the stress-tensor supermultiplet has been a recent frontier of study. It was first computed in the limit where the operator momentum $q$ becomes lightlike (referred to as the lightlike form factor) using the bootstrap method with master integrals \cite{Guo:2022qgv}. 
The result of the lightlike form factor was found to exhibit an exact directional dual conformal symmetry.
Furthermore, the FFOPE framework was also generalized to the lightlike form factor in \cite{Guo:2022qgv}, introducing a new type of form factor transition.
For the two-loop four-point form factor with generic off-shell $q$, the symbol of the two-loop remainder was constructed using the symbol-bootstrap method in \cite{Dixon:2022xqh}. Intriguingly, it exhibits an antipodal self-duality, which offers important insights into the antipodal duality between the three-point form factor and the six-point amplitude.
The four-dimensional integrand for the two-loop four-point form factor was also studied in \cite{Gopalka:2023doh}.

The goal of this paper is to provide a first-principle full computation of the two-loop four-point form factors of the stress-tensor supermultiplet in planar $\mathcal{N}=4$ SYM at both the integrand level and the integrated level, the latter of which is not only the symbol but also the full functional result.

It is worth pointing out that the form factor we consider is related to the Higgs plus four-gluon amplitudes in the heavy-top mass limit in QCD.
The related two-loop five-point non-planar master integrals have been recently obtained in \cite{Abreu:2023rco} while the two-loop Higgs-plus-four-gluon amplitudes have not been available in QCD.
Our result provides the first full computation of such an amplitude at the function level and is also a concrete application of the master integral results in \cite{Abreu:2023rco}.

This paper is organized as follows.

In Section~\ref{sec:review}, we first give a brief review of the stress-tensor supermultiplet in ${\cal N}=4$ SYM and its form factor, and we also briefly revisit the on-shell unitarity method.

In Section~\ref{sec:oneloopFF}, we provide the full $D$-dimensional integrand of the one-loop form factor via the unitarity method. In Section~\ref{sec:2lUnitarity}, we further obtain the full $D$-dimensional integrand at two loops.
Compared to the lower-point cases, the four-point form factor contains some new features. First, the four-point form factor contains the parity-odd part, which is proportional to $ \varepsilon(1234)\equiv \varepsilon_{\mu\nu\rho\sigma}p_1^\mu p_2^\nu p_3^\rho p_4^\sigma$. Moreover, there is a crucial part of the so-called $\mu$-term integrals, which demand the use of $D$-dimensional unitarity cut.

In Section~\ref{sec:subtraction}, we reduce the form factor in a set of uniform-transcendental (UT) integral bases provided by \cite{Abreu:2020jxa, Canko:2020ylt, Abreu:2021smk, Abreu:2023rco}.
We apply the BDS ansatz \cite{Bern:2005iz} to implement the subtraction of infrared (IR) divergence. The cancellation of the IR divergences provides a non-trivial check of our computation.
We obtain the two-loop finite remainder function in the form of the pentagon functions.
At the symbol level, our result reproduces the symbol bootstrap result in \cite{Dixon:2022xqh}, providing an important check of the bootstrap method.
As further non-trivial checks, we perform both the collinear limit and triple collinear limit for the four-point form factor, we find that it reproduces the remainders of the three-point form factor and the six-gluon amplitude, respectively.

In Section~\ref{sec:DDCIintegrand}, we discuss the directional dual conformal symmetry of the form factor that arises in the lightlike limit of $q\rightarrow0$.
We show that the two-loop form factor has the symmetry at the integrand level.

Finally, a summary and discussion are given in Section~\ref{sec:discussion} followed by three appendices.
In Appendix~\ref{app:UT}, we list all one-loop UT (uniformly transcendental) masters that we have used in the paper.
Appendix~\ref{app:consistency} provides some details of further numerical evaluation and checks.
In Appendix~\ref{app:comments}, we provide the two-loop integrand of another similar half-BPS form factor, the four-point two-loop form factor of the length-$3$ half-BPS operator ${\rm tr}(\phi^3)$.

\section{Review and setup}
\label{sec:review}

In this section, we give a brief review of the stress tensor supermultiplet in ${\cal N}$=4 SYM and its tree-level form factor, and the strategy of the unitarity cut method.
This is to set up the notations and prepare for the construction of loop-level form factors in the following sections.

\subsection{Stress-tensor supermultiplet and  the form factor}
\label{sec:RFF}

The explicit form of the chiral stress tensor supermultiplet can be given as \cite{Eden:2011yp}
\begin{align}
{\cal T} (x, \theta^+) &= {\rm tr} ( \phi^{++} \phi^{++}) + i 2 \sqrt{2} \theta_{\alpha}^{+a} {\rm tr} ( \psi_a ^{+ \alpha}  \phi^{++})
\nonumber \\
&+ \theta_{\alpha}^{+a} \epsilon_{ab} \theta^{+b}_{\beta} {\rm tr} \Big( \psi^{+c ( \alpha} \psi_{c}^{ + \beta )} - i \sqrt{2} F^{\alpha \beta} \phi^{++} \Big)
\nonumber \\
&- \theta_{\alpha}^{+a} \epsilon^{\alpha \beta} \theta_{\beta}^b {\rm tr} \Big( \psi^{+ \gamma}_{(a} \psi^{+}_{b) \gamma} - g_{\text{YM}} \sqrt{2} [ \phi^{+C}_{(a} , \bar{\phi}_{C \, +b)} ] \phi^{++} \Big)
\nonumber \\
&- {4\over3} ( \theta^{+}  )^{3\, a}_{\alpha} {\rm tr} \Big( F_{\beta}^{\alpha} \psi_{a}^{+ \beta} + i g_{\text{YM}} [ \phi_a^{+B} ,  \bar{\phi}_{BC} ] \psi^{C \alpha}\Big) + {1\over 3} (\theta^+)^4 \,  {\cal L}(x)
\ .
\label{eq:superST}
\end{align}
Here the $(\theta^+)^0$ component is the scalar half-BPS operator ${\cal T}(x, 0) = {\rm tr} (\phi^{++} \phi^{++})(x)$.
Similar to the translation ${\cal O}(x) = \exp (i {P} x ) {\cal O}(0) \exp (- i  {P} x )$, the supermultiplet \eqref{eq:superST} can be generated via a super-translation from the scalar operator as
\begin{equation}
{\cal T} (x, \theta^{+a}_{\alpha}) =\exp (i Q_{+a}^{\alpha} \theta^{+a}_{\alpha}  ) {\cal T}(x, 0)
\exp (- i Q_{+a}^{\alpha} \theta^{+a}_{\alpha}  ).
\end{equation}
The $(\theta^+)^4$-component ${\cal L}$ is the chiral on-shell Lagrangian
\begin{equation}
{\cal L} = {\rm tr} \left[ - {1\over2} F_{\alpha\beta}F^{\alpha\beta} + \sqrt{2} g_{\rm YM} \psi^{\alpha A} [\phi_{AB}, \psi^B_\alpha] - {1\over8}g_{\rm YM}^2 [\phi^{AB}, \phi^{CD}] [\phi_{AB}, \phi_{CD}] \right] \,.
\label{eq:calL}
\end{equation}

One can define the form factor for the super-operator as the  super Fourier transform of
the matrix element \cite{Brandhuber:2011tv}:
\begin{align}
{\cal F}_{{\cal T},n}^{(0),{\rm MHV}}(1,\ldots, n; q, \gamma_+) = & \int d^4 x \, d^4 \theta^+ \,e^{- i q\cdot x - \gamma_{+}^\alpha \theta^{+}_\alpha}
\langle 1 \ldots n |{\cal T}(x, \theta^+) | 0\rangle \nonumber\\
= & {\delta^{(4)}(q-\sum_i \lambda_i\tilde\lambda_i) \delta^{(4)}(\gamma_+ -\sum_i \lambda_i \eta_{+,i}) \delta^{(4)}(\sum_i \lambda_i \eta_{-i}) \over \langle 12\rangle \langle 23\rangle \ldots \langle n 1 \rangle} \,.
\label{eq:FF-superAll-MHV}
\end{align}
Like that $q$ is the momentum for the operator, the $\gamma_{+}$ plays the role of super-momentum for the super-operator.
To obtain different components of the super-operator, we can expand the fermionic delta function and take different coefficients in the $\gamma$ expansion. For example:
\begin{align}
(\gamma)^4\textrm{ - term}: \qquad & {\cal F}_{{\rm tr}(\phi^2),n}^{(0),{\rm MHV}} = {\delta^{(4)}(q-\sum_i \lambda_i\tilde\lambda_i) \delta^{(4)}(\sum_i \lambda_i \eta_{i}^+) \over \langle 12\rangle \langle 23\rangle \ldots  \langle n 1 \rangle} \,, \\
(\gamma)^0\textrm{ - term}: \qquad & {\cal F}_{{\cal L},n}^{(0),{\rm MHV}} = {\delta^{(4)}(q-\sum_i \lambda_i\tilde\lambda_i) \delta^{(8)}(\sum_i \lambda_i \eta_i^A) \over \langle 12\rangle \langle 23\rangle \ldots \langle n 1 \rangle} \,.
\label{eq:FF-super-component-MHV}
\end{align}
In practical computation for the loop form factors, it is convenient and enough to focus on the scalar operator ${\rm tr}(\phi^2)$ or the chiral Lagrangian ${\cal L}$. 
For simplicity, one can choose an explicit projection of the Harmonic super space index `$+$', for example, the half-BPS scalar operator can be chosen as ${\rm tr}(\phi^{12}\phi^{12})$.

The non-MHV tree-level form factors can be computed using MHV rules or BCFW recursion methods \cite{Brandhuber:2011tv}. In the computation of $D$-dimensional unitarity cuts, one can compute the tree-level form factor (carrying $D$-dimensional cut momenta) using Feynman diagrams.

For the reader's convenience, below we collect some explicit results for the stress-tensor form factor beyond the tree level.
\begin{itemize}

\item
General MHV one-loop \cite{Brandhuber:2010ad}; NMHV one-loop \cite{Bork:2012tt, Bianchi:2018rrj}.

\item 
Sudakov form factor: two-loop \cite{vanNeerven:1985ja}; three-loop \cite{Gehrmann:2011xn}; four-loop \cite{Boels:2012ew, Boels:2017ftb, Huber:2019fxe}; five-loop integrand \cite{Yang:2016ear}; 

\item
Three-point: two-loop \cite{Brandhuber:2012vm}; three- to eight-loops (large $N_c$ limit) \cite{Dixon:2020bbt, Dixon:2022rse}; 
three-loop (full-color) \cite{Lin:2021kht, Lin:2021qol, Guan:2023gsz}; four-loop full-color integrand \cite{Lin:2021lqo};

\item
Strong coupling: general picture \cite{Alday:2007he}; Y-system in $AdS_3$ \cite{Maldacena:2010kp}; Y-system in $AdS_5$ \cite{Gao:2013dza};

\item
FFOPE with $q^2\neq0$ \cite{Sever:2020jjx, Sever:2021nsq, Sever:2021xga}; lightlike FFOPE with $q^2=0$ \cite{Guo:2022qgv};

\item
Four-point MHV at two-loop: symbol and function (for $q^2=0$) \cite{Guo:2022qgv}; symbol (for general $q$) \cite{Dixon:2022xqh};
integrand and integrated function (for general $q$) by the present paper.

\end{itemize}

\subsection{Unitarity-cut method}

The unitarity-cut method is an efficient way of computing sets of ordinary loop Feynman diagrams \cite{Bern:1994cg, Bern:1994zx, Britto:2004nc}. 
It is intended to construct the integrand of a loop amplitude or form factor by sewing together gauge-invariant tree building blocks. Practically, one can cut a set of internal propagators, represented by loop momenta $\{\ell_a\}$
\begin{equation}
	\frac{i}{\ell_a^{2}}\xrightarrow{\text{cut}} 2\pi\delta_+(\ell_a^{2})\ ,
\end{equation}
such that an $l$-loop $n$-point form factor factorizes as products of tree-level form factors and amplitudes as
\begin{align}
	\label{eq:FFcut2}
	{\cal F}_n^{(l)}\Big|_{\textrm{cut-}\{\ell_a\}} =\sum_{\textrm{physical states of}\ \{\ell_a\}}\prod (\text{tree-level building blocks}) \,.
\end{align}
The product of tree blocks corresponds to the residues of poles in the integrand, where the poles are cut propagators. If the integrand has all consistent residues for all possible cuts (a spanning set of cuts), then the integrand is guaranteed to have correct physical results.

In dimensional regularization, the loop momentum $\ell$ lives in $D$-dimensional spacetime with $D=4-2\epsilon$, and one has
\begin{equation}\label{eq:ell4dDd}
\ell^2 = \ell_{(4)}^2 - \ell_{(-2\epsilon)}^2 = \ell_{(4)}^2 - \mu_{\ell\ell}\,,
\end{equation}
where $\ell$ is divided into the four-dimensional component $\ell_{(4)}$, and the extra-dimensional component $\ell_{(-2\epsilon)}$. In the $D$-dimensional case, with the cut condition $\ell^2=0$, the four-dimensional part $\ell_{(4)}$ is no longer massless but has an effective mass $\mu_{\ell\ell}$.
For the four-dimensional cut, one identifies that $\ell \sim \ell_{(4)}$, therefore, the information of $\mu_{\ell\ell}$ may not be detected.

The integrals containing such $\mu$ terms in general play important roles in the physical results. 
Beyond one loop, these $\mu$ term integrals can even contribute to divergent parts of the amplitudes or form factors. Therefore, it is crucial to determine their contribution.

We will apply $D$-dimensional unitarity cuts to capture the $\mu$ terms. In $D$-dimensional unitarity cuts, one can utilize the Feynman diagrams to construct the tree amplitudes and form factors which are the tree blocks in \eqref{eq:FFcut2}. In such expressions, the full $D$-dimensional information for the loop momenta is kept. 
The helicity sum can be expressed as the contraction of polarization vectors/spinors for the gluons/fermions associated with the cut lines as 
\begin{equation}
\sum_{\textrm{physical states of}\ \ell} \epsilon_{\mu}(\ell) \epsilon^*_{\nu}(-\ell) = \eta_{\mu\nu}-\frac{k_{\mu}\ell_{\nu}+k_{\nu}\ell_{\mu}}{\ell\cdot k} \,,  \quad \sum_{\textrm{physical states of}\ \ell} u_s(\ell) \bar{u}_s(-\ell) = \ell \mkern -8 mu \slash \label{formula:sumfermion}\,,
\end{equation}
where $\epsilon_\mu$ denotes the internal gluon, $u_s$ denotes the internal fermion and $k$ is an arbitrary reference momentum. The final result is independent of the reference momentum $k$ due to the gauge invariance.

We will apply the four-dimensional helicity (FDH) scheme, see \cite{Bern:2002zk};
the external lines are set in the four-dimension and the internal momenta are set in $D=4-2\epsilon$ ($D>4$), while the internal helicity states are set to be in four-dimension. The scheme can guarantee the supermultiplets of the cut internal states are the same as those of the external states:
\begin{align}\label{formula:superfield}
	\mathrm{\Phi}(p, \eta) = & g_{+}(p)+\eta^{A} \bar\psi_{A}(p)+\frac{1}{2!} \eta^{A} \eta^{B} \phi_{A B}(p) +\frac{1}{3!} \eta^{A} \eta^{B} \eta^{C} \varepsilon_{A B C D} \psi^{D}(p)\notag \\
	&+\frac{1}{4!} \eta^{A} \eta^{B} \eta^{C} \eta^{D} \varepsilon_{A B C D} g_{-}(p) \, .
\end{align}
We also use $\psi_{ABC}(p) \equiv \varepsilon_{A B C D} \psi^{D}(p)$ to denote the four-dimensional fermion field carrying the negative helicity. Hence, the supersymmetry is preserved in this scheme.\footnote{We recall that in the four-dimensional cuts, the helicity sum can be performed by integrating over the Grassmann numbers of internal lines $\sum_{\textrm{physical states of}\ \ell}\xrightarrow{4-\text{dim}}\int \text{d}^4 \eta_{\ell}$.}

We will discuss the $D$-dimensional cuts in more detail in the next sections. Here we clarify some conventions that will be used later.
First, the standard four-dimensional $\gamma$ trace ${\rm tr}_-(i,j,\cdots,k)$ is defined by
\begin{align}
	\label{formula:traceid}
	&{\rm tr}_-(i, j,\cdots,k) \equiv \frac{1}{2}{\rm tr}[(1-\gamma_5)\gamma_\mu\gamma_\nu \cdots \gamma_\rho] p_i^\mu p_j^\nu\cdots p_k^\rho \,, \\
	&{\rm tr}(i,j,\cdots, k) \equiv {\rm tr}[\gamma_\mu \gamma_\nu \cdots \gamma_\rho ] p_i^{\mu} p_j^{\nu} \cdots p_k^\rho \,,
\end{align}
where $\mu, \nu, \cdots, \rho \in \{0, \ldots, 3\}$.
For the case with four momenta, we have
\begin{align}
	\label{formula:traceidlength4}
	{\rm tr}_-(i, j,k,l) \equiv 
	 {1\over2}{\rm tr}(i,j,k,l) +2i \varepsilon(i j k l) \,,
\end{align}
where
\begin{equation}
\varepsilon(ijkl)\equiv \varepsilon_{\mu\nu\rho\sigma}p_i^\mu p_j^\nu p_k^\rho p_l^\sigma \,,
\end{equation}
is the parity-odd contraction.

Second, to deal with fermions in $D$-dimensional cuts, the $\gamma$ matrices should be replaced by $\Gamma$ matrices $\Gamma^M_{\alpha\beta}$, which has $D$-dimensional Lorentz index $M$ (corresponding to loop momenta $\ell^M$) and 4-dimensional spin indices $\alpha,\beta$ (corresponding to particle states). The $\Gamma$ algebra is
 \begin{align}
 	(\Gamma^{M}\Gamma^{N}+\Gamma^{N}\Gamma^{M})_{\alpha\beta}=2\eta^{MN}{\bf I}_{\alpha\beta}\,,
 \end{align} 
 and the ${\rm tr}({\bf I}_{\alpha\beta})$ is set as $4$. 
We introduce the $D$-dimensional trace ${\bf tr}(i, j, \cdots, k)$ as
\begin{align}
	\label{formula:traceiD}
	{\bf tr}(i, j,\cdots,k) \equiv {\rm tr}[\Gamma_M \Gamma_N \cdots \Gamma_L] p_i^M p_j^N \cdots p_k^L \,,
\end{align}
where $M, N, \cdots, L \in \{0, 1, \ldots, D-1\}$. 
The difference between ${\bf tr}(i,j,k,l)$ and ${\rm tr}(i,j,k,l)$ for $D$-dimensional momenta is
\begin{align}
	{\bf tr}(i,j,k,l) - {\rm tr}(i,j,k,l) = & 4 \mu_{ij} \mu_{kl} +4 \mu_{il}\mu_{jk} -4 \mu_{ik}\mu_{jl} -2 s_{12} \mu_{34} +2 s_{13} \mu_{24} -2 s_{14} \mu_{23} \notag \\
	& -2 s_{23} \mu_{14} +2 s_{24} \mu_{13} -2 s_{34} \mu_{12} \,,
\end{align}
where $\mu_{ij} \equiv {p_{i,(-2\epsilon)}} \cdot p_{j,(-2\epsilon)}$.


\section{One-loop integrand via unitarity}
\label{sec:oneloopFF}
	
In this section, we revisit the construction of the one-loop four-point MHV form factor by the unitarity method.
We first give the full results of the integrand and then provide some details for the related unitarity cuts.
We will focus on the $D$-dimensional unitarity cuts, from which we determine the new contribution of pentagon integrals.

\subsection{Summary of the one-loop integrand}

We first present the one-loop integrand result that is valid in general $D$-dimensional cuts as the following form
\begin{align}
	{\cal F}^{(1)}_4 = {\cal F}^{(0)}_4 {\cal I}_4^{(1)} \,,
\end{align}
where ${\cal I}_4^{(1)}$ is expanded by
\begin{align}
	\label{eq:calI1loop}
	{\cal I}_4^{(1)} = e^{\epsilon\gamma_\mathrm{E}} \int\frac{d^D l}{i\pi^{\frac{D}{2}}} \left[ \mathbb{I}_{1}^{(1)}+ \mathbb{I}_{2}^{(1)}+ \mathbb{I}_{3}^{(1)} + {\rm cyclic}(p_1,p_2,p_3,p_4) \right] \,,
\end{align}
and $\mathbb{I}_{i}^{(1)}$ is the one-loop integrand defined by
\begin{align}
	\mathbb{I}_{i}^{(1)} = { N_i^{(1)} \over \prod_\alpha D^{(1)}_{i,\alpha} }\,,
\end{align}
with the propagators $D^{(1)}_{i,\alpha}$ and the numerators $N_i^{(1)}$ given in Table~\ref{tab:oneloopintegrand}.

\begin{table}[h]
	\centering
	\begin{tabular}{| l | l | l |}
		\hline
		\makecell*[c]{$N_1^{(1)}$}	&\makecell*[c]{\includegraphics[scale=0.27]{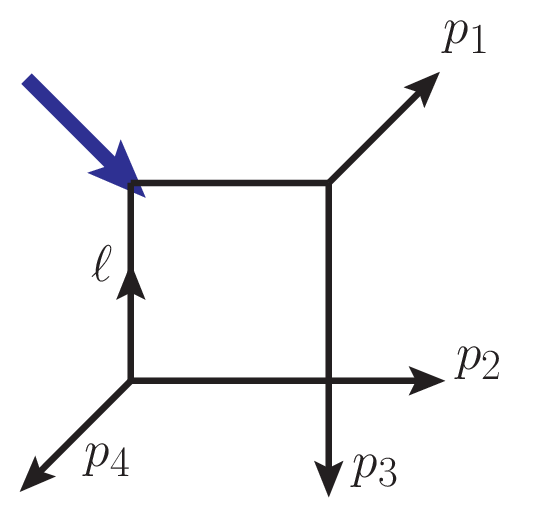}}&\makecell*[c]{${1\over2} \left( -\text{tr}_{-}(\ell,p_2+p_3,1,4)  -(\ell+p_4)^2 s_{14} \right)$\\+$(p_1 \leftrightarrow p_4, p_2\leftrightarrow p_3,\ell\leftrightarrow-\ell-p_1-p_2-p_3-p_4)$}\\ \hline
		\makecell*[c]{$N_2^{(1)}$}	&\makecell*[c]{\includegraphics[scale=0.25]{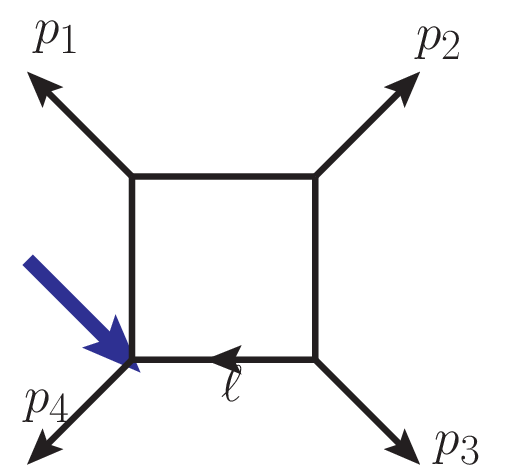}} & \makecell*[c]{ $\frac{1}{2} \left( -{\rm tr}_{-}(1,3,\ell,2) -\ell^2 s_{12}- (\ell +p_3)^2 s_{13} \right) $\\ +$(p_1 \leftrightarrow p_3,\ell\leftrightarrow-p_1-p_2-p_3-\ell)$}  \\	\hline
		\makecell*[c]{$N_3^{(1)}$}	&\makecell*[c]{\includegraphics[scale=0.25]{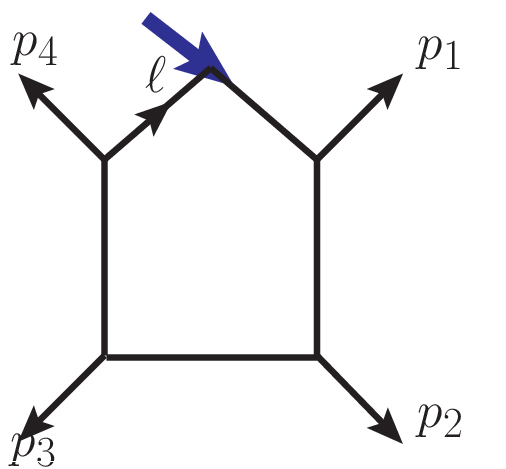}}&\makecell*[c]{${\rm tr}_{-}(4,1,2,3) \mu_{\ell\ell} $}\\ \hline
	\end{tabular}
	\caption{
		{The topologies and numerators of the one-loop form factor, where the integer $i$ in ${\rm tr}_-$ denotes the external momentum $p_i$, etc., in this paper.}
		\label{tab:oneloopintegrand}
	}
\end{table}
	
We comment here that the box integrands $\mathbb{I}_1^{(1)}$ and $\mathbb{I}_2^{(1)}$ are equivalent to the results from four-dimensional cuts in \cite{Brandhuber:2010ad}.
Using $D$-dimensional cuts, we also obtain the new pentagon integrand $\mathbb{I}_3^{(1)}$, which is proportional to the extra-dimensional component $\mu_{\ell\ell}$.
We note that the contribution of $\mathbb{I}_3^{(1)}$ starts at ${\cal O}(\epsilon^1)$ order, but it will provide non-trivial two-loop IR subtraction terms as will be discussed in Section~\ref{sec:subtraction}.

\subsection{$D$-dimensional unitarity cuts}

\begin{figure}[tb]
	\centering
	\subfigure[$q^2$-cut]{\includegraphics[scale=0.8]{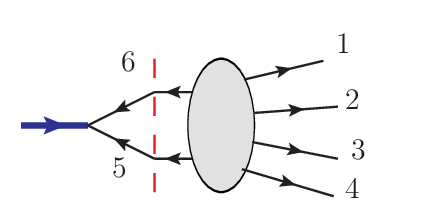} \label{fig:oneLoopFourptcut}}
	\subfigure[$s_{12}$-cut]{\includegraphics[scale=0.8]{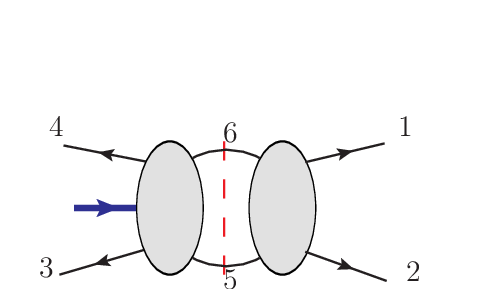} \label{fig:oneLoopFourptcut1}}
	\caption{$q^2$-cut and $s_{12}$-cut channels for the one-loop four-point form factor of ${\rm tr}(\phi_{12}^2)$.}
	
	\end{figure}

The unitarity cut in the four-dimension has been discussed in detail in \cite{Brandhuber:2010ad}. Here we mainly focus on the $D$-dimensional cuts, which lead to the new contribution $\mathbb{I}_3^{(1)}$. We discuss two kinds of cuts shown in Figure~\ref{fig:oneLoopFourptcut} and Figure~\ref{fig:oneLoopFourptcut1} as examples.

\paragraph{$q^2$-cut.}

The $q^2$-cut channel shown in Figure~\ref{fig:oneLoopFourptcut} with external states as $(1^+,2^+,3^{\phi_{12}},4^{\phi_{12}})$ is given by\footnote{We use the abbreviation $i^+$ to denote $i^{g_+}$, which carries the positive helicity. }
\begin{equation}
	{\cal F}^{(1)}_4 \big|_{\text{$q^2$-cut}} = {\cal F}_2^{ (0)}(\bar{6}^{\phi_{12}}, \bar{5}^{\phi_{12}}) \,{\cal A}_6^{ (0)}(1^+, 2^+, 3^{\phi_{12}}, 4^{\phi_{12}}, 5^{\phi_{34}}, 6^{\phi_{34}}) \,,
\end{equation}
where $\bar{5}$ denotes $-p_5$, etc., and ${\cal F}_2^{ (0)}(\bar{6}^{\phi_{12}}, \bar{5}^{\phi_{12}})$ equals to $-1$. The $D$-dimensional information of the remaining tree-level six-point amplitude is important for the $D$-dimensional unitarity cut. Therefore, we obtain the amplitude by the Feynman diagrams as
\begin{align}
	\label{formula:sixpointamp}
	& {\cal A}_6^{(0)}(1^+, 2^+, 3^{\phi_{12}}, 4^{\phi_{12}}, 5^{\phi_{34}}, 6^{\phi_{34}}) = 2 \Big[ \frac{(\epsilon_1\cdot p_2) (\epsilon_2\cdot p_4)-(\epsilon_1\cdot p_4) (\epsilon_2\cdot p_1)}{s_{12} s_{45}} -\frac{(\epsilon_1\cdot p_6) (\epsilon_2\cdot p_4)}{s_{16} s_{45}} \notag \\
	& \hskip 2cm -s_{34}\frac{ (\epsilon_1\cdot p_6) (\epsilon_2\cdot p_1)-(\epsilon_1\cdot p_2) (\epsilon_2\cdot p_6)}{s_{12} s_{45} s_{126}}  -\frac{s_{34} (\epsilon_1\cdot p_6) (\epsilon_2\cdot (p_1+p_6))}{s_{16} s_{45} s_{126}} \Big] \,,
\end{align}
where $p_5$ and $p_6$ should be understood as $D$-dimensional loop momenta.

Now we show that the $\mu$-term contribution arises only from the term $(\epsilon_1\cdot p_6)(\epsilon_2\cdot p_6)$, which appears in the last quadratic term of the $D$-dimensional loop momentum $p_6$ on the RHS of \eqref{formula:sixpointamp}.

First, we can specify the polarization vectors of the external gluons using four-dimensional spinor helicity as
\begin{align}
	\label{eq:gaugechoice}
	\epsilon_{1,\mu}^+ = \frac{\langle3|\gamma_\mu|1\rbrack}{\sqrt{2}\langle13\rangle}\,, \qquad \epsilon_{2,\mu}^+=\frac{\langle3|\gamma_\mu|2\rbrack}{\sqrt{2}\langle23\rangle} \,,
\end{align}
where $\mu \in \{0,1,2,3\}$, and the extra-dimensional components of $\epsilon_{1}^+$ is treated as $0$, etc. Then we can factorize the tree-level form factor and transform the term to a four-dimensional gamma trace variable as
\begin{align}
	\label{eq:mutermorigin}
	(\epsilon_1 \cdot p_6) (\epsilon_2 \cdot p_6) = \frac{\langle 34\rangle}{\langle12\rangle \langle14\rangle \langle23\rangle} \frac{{\rm tr}_{-}(4, 1, 3, 6, 2, 1, 6, 3) }{2 s_{13} s_{34}}  \,,
\end{align}
where the momenta in ${\rm tr}_-$ expression is contracted only for their four-dimensional part.
The length-$8$ single gamma trace will produce a $\mu$-term variable if we expand it to lower length ones as
\begin{align}
	{\rm tr}_{-}(4,1,3,6,2,1,6,3) = & {\rm tr}_{-}(4,1,2,3) s_{13} \, p_{6, \text{4d}}^2 +{\rm tr}_{-}(4,1,6,3,1,3)s_{26} \\
	& +{\rm tr}_{-}(4,1,3,6,2,3)s_{16} \,, \notag
\end{align}
where $p_{6,\text{4d}}^2$ is equal to $\mu_{66}$ due to the $p_6$ is a $D$-dimensional on-shell momentum. In other words, $p_6^2 = p_{6,\textrm{4d}}^2 -\mu_{66} = 0$, as mentioned in Section~\ref{sec:review}.
We note here that this $\mu$-term contribution corresponds precisely to the numerator $N_3^{(1)}$ in Table~\ref{tab:oneloopintegrand}.

\paragraph{$s_{12}$-cut channel.}
The $q^2$-cut channel involves only the cut scalar states. A more complicated case is the $s_{12}$-cut channel shown in Figure~\ref{fig:oneLoopFourptcut1}, which contains gluonic and fermionic contributions and provides additional checks for our result.
We consider the external states as $(1^+,2^{\phi_{12}},3^{\phi_{12}},4^+)$, then the cut result is
\begin{align}
	{\cal F}_4^{(1)} \big|_{\text{$s_{12}$-cut}} = \sum_{\{\Phi_5, \Phi_6\}} {\cal F}_4^{ (0)}(3^{\phi_{12}}, 4^{+}, \bar{6}^{\Phi_{\bar{6}}}, \bar{5}^{\Phi_{\bar{5}}}) {\cal A}_4^{ (0)}(1^+, 2^{\phi_{12}}, 5^{\Phi_5}, 6^{\Phi_6}) \,,
\end{align}
where the subscript $\Phi_i$ denotes the physical state associated with $p_i$. The possible physical states $(\Phi_5, \Phi_6)$ are $(g^-, \phi_{34})$, $(\phi_{34}, g^-)$, $(\psi_{134}, \psi_{234})$, and $(\psi_{234}, \psi_{134})$.
The most complicated terms are $(\psi_{134}, \psi_{234})$ and $(\psi_{234}, \psi_{134})$ that involve the fermionic tree-level results.
In this case, the two tree-level building blocks can be also computed by Feynman diagrams to capture the $D$-dimensional information, as
\begin{align}
	{\cal F}_4^{(0)}(3^{\phi_{12}}, 4^{+}, \bar{6}^{\psi_2}, \bar{5}^{\psi_1}) = & \frac{\bar{u}_s(-p_6) \epsilon\mkern -8 mu \slash _4 (p\mkern -8 mu \slash _6- p\mkern -8 mu \slash_4) v_s(p_5)}{\sqrt{2} (p_6-p_4)^2 (-p_4+p_5+p_6)^2}+\frac{\sqrt{2} (p_3 \cdot \epsilon_4 ) \bar{u}_s(-p_6) v_s(p_5)}{(p_3+p_4)^2 (p_5+p_6)^2} \notag \\
	& +\frac{\sqrt{2} (p_5+p_6) \cdot \epsilon_4 \bar{u}_s(-p_6) v_s(p_5)}{(p_5+p_6)^2 (-p_4+p_5+p_6)^2}\,, \notag \\
	{\cal A}_4^{(0)}(1^+, 2^{\phi_{12}}, 5^{\psi_{234}}, 6^{\psi_{134}}) = & \frac{\bar{v}_s(-p_5) p\mkern -8 mu \slash_{16} \epsilon\mkern -8 mu \slash _1 u_s(p_6)}{\sqrt{2} (p_1 +p_6)^2} -\frac{\sqrt{2} (p_2\cdot \epsilon_1) \bar{v}_s(-p_5) u_s(p_6)}{(p_5 +p_6)^2} \,, 
\end{align}
where $p\mkern -8 mu \slash _6 = p_6^M \Gamma_M$ is contracted with the $D$-dimensional gamma matrix, etc., and $u_s$ and $\bar{u}_s$ are the spinors of the external momenta. Then, by summing over all spin configurations as \eqref{formula:sumfermion}, we obtain
\begin{align}
	& \sum_{\psi} {\cal F}_4^{(0)}(3^{\phi_{12}}, 4^{+}, \bar{6}^{\psi_2}, \bar{5}^{\psi_1}) {\cal A}_4^{(0)}(1^+,2^{\phi_{12}},5^{\psi_{234}},6^{\psi_{134}}) \notag\\
	= & 2\Big[ \frac{\text{\bf tr}(p_6,p_5) (p_2\cdot\epsilon _1) (\epsilon_4\cdot p_3)}{s_{34} s_{56}^2} +\frac{\text{\bf tr}(p_6,p_5) (p_2\cdot\epsilon_1)  (\epsilon_4\cdot p_{56})}{s_{56}^2 (-p_4+p_5+p_6)^2} \Big] \notag \\
	& +\frac{ (p_2\cdot\epsilon _1) \text{\bf tr}(p_6,p_5,p_6-p_4,\epsilon_4)}{(p_6-p_4)^2 s_{56} (-p_4+p_5+p_6)^2}-\frac{(\epsilon_4\cdot p_3) \text{\bf tr}(p_6,\epsilon_1,p_{16},p_5)}{s_{34} s_{16} s_{56}}\notag\\
	& -\frac{(\epsilon _4\cdot p_{56}) \text{\bf tr}(p_6,\epsilon_1, p_{16},p_5)}{s_{16} s_{56} (-p_4+p_5+p_6)^2} +\frac{\text{\bf tr}(p_6,\epsilon_1,p_{16},p_5,p_4-p_6,\epsilon_4)}{2 s_{16} (p_6-p_4)^2 (-p_4+p_5+p_6)^2} \,,
\end{align}
where $p_{56}$ denotes $p_5 +p_6$, and the $D$-dimensional gamma traces ${\bf tr}$ defined in \eqref{formula:traceiD} keep the $D$-dimensional information of the loop momenta, as we discussed in Section~\ref{sec:review}.
We mention here that it is practically convenient to compare them with the cut integrand numerically.\footnote{
We give a brief comment on the numerical method for $D$-dimensional kinematics.
Any massless momentum can be expressed as $\ell = (\ell^{0}, \ell^{0} \hat{n}_{D-1})$,
where $\ell^{0}$ is the zeroth component, and $\hat{n}_{D-1} = (\cos\theta_1\,,\ \ldots \,,\ \sin\theta_1 \cdots \sin\theta_{D-2})$ is a $(D-1)$-dimensional unit vector. We consider the $D > 4$ to capture the information beyond the four-dimension. This form takes the advantage that the momentum conservation equations and on-shell conditions are linear for $\ell^0$ (except the three-point case). Then it is convenient to apply the numerical evaluation with specific dimensions such as $D=6$, which is enough for the two-loop four-point case.}

\section{Two-loop integrand via unitarity}
\label{sec:2lUnitarity}

In this section, we consider the two-loop four-point planar MHV form factor. As in the one-loop case, we first give the full results of the integrand and then 
provide some details for the related unitarity cuts.

\subsection{Summary of the two-loop integrand}\label{subsec:2loopsec1}

\begin{figure}[t]
\centering
\subfigure[$\mathbb{I}_{1}$]{\includegraphics[scale=0.36]{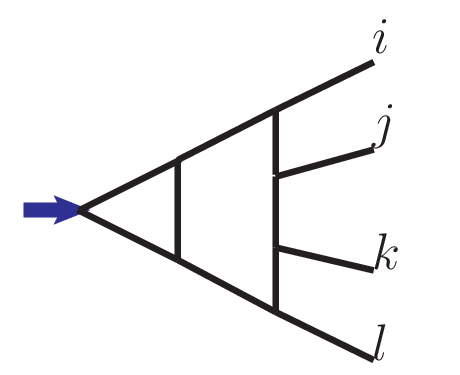}\label{Fig:Twodoublecutstop1}} \qquad
\subfigure[$\mathbb{I}_2$]{\includegraphics[scale=0.35]{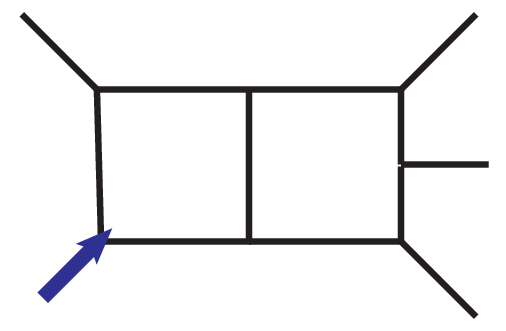}\label{Fig:Twodoublecutstop6}} \qquad
\subfigure[$\mathbb{I}_{3}$]{\includegraphics[scale=0.35]{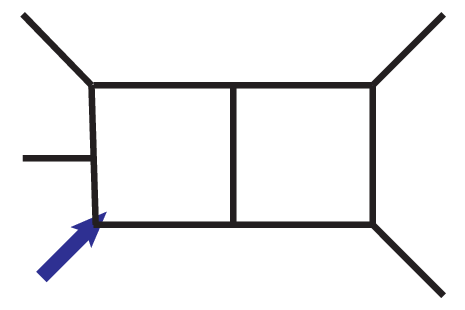}\label{Fig:Twodoublecutstop4}}

\subfigure[$\mathbb{I}_{4}$]{\includegraphics[scale=0.34]{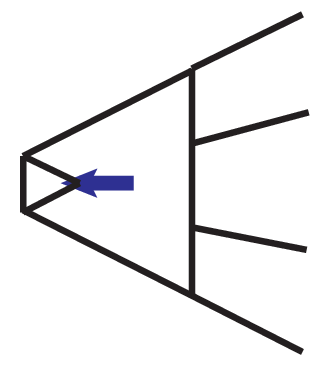}\label{Fig:Twodoublecutstop2}} \qquad\quad
\subfigure[$\mathbb{I}_5$]{\includegraphics[scale=0.35]{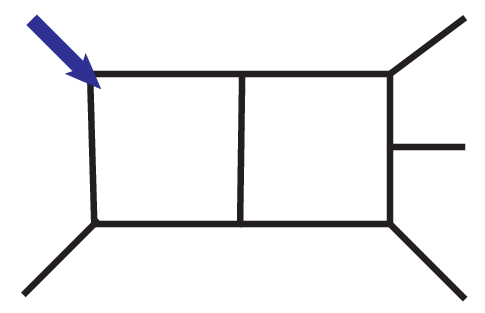}\label{Fig:Twodoublecutstop7}} \qquad
\subfigure[$\mathbb{I}_{6}$]{\includegraphics[scale=0.35]{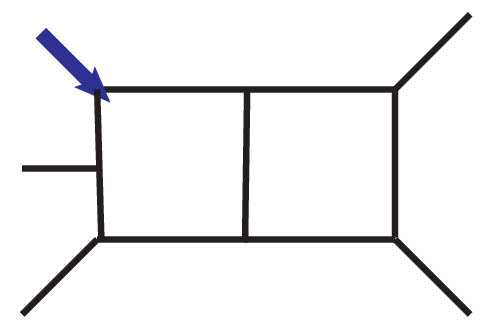}\label{Fig:Twodoublecutstop5}}

\subfigure[$\mathbb{I}_{7}$]{\includegraphics[scale=0.35]{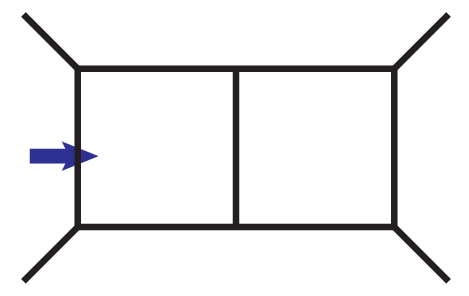}\label{Fig:Twodoublecutstop3}} \qquad
\subfigure[$\mathbb{I}_{8}$]{\includegraphics[scale=0.35]{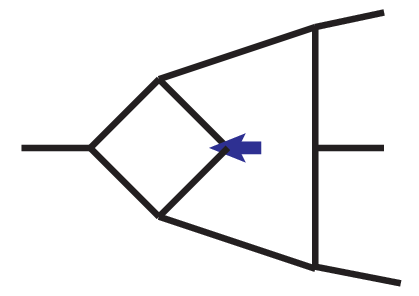}\label{Fig:Twodoublecutstop8}} \qquad
\subfigure[$\mathbb{I}_{9}$]{\includegraphics[scale=0.35]{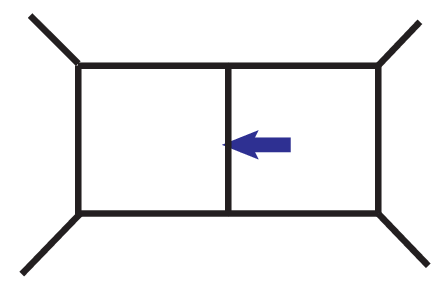}\label{Fig:Twodoublecutstop9}}

\caption{Topologies of nine planar integrals for two-loop four-point form factor.}
\label{Fig:TwoLoopFourptwithLength2}
\end{figure}

The two-loop form factor can be given as
\begin{equation}
	{\cal F}^{(2)}_4={\cal F}^{(0)}_4 {\cal I}^{(2)}_4 \,,
\end{equation}
where the loop correction ${\cal I}_4^{(2)}$ is given by summing all nine two-loop integrands $\{\mathbb{I}_{i}\}$ related to the topologies shown in Figure~\ref{Fig:TwoLoopFourptwithLength2}, which are defined by\footnote{We abbreviate $\mathbb{I}^{(2)}_{i}$ as $\mathbb{I}_{i}$ for simplicity. The topologies are all maximal topologies with eight propagators.}
\begin{align}
	\mathbb{I}_{i} = \frac{N_i}{\prod_{\alpha=1}^{8} D_{i,\alpha} }\,,
\end{align}
where the subscript $i$ denotes the $i$-th topology. More explicitly, ${\cal I}_4^{(2)}$ is
\begin{align}
	\label{eq:calI2loop}
	{\cal I}^{(2)}_4 = e^{2\epsilon \gamma_\mathrm{E}} \int \frac{d^D l_a}{i \pi^{\frac{D}{2}}} \frac{d^D l_b}{i\pi^{\frac{D}{2}}} \left( \sum_{i=1}^{8} \mathbb{I}_i +{1\over2}\mathbb{I}_9 +{\rm cyclic}(p_1, p_2, p_3, p_4) \right) \,,
\end{align}
where the factor $1/2$ of $\mathbb{I}_9$ comes from the symmetry of the ninth topology in Figure~\ref{Fig:TwoLoopFourptwithLength2}.

We provide the explicit expressions of numerators $N_i$ for each integrand in Table~\ref{tab:integrand}. Note that $N_4$ and $N_1$ are equal,\footnote{It is important to distinguish them in the unitarity cuts at the integrand level, since they give different cut integrands.}
and $N_5$ (and $N_6$) can be obtained from $N_2$ (and $N_3$) by flipping momenta.

\begin{center}
	\begin{longtable}{|c|c|c|}
	
		\hline \makecell*[c]{$N_1$}	&\makecell*[c]{\includegraphics[scale=0.3]{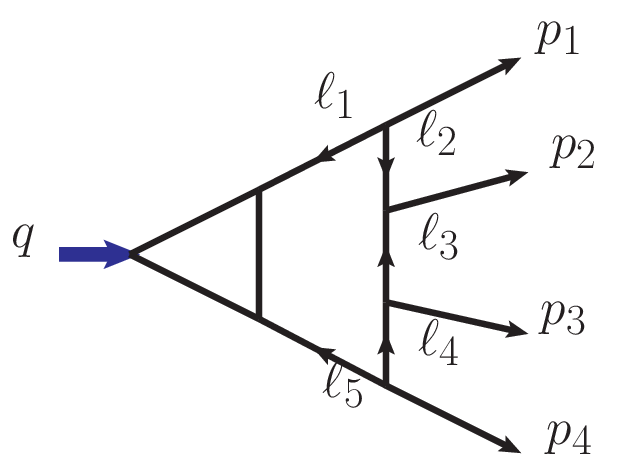}}&\makecell*[c]{$\Big[\frac{1}{2} q^2  \ell_3^2 [\ell_1^2 (s_{234}-s_{23} )  +s_{14} \ell_2^2 ]+ n_1(1,2,3,4,\ell_3)\Big]$  \\ 
		$+(p_1 \leftrightarrow p_4, p_2 \leftrightarrow p_3,\ell_1\leftrightarrow\ell_5,\ell_2\leftrightarrow\ell_4,\ell_3\rightarrow-\ell_3)$}\\ \hline
		\makecell*[c]{$N_2$}	&\makecell*[c]{\includegraphics[scale=0.3]{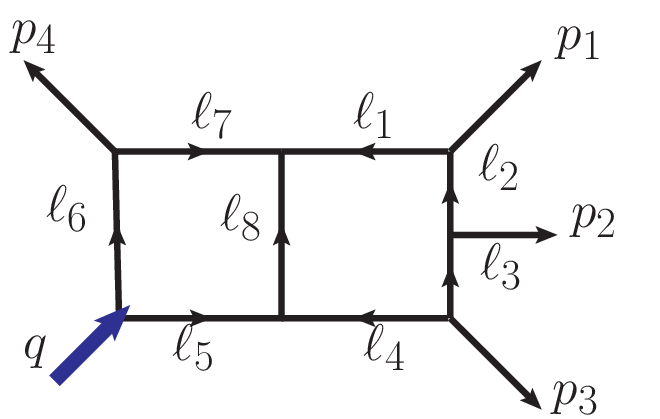}} & \makecell*[c]{ $	n_2(1,2,3,4,\ell_1,\ell_7)+\ell_1^2 n_{3}+\ell_4^2 n_{4}+\ell_6^2 n_{5}$\\$+\ell_5^2 n_{6}-\frac{1}{2} \left(s_{123}-s_{12}\right) q^2 \ell_7^2 \ell_2^2$}  \\	\hline
		\makecell*[c]{$N_3$}	&\makecell*[c]{\includegraphics[scale=0.25]{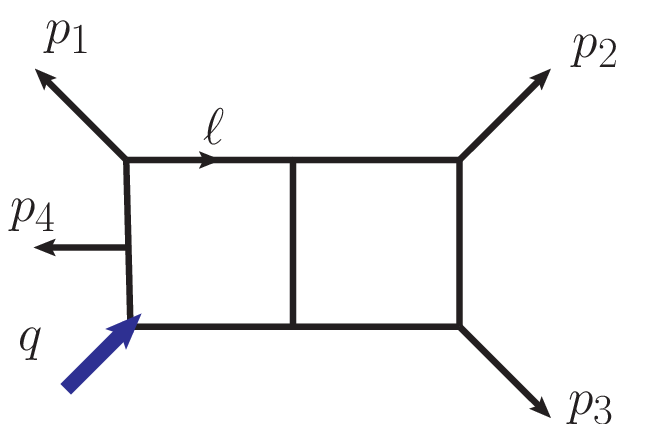}}&\makecell*[c]{$-s_{23} \mu _{\ell\ell}  \text{tr}_-(2,1,4,3)$}\\ \hline
		\makecell*[c]{$N_7$}	&\makecell*[c]{\includegraphics[scale=0.25]{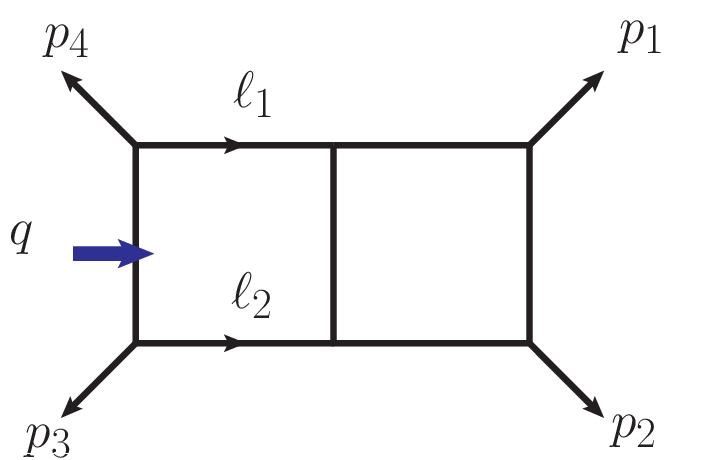}}&\makecell*[c]{$\frac{1}{2} s_{12}
		\text{tr}_-(4,1,2,3,\ell_2,\ell_1) +(p_1 \leftrightarrow p_2, p_3 \leftrightarrow p_4,\ell_1\leftrightarrow\ell_2)$}\\ \hline
		\makecell*[c]{$N_8$}&	\makecell*[c]{\includegraphics[scale=0.3]{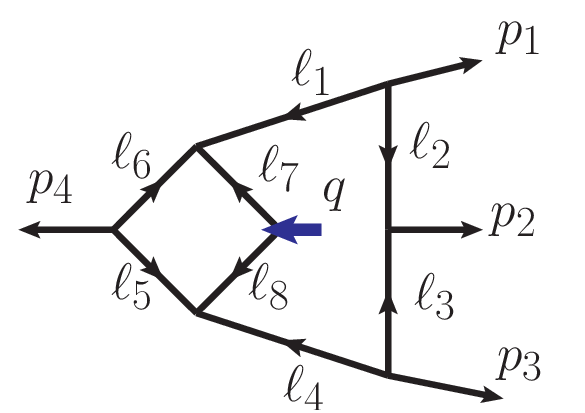}}&\makecell*[c]{$n_7+(p_1 \leftrightarrow p_3,\ell_1\leftrightarrow \ell_4,\ell_2\leftrightarrow \ell_3,\ell_5\leftrightarrow \ell_6,\ell_7\leftrightarrow \ell_8)$}\\ \hline
		\makecell*[c]{$N_9$}	&\makecell*[c]{\includegraphics[scale=0.3]{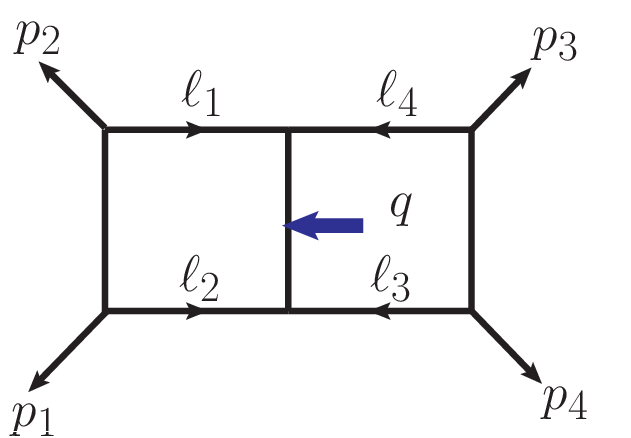}}&\makecell*[c]{
		$\Big\{\text{tr}_-(4,1,2,3)\Big[\big[(\ell_1\cdot \ell_2)(\ell_3\cdot \ell_4)+(\ell_1\cdot \ell_4)(\ell_2\cdot \ell_3)-(\ell_1\cdot \ell_3)(\ell_2\cdot \ell_4)\big]$\\$+i \varepsilon(\ell_1 \ell_2 \ell_3 \ell_4)-\mu _{\ell_1\ell_1} s_{34}+\mu _{\ell_1\ell_4} ({1\over2}q^2-s_{12} )\Big] -\ell_1^2 \ell_3^2\text{ tr}(1,2,3,4)/2\Big\}$\\$+(p_1 \leftrightarrow p_4, p_2 \leftrightarrow p_3, \ell_1\leftrightarrow \ell_4,\ell_2\leftrightarrow\ell_3)$
		}\\ \hline
		\caption{The topologies and numerators for the two-loop form factor.\label{tab:integrand}} 
	\end{longtable}
\end{center}

The functions $n_i$ for $i=1,\ldots,7$ introduced in Table~\ref{tab:integrand} are defined as follows:
\begin{align}
	\label{eq:nfun}
	n_1
	= & \frac{1}{2} q^2 \left[ \ell_3^2 \, \text{tr}_-(1,4,p_2+p_3,\ell_3-p_2)-\mu _{\ell_3\ell_3} \text{tr}_-(4,1,2,3) \right] , \\
	n_2
	= & s_{123} \left[ s_{12} \text{tr}_-(1,\ell_7,4,3)-s_{12\ell_1} \text{tr}_-(1,\ell_7,4,3)+\text{tr}_-(1,\ell_7,4,3,\ell_1,2) \right] \notag \\
	& -\ell_7^2 \left[ s_{34} \text{tr}_-(1,3,p_{1}+\ell_1,2)-s_{12\ell_1} \text{tr}_-(1,p_{23},4,3)+s_{12} \text{tr}_-(4,3,p_{1}+\ell_{1},2) \right] \,, \notag \\
	n_{3}= & -s_{23} \text{tr}_-(p_{12},\ell_7,4,3)-s_{34}\left( \text{tr}_-(2,4,\ell_7,3) +\text{tr}_-(2,\ell_7,4,3) \right) \,, \notag \\
	n_{4}= & - s_{12} \left[
	\text{tr}_-\left(1,\ell_7,4,p_{23}\right)
	+\frac{1}{2}\ell_7^2  (s_{12} -s_{34}) \right] \,, \notag \\
	n_{5} = & \frac{1}{2} \Big[\ell_2^2 s_{123} \left(s_{123}-s_{12}\right)- s_{123} \text{tr}_-\left(1,2,3,\ell_1\right)- \ell_1^2 s_{23} \left(s_{123}+s_{124}-q^2\right) \notag \\
	&+ \ell_4^2 s_{12} \left(s_{14}-s_{23}+s_{123}\right)\Big] \,, \notag \\
	n_{6}= & -\frac{s_{23} s_{24}}{2}\ell_1^2 -\frac{1}{2} s_{123} \text{tr}_-(1,2,3,4)+\frac{1}{2} \ell_2^2 \left(s_{123} s_{124}-s_{12} q^2\right)
	- \frac{1}{2} \text{tr}_-(1,2,4,3)\ell_4^2 \notag \\
	& +\frac{1}{2} \left(s_{123}-q^2\right) \text{tr}_-\left(1,2,3,\ell_1\right)+\frac{1}{2} s_{23} \text{tr}_-\left(1,\ell_1,4,3\right)+\frac{1}{2} \left(s_{23}-s_{123}\right) \text{tr}_-\left(2,\ell_1,4,3\right) \,, \notag
\end{align}
\begin{align}
	n_{7} = & \ell_1^2\Big\{-\frac{1}{2}s_{14} s_{23}\ell_8^2
	+\frac{1}{2}\ell_7^2  s_{23} \left(s_{14}+2 s_{124}-q^2\right)+\frac{1}{2} \ell_3^2 \left(s_{24}+s_{34}\right) q^2 \notag \\
	&+\ell_4^2 \left[\frac{1}{4} s_{24} \left(s_{24}-2 s_{124}\right)-\frac{1}{2} \text{tr}_-(1,2,4,3)\right]-s_{23} \text{tr}_-\left(1,4,\ell_7,3\right)-\frac{1}{2} q^2 \text{tr}_-\left(2,3,4,\ell_1\right)\notag\\&-s_{23} \text{tr}_-\left(2,4,\ell_7,3\right)+\frac{1}{2} q^2 \Big(\text{tr}_-(1,4,2,3)+s_{23} s_{24}\Big)+\frac{1}{2} s_{23} \left(q^2-2 s_{124}\right) \left(\ell_7+p_{4}\right){}^2\Big\} \notag \\
	&+{1\over2}n_1\left(4,1,2,3,\ell_1+p_1\right)+{1\over2}n_1\left(3,2,1,4,-\ell_1-p_1\right)-{1\over2}n_2\left(1,2,3,4,\ell_1,\ell_7-p_4\right)\notag\\&+\frac{1}{4} q^2 \Big[\ell_2^2 (\ell_3^2-\ell_1^2) s_{34}+\ell_2^2 \ell_6^2 \left(s_{12}-s_{123}\right)+\ell_5^2 \Big(\ell_3^2 \left(s_{123}-s_{23}\right)+2 \text{tr}_-\left(3,2,1,\ell_3\right)\Big)\Big] \,. \notag
\end{align}

\paragraph{Comment on the $\mu$-term contribution.}

Our results contain the full $D$-dimensional information that is characterized by the $\mu$-term contribution.
For example, the length-6 trace variable ${\rm tr}_-(1, \ell_7, 4, 3, \ell_1, 2)$ in $n_2$ given in \eqref{eq:nfun}, it can be expanded to lower length ones as
\begin{align}
	\label{eq:hightracereduction}
	{\rm tr}_-(1,\ell_7,4,3,\ell_1,2) = & (\ell_7\cdot\ell_1)_\text{4d} {\rm tr}_- (4,1,2,3) +\cdots \\
	= & (\ell_7\cdot\ell_1 +\mu_{\ell_1\ell_7}) {\rm tr}_- (4,1,2,3) +\cdots \,, \nonumber
\end{align}
where the four-dimensional Lorentz product $(\ell_7\cdot\ell_1)_{\rm 4d}$ leads to the $\mu$-term variable $\mu_{\ell_1\ell_7}$, if we avoid the four-dimensional Lorentz product in the finite result.
Such $\mu$-term contribution is not sufficient to be detected by The four-dimensional unitarity cuts.

We comment that there is an ambiguity for the choice of $\mu$-term contribution.
First, the ambiguity for the even part may appear if the power of numerators is at least ten. 
For example, it can come from the Gram determinant as
\begin{equation}
	{\rm Gram}(\{\ell_i, p_1, p_2, p_3, p_4\}, \{\ell_j, p_1, p_2, p_3, p_4\}) = -\mu_{ij} {\rm Gram}(p_1, p_2, p_3, p_4) \,.
\end{equation}
Since the power of our numerators is eight, there is no ambiguity for the parity-even part of $\mu$-term contribution in our case.
On the other hand, there is still an ambiguity in the parity-odd part of the $\mu$-term contribution.
For example, $\mu_{ij}$ can arise from a combination without $\mu_{ij}$ as\footnote{Noticing $\varepsilon(1234) = {\rm tr}_-(1,2,3,4) -{\rm tr}_-(4,1,2,3)$, etc., therefore it is possible to add expressions of ${\rm tr}_-$ to modify the parity-odd part of the $\mu$-term contribution.}
\begin{align}
	\label{eq:ambiguity}
	\mu_{ij} \varepsilon(1234) = & (p_1 \cdot \ell_i)\varepsilon(\ell_j234) +(p_2 \cdot \ell_i) \varepsilon(1\ell_j34) +(p_3\cdot\ell_i)\varepsilon(12\ell_j4) \\
	& +(p_4\cdot\ell_i) \varepsilon(123\ell_j) -(\ell_i \cdot \ell_j)\varepsilon(1234) \,. \notag
\end{align}
In the Section~\ref{sec:DDCIintegrand}, we will provide a choice of the decomposition for the two-loop integrand with $\mu$ terms explicitly given when discussing the dual conformal symmetry.

\subsection{More details in the unitarity cuts}

We give more details both for the four- and $D$-dimensional unitarity cuts. 

\begin{figure}[t]
\centering
\subfigure[]{\includegraphics[scale=0.7]{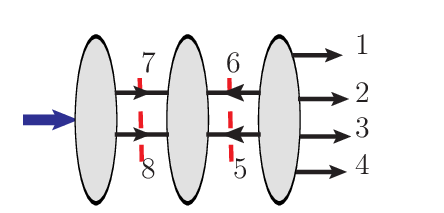}\label{Fig:TwodoublecutsinLength2cut1}}
\subfigure[]{\includegraphics[scale=0.7]{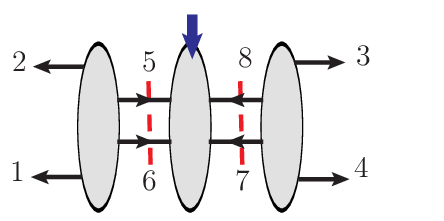}\label{Fig:TwodoublecutsinLength2cut9}}

\subfigure[]{\includegraphics[scale=0.7]{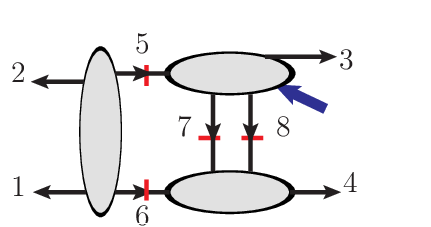}\label{Fig:Ddimcut2}}
\subfigure[]{\includegraphics[scale=0.65]{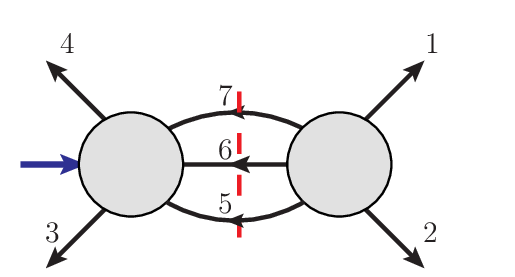}\label{Fig:triplecut}}
\caption{The unitarity cuts as illustrations for the two-loop four-point form factor of $\text{tr}(\phi_{12}^2)$.\label{Fig:TwodoublecutsFirst}}
\end{figure}

\paragraph{Four-dimensional cuts.}
The four-dimensional cuts can determine most of the numerators $N_1$ and $N_9$ in Table~\ref{tab:integrand}. We give the two examples in Figure~\ref{Fig:TwodoublecutsFirst}.

The two-double-cut shown in Figure~\ref{Fig:TwodoublecutsinLength2cut1} corresponds to
\begin{align}
	\label{formula:cut1}
	{\cal F}_4^{(2)} \big|_{\text{cut-(a)}} = & \int \prod_{a=5}^8 {\rm d}^4\eta_a \big( {\cal A}_6^{\text{MHV},(0)}(1,2,3,4,5,6) {\cal A}_4^{\text{MHV},(0)}(\bar{6}, \bar{5}, \bar{8}, \bar{7}) {\cal F}_2^{\text{MHV},(0)}(7,8) \big) \notag \\
	= & {\cal F}_4^{(0)}(1^{+}, 2^{+}, 3^{\phi_{12}}, 4^{\phi_{12}}) \frac{\langle 41\rangle  \langle 56\rangle ^2 \langle 78\rangle }{\langle 54\rangle  \langle 58\rangle  \langle 61\rangle  \langle 76\rangle } \notag \\
	= & {\cal F}_4^{(0)}(1^{+}, 2^{+}, 3^{\phi_{12}}, 4^{\phi_{12}}) \frac{q^2\text{tr}_-(p_1,p_4,p_{123},p_6) }{s_{54}s_{16}  s_{58}} \,.
\end{align}
This unitarity cut has only contribution from $N_1$, since only the topology in Figure~\ref{Fig:Twodoublecutstop1} contributes to this cut.\footnote{
The integrand based on four-dimensional cuts has been considered in \cite{Gopalka:2023doh}. There seems no contribution of topology associated with $N_1$ in the result of \cite{Gopalka:2023doh}, which would be inconsistent with the cut of Figure~\ref{Fig:TwodoublecutsinLength2cut1} even in the four-dimension. The $N_1$ part has the non-trivial contribution in the IBP reduced form and to the remainder function as will be considered in the next section.
}

Another two-double-cut shown in Figure~\ref{Fig:TwodoublecutsinLength2cut9} is
\begin{align}
	{\cal F}^{(2)}_4 \big|_{\text{cut-(b)}} = & \int \prod_{a=5}^8 {\rm d}^4\eta_a \big( {\cal A}_4^{\text{MHV},(0)}(1,2,5,6) {\cal A}_4^{\text{MHV},(0)}(3,4,7,8) {\cal F}_4^{\text{MHV},(0)}(\bar{6}, \bar{5}, \bar{8}, \bar{7}) \big) \notag \\
	= & {\cal F}_4^{(0)}(1^{+}, 2^{+}, 3^{\phi_{12}}, 4^{\phi_{12}}) \frac{\langle 14\rangle  \langle 23\rangle  \langle 56\rangle ^2 \langle 78\rangle ^2}{\langle 16\rangle  \langle 25\rangle  \langle 38\rangle  \langle 47\rangle  \langle 58\rangle  \langle 67\rangle } \notag \\
	= & {\cal F}_4^{(0)}(1^{+}, 2^{+}, 3^{\phi_{12}}, 4^{\phi_{12}}) \frac{\text{tr}_-(2,3,4,1)\text{tr}_-(5,6,7,8)}{s_{16}s_{38}s_{58}s_{67}} \,.
\end{align}
This unitarity cut has only contribution from $N_9$ since only the topology in Figure~\ref{Fig:Twodoublecutstop9} contributes to this cut.

To fix the remaining $\mu$-term contributions, we need to consider the $D$-dimensional cuts.

\paragraph{$D$-dimensional cuts.}

The $D$-dimensional version of the unitarity cut in Figure~\ref{Fig:TwodoublecutsinLength2cut1} with external states as $(1^+,2^+,3^{\phi_{12}},4^{\phi_{12}})$  is 
\begin{align}
	{\cal F}^{(2)}_4 \big|_{\text{cut-(a)}} & = \sum_{\{\Phi_i\}} {\cal A}_6^{(0)}(1^{+}, 2^{+}, 3^{\phi_{12}}, 4^{\phi_{12}}, 5^{\Phi_5}, 6^{\Phi_6}) {\cal A}_4^{(0)}(\bar{6}^{\Phi_{\bar{6}}}, \bar{5}^{\Phi_{\bar{5}}}, \bar{8}^{\Phi_{\bar{8}}}, \bar{7}^{\Phi_{\bar{7}}}) {\cal F}_2^{(0)}(7^{\Phi_7}, 8^{\Phi_8})  \notag \\
	& = {\cal A}_6^{(0)}(1^{+}, 2^{+}, 3^{\phi_{12}}, 4^{\phi_{12}}, 5^{\phi_{34}}, 6^{\phi_{34}}) {\cal A}_4^{(0)}(\bar{6}^{\phi_{12}}, \bar{5}^{\phi_{12}}, \bar{8}^{\phi_{34}}, \bar{7}^{\phi_{34}}) {\cal F}_2^{(0)}(7^{\phi_{12}}, 8^{\phi_{12}}) .
\end{align}
This cut is similar to the one-loop $q^2$-cut discussed in Section~\ref{sec:oneloopFF} and determines the $D$-dimensional ambiguity of $N_{1}$.

The $D$-dimensional unitarity cut as shown as Figure~\ref{Fig:TwodoublecutsinLength2cut9} with external states as $(1^+,2^+,3^{\phi_{12}},4^{\phi_{12}})$  involves the two-gluons cut states as
\begin{align}
{\cal F}^{(2)}_4 \big|_{\text{cut-(b)}} = &\sum_{\{\Phi_i\}} {\cal A}_4^{(0)}(1^{+}, 2^{+}, 5^{\Phi_5}, 6^{\Phi_6}) {\cal A}_4^{(0)}(3^{\phi_{12}}, 4^{\phi_{12}}, \bar{7}^{\Phi_{\bar{7}}},\bar{8}^{\Phi_{\bar{8}}}) {\cal F}_2^{(0)}(\bar{6}^{\Phi_{\bar{6}}}, \bar{5}^{\Phi_{\bar{5}}}, 8^{\Phi_8},7^{\Phi_7} ) \notag\\
=& {\cal A}_4^{(0)}(1^{+}, 2^{+}, 5^{-}, 6^{-}) {\cal A}_4^{(0)}(3^{\phi_{12}}, 4^{\phi_{12}}, \bar{7}^{\phi_{34}},\bar{8}^{\phi_{34}}) {\cal F}_2^{(0)}(\bar{6}^{+}, \bar{5}^{+}, 8^{\phi_{12}},7^{\phi_{12}} )\,.
\end{align}
Note that the external states of the cut legs for $p_5$ and $p_6$ must be gluon states $g^-$, and then the cut legs for $p_7$ and $p_8$ can only be scalar states $\phi_{12}$.

Another two-double-cut example similar to the one-loop case is shown in Figure~\ref{Fig:Ddimcut2} with external states as $(1^{\phi_{12}},2^{\phi_{12}},3^+,4^+)$ 
\begin{equation}
	\label{eq:cofigurationfermion}
	{\cal F}_4^{(2)} \big|_{\text{cut-(c)}} =  {\cal A}_4^{(0)}(1^{\phi_{12}}, 2^{\phi_{12}}, 5^{\phi_{34}}, 6^{\phi_{34}}) 
 \sum_{\{\Phi_{i}\}} {\cal F}_4^{(0)}(\bar{6}^{\phi_{12}}, 3^{+}, 7^{\Phi_7}, 8^{\Phi_8}) {\cal A}_4^{(0)}(4^+, \bar{5}^{\phi_{12}}, \bar{8}^{\Phi_{\bar{8}}}, \bar{7}^{\Phi_{\bar{7}}}) ,
\end{equation}
where the states $(\Phi_{7}, \Phi_{8})$ can be $(\phi_{12}, g^+)$, $(g^+, \phi_{12})$, $(\psi_{1}, \psi_{2})$, and $(\psi_{2}, \psi_{1})$.

This cut is similar to the one-loop $s_{12}$-cut discussed in Section~\ref{sec:oneloopFF}.

\begin{figure}[t]
	\begin{center}
		\includegraphics[width=1\linewidth]{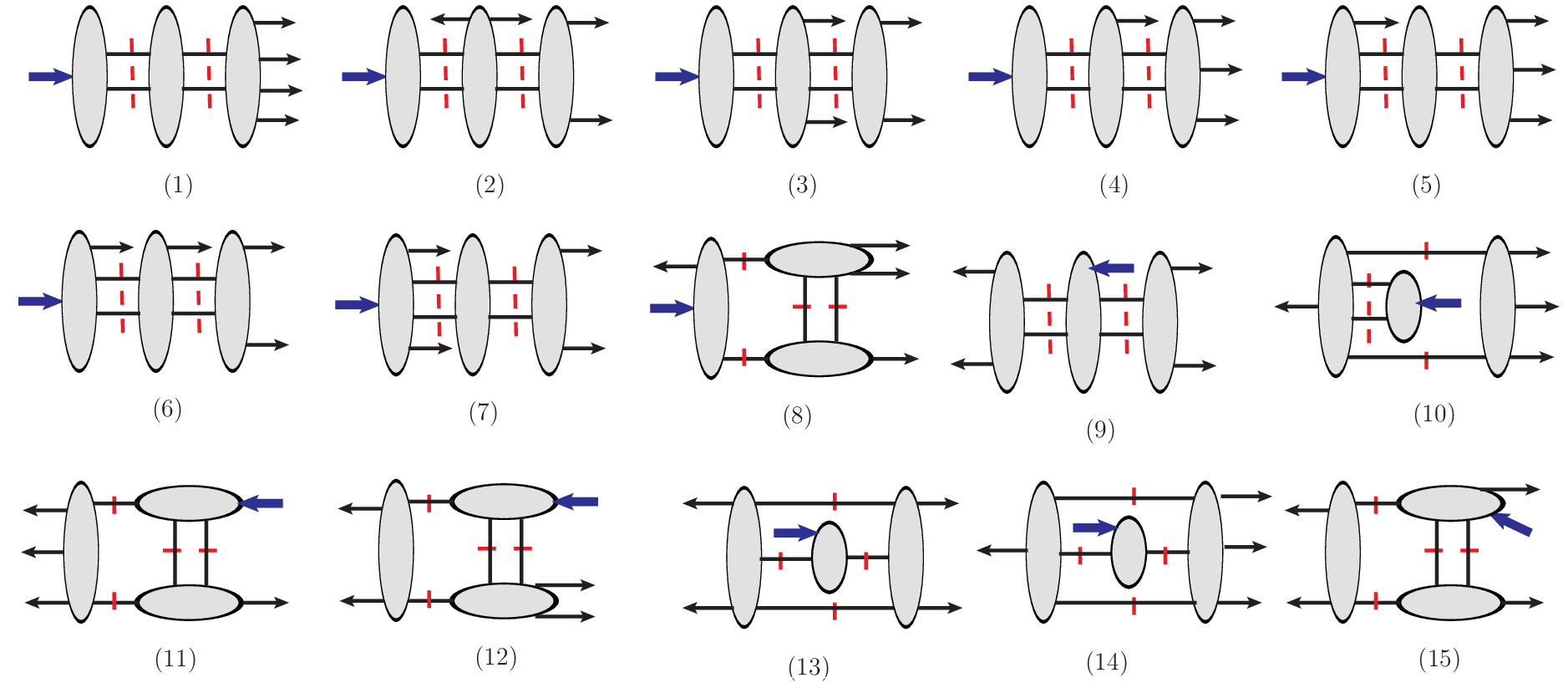}
		\caption{The two-double cuts for the two-loop form factor.}
		\label{Fig:TwodoublecutsinLength2}
	\end{center}
\end{figure}

We apply a series of two-double cuts as listed in Figure~\ref{Fig:TwodoublecutsinLength2}. 
We mention that the $\mu$-term contributions appear only in the maximal topologies, due to the loop momenta taking only quadratic power in the numerators (see Section~\ref{sec:subtraction}). Consequently, we find that the simple two-double cuts $(1)$, $(2)$, $(4)$, $(7)$, and $(10)$ in Figure~\ref{Fig:TwodoublecutsinLength2} are sufficient to determine the $\mu$-term contributions. 
Other cuts provide also crosschecks that there are not $\mu$-term contributions in the sub-topologies.

\begin{figure}[t]
	\begin{center}
	\includegraphics[width=.95\linewidth]{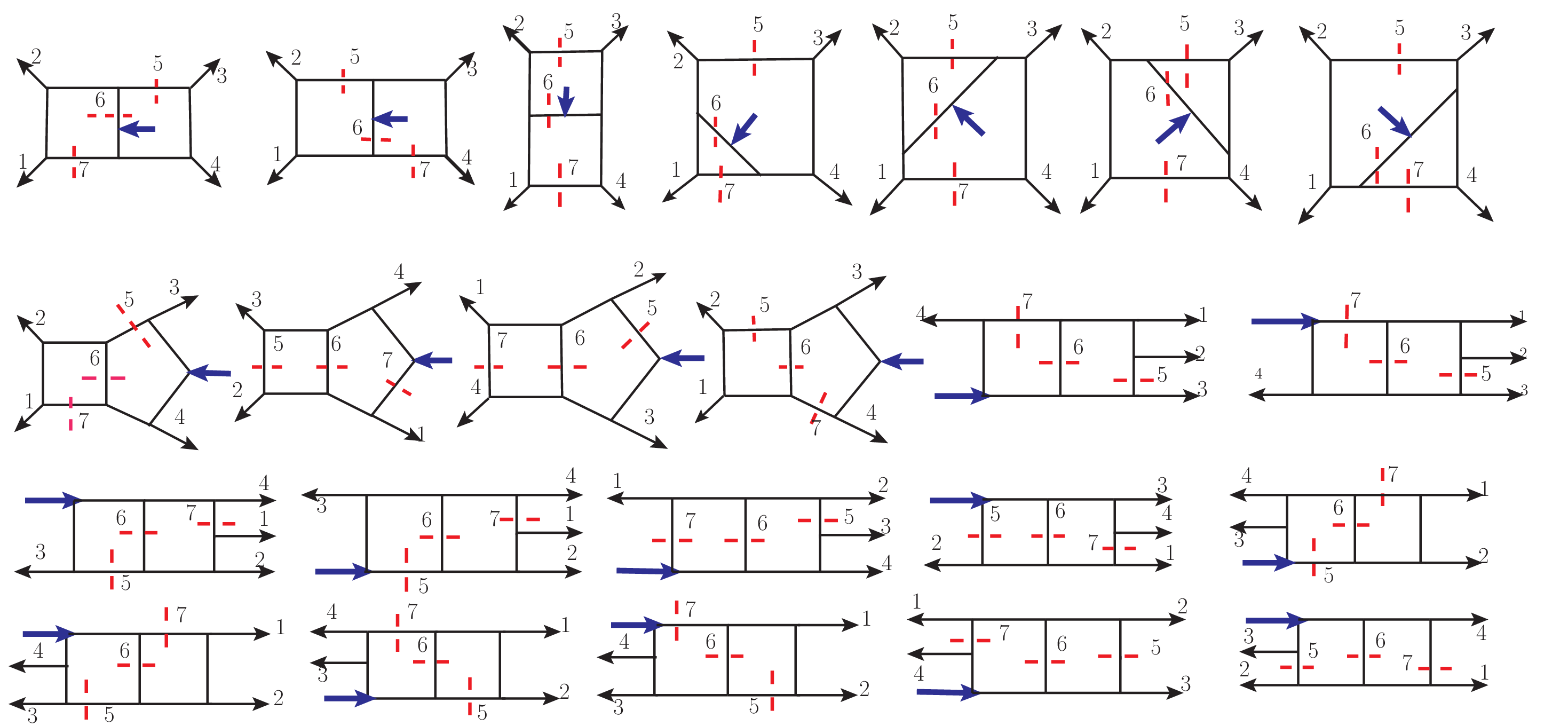}
	\caption{The 23 topologies of cut integrals detected by the triple cut.}
	\label{Fig:triplecutphisq}
	\end{center}
	\end{figure}

\begin{figure}
\begin{center}
\includegraphics[width=.9\linewidth]{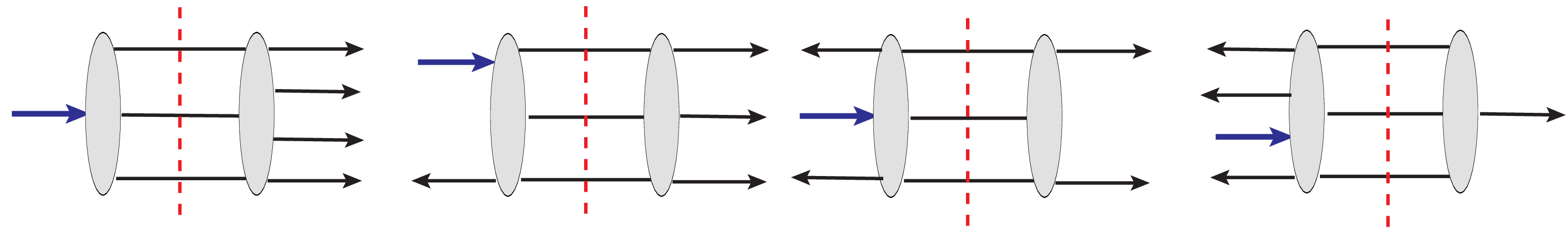}
\caption{The triple cuts.}
\label{Figtriplecut}
\end{center}
\end{figure}

Finally, the triple cut will provide the most stringent crosscheck for the integrand result. We find that all triple cuts pass directly given the result derived by the two-double cuts.
Consider for example the cut shown in Figure~\ref{Fig:triplecut} with external states as $(1^+,2^+,3^{\phi_{12}},4^{\phi_{12}})$:
\begin{align}
	\label{for1}
	{\cal F}_4^{(2)} \big|_{\text{cut-(d)}} = \sum_{\{{\Phi_i}\}} {\cal F}_5^{(0)}(3^{\phi_{12}}, 4^{\phi_{12}}, \bar{7}^{\Phi_{\bar{7}}}, \bar{6}^{\Phi_{\bar{6}}}, \bar{5}^{\Phi_{\bar{5}}}) {\cal A}_5^{(0)}(5^{\Phi_5}, 6^{\Phi_6}, 7^{\Phi_7}, 1^+, 2^+)  \,.
\end{align}
This cut involves 23 cut integrands as shown in Figure~\ref{Fig:triplecutphisq}.
The full set of triple cuts is shown in Figure~\ref{Figtriplecut}.

\section{Integrated results}
\label{sec:subtraction}

In this section, we express the two-loop form factor in terms of the UT master integrals and obtain the finite remainder of the form factor at the function level. 
The correct IR divergences and the collinear factorization properties of finite remainder provide non-trivial consistency checks. We also show that in the triple-collinear limit, the remainder function of the four-point MHV form factor gives the remainder of the two-loop six-gluon MHV amplitudes.

As for the notation, we define the loop correction of the $n$-point MHV form factor of the stress-tensor supermultiplet as
\begin{equation}
	{\cal F}^{(\ell)}_n={\cal F}^{(0)}_n {\cal I}^{(\ell)}_{{\cal F}, n} \,.
\end{equation}
The two-loop finite remainder functions of the $n$-point MHV form factor and $n$-point MHV amplitude are denoted as ${\cal R}_{{\cal F}, n}$ and ${\cal R}_{{\cal A}, n}$, respectively.

\subsection{The IBP reduction of the form factor}

It is well known that an arbitrary Feynman integral can be reduced into a finite set of linear-based integrals, which is so-called as \emph{master integral}, by utilizing the integration-by-part (IBP) relations \cite{Chetyrkin:1981qh, Tkachov:1981wb}.

The integrand expression obtained in previous sections can be expressed in the following form by introducing a set of linearly independent propagators
\begin{align}
	\mathbb{I}_{i} = \sum_{a} c_{i,a} \mathbb{I}_{i,a}^{\text{IBP-input}} \,,
\end{align}
where $\mathbb{I}_{i,a}^{\text{IBP-input}}$ are the IBP-input integrals, defined as
\begin{equation}
	\mathbb{I}_{i,a}^{\text{IBP-input}} = \prod_{k} D_{i,k}^{-\alpha_{i,a,k}} \,,
\end{equation}
where $D_{i,k}$ is the $k$-th propagator of $i$-th topology (including irreducible numerators).

For the two-loop form factor, after expanding the numerators in the propagator bases, we find that the numerators are only quadratic in loop momenta, which is due to the special UV property of the half-BPS form factor in ${\cal N}=4$ SYM.

The special comments are to the $\mu$-term variables $\mu_{ij}$ arisen from higher-length trace reduction as in \eqref{eq:hightracereduction} and the parity-odd part where the internal momenta appear in the Levi-Civita tensor. To express them in terms of propagator bases, \emph{i.e.} Lorentz products, we can apply the following equation
\begin{equation}
	\label{eq:muproduct}
	\mu_{ij} = \frac{{\rm Gram}(\{\ell_i, p_1,p_2,p_3,p_4\}, \{\ell_j, p_1,p_2,p_3,p_4\})}{\varepsilon(1234)^2} \,,
\end{equation}
and
\begin{equation}
	\label{eq:varepsilonproduct}
	\varepsilon(a, b, c, d) = -\frac{{\rm Gram}(\{a, b, c, d\}, \{p_1, p_2, p_3, p_4\})}{\varepsilon(1234)}\,,
\end{equation}
where $\varepsilon(a,b,c,d)$ contains loop momenta.

The IBP-input integrals are ready to be reduced to a finite set of master integrals by using IBP relations.
We perform the IBP reduction by the program {\bf Kira} \cite{Maierhofer:2017gsa, Klappert:2020nbg} in combination with {\bf Fermat} \cite{Lewis} and {\bf FireFly} \cite{Klappert:2019emp, Klappert:2020aqs}.
The coefficients of master integrals in the result can be simplified significantly for the half-BPS form factor if choosing the master integrals to have a uniform transcendental degree. The related UT master integrals, which will be denoted by $I^{(\ell), \text{UT}}$, are given in \cite{Abreu:2020jxa, Canko:2020ylt, Abreu:2021smk, Abreu:2023rco}. 

We find the one- and two-loop form factor loop corrections take the same compact form as
\begin{equation}
	{\cal I}_{{\cal F}, 4}^{(\ell)} = \sum_{i} \left( a_{i}^{(\ell)} + B b_{i}^{(\ell)} \right) I_{i}^{(\ell), \text{UT}} \,,
\end{equation}
where $a_i^{(\ell)}$ and $b_i^{(\ell)}$ are rational numbers, and $B$ is the parity-odd factor defined as
\begin{align}\label{eq:Bfactor}
	B = \frac{s_{12} s_{34} + s_{14} s_{23} - s_{13} s_{24}}{4i\varepsilon(1234)} \,.
\end{align}
The result can be reorganized into two parts
\begin{equation}
	{\cal I}_{{\cal F}, 4}^{(\ell)} = {\cal G}_{1}^{(\ell)} + B {\cal G}_{2}^{(\ell)} \,,
\end{equation}
with
\begin{equation}
	{\cal G}_1^{(\ell)} = \sum_{i}  a_{i}^{(\ell)}  I_{i}^{(\ell), \text{UT}}\,, \qquad 
	{\cal G}_2^{(\ell)} = \sum_{i}  b_{i}^{(\ell)}  I_{i}^{(\ell), \text{UT}}\,.
\end{equation}

\paragraph{One-loop result.} 
The one-loop result can be given explicitly as
\begin{align}
	\label{eq:oneloopibpresult}
	{\cal G}_1^{(1)} = & I_{\text{Bub}}^{(1)}(1,2,3) -I_{\text{Bub}}^{(1)}(1,2,3,4) -\frac{1}{2} I_{\text{Box}}^{(1)}(1,2,3) -\frac{1}{2} I_{\text{Box}}^{(1)}(1,2+3,4) -\frac{1}{2}I_{\text{Pen}}^{(1)}(1,2,3,4)  \nonumber \\
	&+\text{cycling $(p_1,p_2,p_3,p_4)$} \,, \\
	{\cal G}_2^{(1)} = & \frac{1}{2} I_{\text{Pen}}^{(1)}(1,2,3,4) +\text{cycling $(p_1,p_2,p_3,p_4)$} \,, \nonumber
\end{align}
where the integer $i$ in $I_{\text{Bub}}^{(1)}$ denotes the external momentum $p_i$, etc., and the explicit definitions of the UT master integrals are given in Appendix~\ref{app:UT}. The result can be also arranged into the parity-even and parity-odd parts as
\begin{align}
	{\cal I}_{{\cal F}, 4}^{(1)} \big|_{\text{even}} = & I_{\text{Bub}}^{(1)}(1,2,3) -I_{\text{Bub}}^{(1)}(1,2,3,4) -\frac{1}{2} I_{\text{Box}}^{(1)}(1,2,3) -\frac{1}{2} I_{\text{Box}}^{(1)}(1,2+3,4) \\
	& +\frac{B}{2} I_{\text{Pen}}^{(1)}(1,2,3,4) +\text{cycling $(p_1,p_2,p_3,p_4)$} \,, \notag \\
	{\cal I}_{{\cal F}, 4}^{(1)} \big|_{\text{odd}} = & -\frac{1}{2} I_{\text{Pen}}^{(1)}(1,2,3,4) +\text{cycling $(p_1,p_2,p_3,p_4)$} \,. \notag
\end{align}
The parity-odd part is defined by the terms that are proportional to $\varepsilon(1234)$. Note that the $\varepsilon(1234)$ variable appears in both the factor $B$ and the definition of the parity-odd integral $I_{\text{Pen}}^{(1)}$. We mention that $I_{\text{Pen}}^{(1)}$ contributes only to the $\mathcal{O}(\epsilon)$ order, and so does ${\cal G}_2^{(1)}$ and ${\cal I}_{{\cal F}, 4}^{(1)} \big|_{\text{odd}}$.

\paragraph{Two-loop result.}
The two-loop correction is much more complicated and involves $497$ UT master integrals as follows
\begin{equation}
	{\cal G}_1^{(2)} = \sum_{i=1}^{497}  a_{i}^{(2)}  I_{i}^{(2), \text{UT}}\,, \qquad 
	{\cal G}_2^{(2)} = \sum_{i=1}^{497}  b_{i}^{(2)}  I_{i}^{(2), \text{UT}}\,.
\end{equation}
Details of the UT masters and the expansion of the two-loop form factor are provided in the ancillary files. 

\subsection{Two-loop remainder function}\label{subsec:remainder}

The divergence of the two-loop form factor is determined by the one-loop correction and two-loop cusp and collinear anomalous dimensions, which can be captured nicely via the BDS ansatz form \cite{Bern:2005iz} as
\begin{align}
	\label{eq:F2loopIR}
	{\cal I}_{{\cal F}, n}^{(2)}(\epsilon) = \frac{1}{2} \left( {\cal I}_{{\cal F}, n}^{(1)}(\epsilon) \right)^{2} + f^{(2)}(\epsilon) {\cal I}_{{\cal F}, n}^{(1)}(2 \epsilon) + {\cal R}_{{\cal F}, n}^{(2)} + {\cal O}(\epsilon)  \,,
\end{align}
where $f^{(2)}(\epsilon) = -2 \zeta_{2} -2 \zeta_{3} \epsilon -2 \zeta_{4} \epsilon^{2}$, and ${\cal R}_{{\cal F}, n}^{(2)}$ is the remainder function.
We comment here that the definition of the remainder function differs from the expression in \cite{Dixon:2021tdw, Dixon:2022xqh} by the constant $C^{(2)} = \zeta_4$.

The master integrals are evaluated in a set of algebraically independent \emph{pentagon functions} $\{f_i^{(n)}\}$  in \cite{Chicherin:2020oor, Chicherin:2021dyp, Abreu:2023rco}, where $n$ denotes the transcendental degree and $i$ labels the function basis. The pentagon functions are the $\mathbb{Q}$-linear combination of a set of the Chen iterated path integrals \cite{Chen:1977oja} defined as
\begin{align}
	[w_1]_{\boldsymbol{x}_0}(\boldsymbol{x}) = & \int_{\boldsymbol{x}_0}^{\boldsymbol{x}} d\log(w_1(\boldsymbol{t})) \,, \\
	\qquad [w_1, \ldots, w_{n-1}, w_n]_{\boldsymbol{x}_0}(\boldsymbol{x}) = & \int_{\boldsymbol{x}_0}^{\boldsymbol{x}} d\log(w_n(\boldsymbol{t})) [w_1, \ldots, w_{n-1}]_{\boldsymbol{x}_0}(\boldsymbol{t}) \,, \nonumber
\end{align}
where $\{w_a\}$ are the function letters depending on the kinematics.
They can be evaluated numerically with the start point $\boldsymbol{x}_0$ as given in \cite{Abreu:2023rco} as
\begin{align}
	\boldsymbol{x}_0 = (q^2 = 1, s_{234} = 3, s_{12} = 2, s_{23} = -2, s_{34} = 7, s_{123} = -2) \,.
\end{align}
Furthermore, to extract the symbol \cite{Goncharov:2010jf}, the Chen iterated path integral can be converted directly to the symbol  as
\begin{align}
	{\cal S}([w_1, w_2, \cdots, w_n]_{\boldsymbol{x}_0}(\boldsymbol{x})) = w_1 \otimes w_2 \otimes \cdots \otimes w_n \,,
\end{align}
where the arguments $\{w_a\}$ are also the symbol letters of the symbol.

As a result, we obtain the remainder function ${\cal R}_{{\cal F}, 4}^{(2)}$ which is parity-even and depends on the following ratios
\begin{equation}
	u_{i} = \frac{s_{i,i+1}}{q^2} \,, \qquad v_{j} = \frac{s_{j,j+1,j+2}}{q^2} \,,
\end{equation}
where $i,j = 1,2,3,4$ and note that only five of them are independent. We find that the remainder function in terms of the Chen iterated path integrals contains only 34 function letters $\{m_i\}$ which are the same as the symbol letters. 
We have checked that our result is consistent with the symbol expression of the same form factor obtained via symbol bootstrap in \cite{Dixon:2022xqh}.

We provide both the two-loop remainder function in terms of the pentagon functions and the Chen iterated path integrals in the ancillary files.

\paragraph{Cancellations.}

Below we comment on several interesting cancellations in computing the remainder function.

(1) We find the remainder function is free from the factor $B$ of \eqref{eq:Bfactor} that appears in the full two-loop form factor.
The terms in ${\cal G}_{2}^{(\ell)}$ which are proportional to $B$ in the two-loop form factor are canceled at finite order as
\begin{align}
	\label{eq:cancellation}
	{\cal G}_{2}^{(2)}(\epsilon) -{\cal G}_{1}^{(1)}(\epsilon) {\cal G}_{2}^{(1)}(\epsilon) = {\cal O}(\epsilon) \,.
\end{align}
We recall that ${\cal G}_{2}^{(1)}(\epsilon)$ contributes to the ${\cal O}(\epsilon)$ order.
Therefore, the remainder function can be given in terms of ${\cal G}_{1}^{(\ell)}$ as
\begin{align}
	{\cal R}_{{\cal F},4}^{(2)} = {\cal G}_{1}^{(2)}(\epsilon) -\frac{1}{2} \left( {\cal G}_{1}^{(1)}(\epsilon) \right)^2 - f^{(2)}(\epsilon) {\cal G}_{1}^{(1)}(2\epsilon) + {\cal O}(\epsilon) \,.
\end{align}

(2) The parity-odd part is canceled in the finite remainder as
\begin{align}
	& {\cal R}_{{\cal F},4}^{(2)} = {\cal I}_{{\cal F}, 4}^{(2)}(\epsilon) \big|_{\text{even}} -\frac{1}{2} \left( {\cal I}_{{\cal F}, 4}^{(1)}(\epsilon) \big|_{\text{even}} \right)^2 - f^{(2)}(\epsilon) {\cal I}_{{\cal F}, 4}^{(1)}(2\epsilon) \big|_{\text{even}} \,, \\
	& {\cal I}_{{\cal F}, 4}^{(2)}(\epsilon) \big|_{\text{odd}} -\left( {\cal I}_{{\cal F}, 4}^{(1)}(\epsilon) \big|_{\text{even}} \right) \left( {\cal I}_{{\cal F}, 4}^{(1)}(\epsilon) \big|_{\text{odd}} \right) = {\cal O}(\epsilon) \,. \notag
\end{align}
This is also expected from the duality between the form factor and the periodic Wilson loop.

(3) The $\mu$-term UT master integrals are all canceled in the remainder function for the specific choice of the UT basis \cite{Abreu:2020jxa, Canko:2020ylt, Abreu:2021smk, Abreu:2023rco} as\footnote{A similar cancellation but at the integrand level is observed in the two-loop six-point planar amplitudes \cite{Bern:2008ap, Kosower:2010yk}.}
\begin{equation}
	\left( {\cal I}_{{\cal F}, 4}^{(2)}(\epsilon) \big|_{\mu\text{-term UT}} \right) -\left( {\cal I}_{{\cal F}, 4}^{(1)}(\epsilon) \big|_{\mu\text{-term UT}} \right) {\cal I}_{{\cal F}, 4}^{(1)}(\epsilon) = {\cal O}(\epsilon) \,.
\end{equation}
Here, the $\mu$-term UT integrals correspond to the sub-set of master integrals that explicitly contain $\mu_{ij}$ in the numerator. For example, the one-loop $\mu$-term UT integral is only the pentagon master integral $I^{(1)}_{\text{Pen}}$ in \eqref{eq:Pen1loop}, and thus the part that contains only $\mu$-term UT integrals ${\cal I}_{{\cal F}, 4}^{(1)}(\epsilon) \big|_{\mu\text{-term UT}}$ is
\begin{align}
	{\cal I}_{{\cal F}, 4}^{(1)}(\epsilon) \big|_{\mu\text{-term UT}} = \frac{B -1}{2} I_{\text{Pen}}^{(1)}(1, 2, 3, 4) +\text{cycling $(p_1, p_2, p_3, p_4)$} \,.
\end{align}

\subsection{Evaluating the remainder function}

In the previous subsection, we obtain the remainder function in terms of the pentagon functions. In this subsection, we consider evaluating the remainder function and perform several consistency checks in different kinematic limits, including the collinear and lightlike limits.

We will mainly focus on the scattering process $34 \to 12(-q)$, which is in the physical region related to realistic phenomenologies. The process can be understood as a supersymmetric version of the two-parton to Higgs-plus-two-parton scattering.

\begin{figure}[tb]
	\centering
	\includegraphics[scale=0.7]{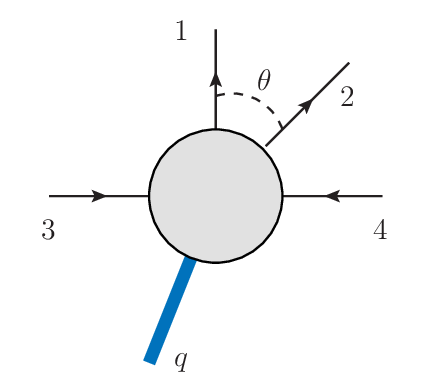}
	\caption{The specific parameterization for scattering process $34 \to 12(-q)$.}
	\label{fig:34to12q}
\end{figure}

Generally, we can parametrize the momenta as shown in Figure~\ref{fig:34to12q} by
\begin{align}
	& p_3 = -\frac{1}{2} (1, 0, 0, 1) \,, \qquad p_4 = -\frac{1}{2} (1, 0, 0, -1) \,, \qquad p_1 = \frac{x_1}{2} (1, 1, 0, 0) \,, \\
	& p_2 = \frac{x_2}{2} (1, \cos(\theta), -\sin(\theta) \sin(\phi), -\sin(\theta) \cos(\phi)) \,. \nonumber
\end{align}
where the ingoing momenta $p_3$ and $p_4$ are placed back-to-back along the $z$-axis and carry a total centre-of-mass energy of $s_{34} = 1$. The scattering angle $\theta$ between the outgoing momenta $p_1$ and $p_2$ is constrained by the on-shell condition $q^2 = m_q^2$ as
\begin{align}
	\cos(\theta) = 1 +\frac{2}{x_1 x_2} \left( 1-x_1 -x_2 -m_q^2 \right) \,,
\end{align}
where $m_q$ should satisfy the relation $m_q^2 < s_{34}^2 = 1$. Hence, the feasible region of the parameters $(x_1, x_2)$ is
\begin{align}
	0 < x_1 < 1 -m_q^2 \,, \qquad 1 -m_q^2 -x_1 < x_2 < \frac{2(1 -m_q^2 -x_1)}{2-x_1} \,,
\end{align}
where the left boundary of $x_2$ corresponds to $\theta = 0$, the right boundary of $x_2$ corresponds to $\theta = \pi$.

\begin{figure}[tb]
	\centering
	\subfigure[Real part]{\includegraphics[scale=0.47]{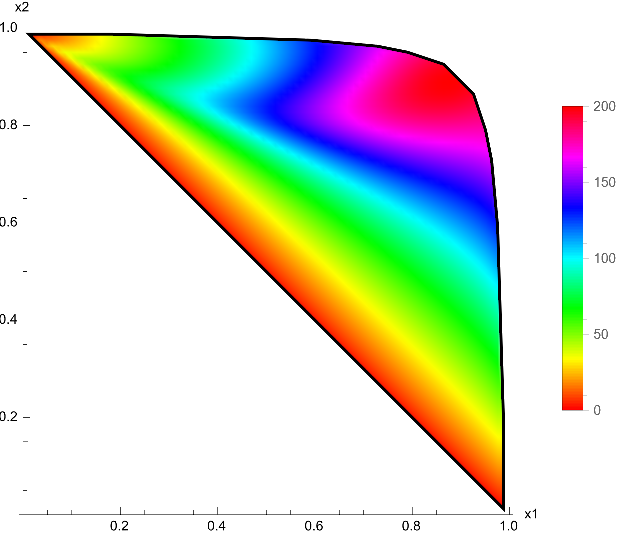}}
	\subfigure[Imaginary part]{\includegraphics[scale=0.47]{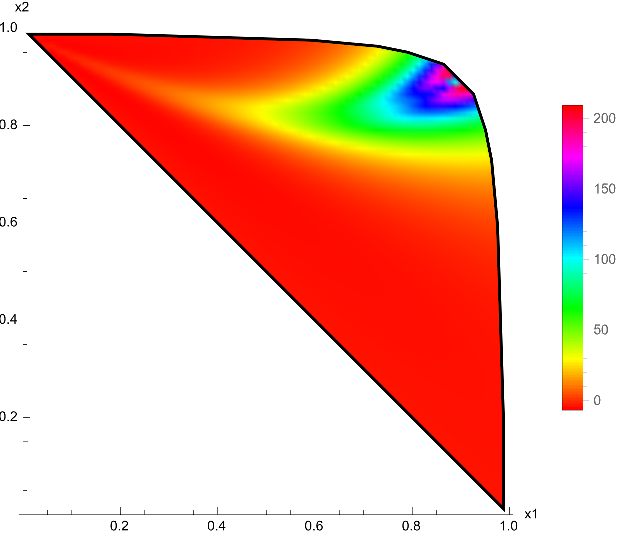}}
	\subfigure[Absolute value]{\includegraphics[scale=0.47]{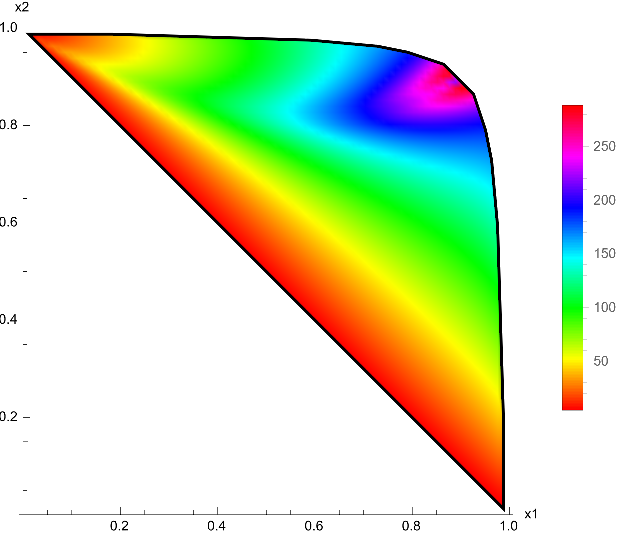}}

	\subfigure[Real part]{\includegraphics[scale=0.44]{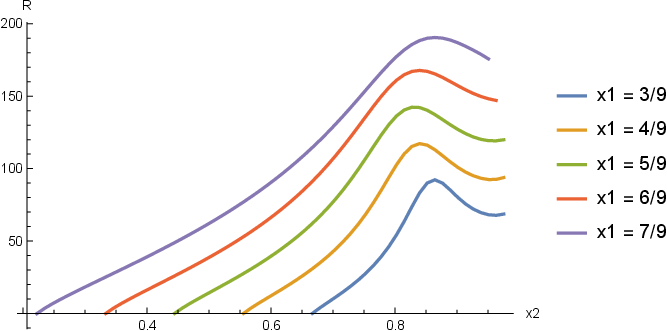}}
	\subfigure[Imaginary part]{\includegraphics[scale=0.44]{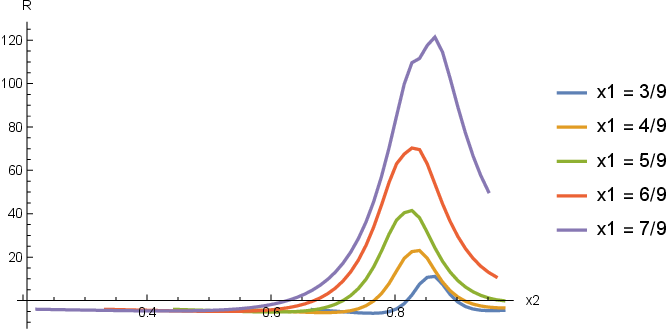}}
	\subfigure[Absolute value]{\includegraphics[scale=0.44]{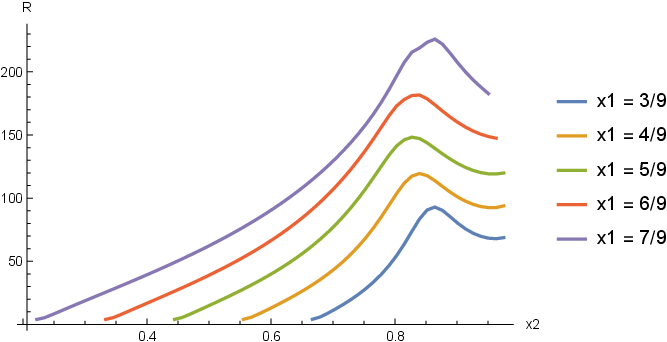}}

	\caption{Evaluating the remainder function for a physical scattering process.
	For the figures in the first row, the lower-left linear boundary and the upper-right curve boundary correspond separately to the outgoing angle $\theta = 0$ and $\theta = \pi$.
	The two boundaries intersect symmetrically at $(0, 1-m_q^2)$ and $(1-m_q^2, 0)$ correspond separately to the soft limit with $p_1 \to 0$ and $p_2 \to 0$, where $m_q = 1/10$.}
	\label{fig:plotremainder}
\end{figure}

It is convenient to use the package {\bf PentagonFunctions++} \cite{Chicherin:2020oor, Chicherin:2021dyp} to evaluate the pentagon functions. The physical region requires the following conditions as
\begin{align}
	\label{eq:physicalregionP34}
	\{(s_{ij}) | & q^2 > 0 \,, \quad p_i^2 = 0 \,, \ \Delta_5 < 0 \,, \ s_{12} > 0 \,, \ (s_{234}\,,\ s_{134}\,,\ s_{34}) > q^2 \,,  \nonumber \\
	& (s_{13} \,,\ s_{23} \,,\ s_{14} \,,\ s_{24} \,,\ s_{124} \,,\ s_{123}) < 0 \} \,,
\end{align}
where $\Delta_5 = -16\varepsilon(1234)^2$. This region is equivalent to ${\cal P}_{45}$ defined by Eq.$(2.18)$ in \cite{Chicherin:2021dyp}. 
In Figure~\ref{fig:plotremainder}, we plot the finite remainder as function of $x_1$ and $x_2$ with $m_q = 1/10$ and $\phi = 1/10$.

Below we apply a series of consistency checks by evaluating numerically the remainder function with {\bf PentagonFunctions++}. The numerical result is of at least $16$-digit accuracy.

\paragraph{Collinear limit.}
The remainder function has the advantage that it is free from both IR and collinear singularities, which has a well-defined smooth collinear limit behavior.
The $n$-point remainder will reduce to $(n-1)$-point remainder in the collinear limit $p_{i} \parallel p_{i+1}$. 
For the four-point form factor, we have
\begin{align}
	{\cal R}_{{\cal F}, 4}^{(2)} \xlongrightarrow{p_i \parallel p_{i+1}} {\cal R}_{{\cal F}, 3}^{(2)} \,.
\end{align}
Concretely, we verify the collinear limit $p_1 \parallel p_2$ by taking momenta $p_1 = t P$, $p_2 = (1-t) P$ and $P^2 \to 0$. We choose the kinematics as
\begin{align}
	s_{12} = P^2 \,, \quad s_{13} = -\frac{1}{6} \,, \quad s_{14} = -\frac{1}{18} \,, \quad s_{23} = -\frac{1}{3} \,, \quad s_{24} = -\frac{1}{9} \,, \quad s_{34} = \frac{5}{3} \,,
\end{align}
where $P^2$ will tend to $0$ in the collinear limit.
Then, we evaluate the form factor at this kinematics point by selecting a series of small $P^2$ and compare the numerical result with the known three-point form factor result. We observe a trend of convergence in the limit process when the parameter $P^2$ close to $0$ but not equal to $0$. The numerical result with $P^2 = 10^{-16}$ is
\begin{align}
	{\cal R}_{{\cal F}, 4}^{(2)} = 4.991676051765813 +0.05741704625875813i \,.
\end{align}
The ${\cal R}_{{\cal F}, 3}$ is evaluated at the corresponding kinematics point at
\begin{equation}
	s_{12}' = -1/2 \,, \quad s_{23}' = 5/3 \,, \quad s_{13}' = -1/6 \,, 
\end{equation}
with $p_1' = p_1+p_2$, $p_2' = p_3$ and $p_3' = p_4$. We find the difference as
\begin{align}
	{\cal R}_{{\cal F}, 4}^{(2)} -{\cal R}_{{\cal F}, 3}^{(2)} = -2.0 \times 10^{-13} -3.1 \times 10^{-14} i \,.
\end{align}

Furthermore, in Figure~\ref{fig:plotremainder}, the remainder function at the lower-left linear boundary corresponds to $\theta=0$. We have checked that this is consistent with taking the collinear limit $p_1 \parallel p_2$. For example, choosing the parameters $x_1 = x_2 = 99/200 +\delta$ and $\delta = 10^{-16}$, the result is
\begin{align}
	{\cal R}_{{\cal F}, 4}^{(2)} = -5.398065883227598-3.949636854837923i \,.
\end{align}
The corresponding kinematics of ${\cal R}_{{\cal F}, 3}^{(2)}$ are 
\begin{equation}
s_{12}' = s_{13}' = -99/200\,, \qquad s_{23}' = 1 \,,
\end{equation} 
with $p_1' = p_1+p_2$, $p_2' = p_3$ and $p_3' = p_4$, and the difference is
\begin{align}
	{\cal R}_{{\cal F}, 4}^{(2)} -{\cal R}_{{\cal F}, 3}^{(2)} = 4.6 \times 10^{-7} -6.6 \times 10^{-10} i \,.
\end{align}
The above two examples also show that the speed of convergence is not the same under different parameterizations.

\paragraph{Lightlike limit}
The lightlike form factor (with $q^2=0$) for the stress-tensor supermultiplet has been obtained using the master-integral bootstrap method in \cite{Guo:2022qgv}, which is shown to have the directional dual conformal (DDC) symmetry and depends on only the three conformal ratios
\begin{equation}
	\tilde{u}_1 = \frac{s_{12}}{s_{34}} \,, \qquad \tilde{u}_2 = \frac{s_{23}}{s_{14}} \,, \qquad \tilde{u}_3 = \frac{s_{123} s_{134}}{s_{234} s_{124}} \,,
\end{equation}
with $q^2 = 0$. The remainder function should have a smooth behavior in the lightlike limit. We verify the limit with the Mandelstam variables as
\begin{align}
	& s_{12} = \frac{281}{149} \,, \qquad  s_{13} =  -\frac{4079759}{3788623} \,, \qquad s_{14} =  \delta -\frac{14161027}{6534739} \,, \\
	& s_{23} = -\frac{113}{47} \,, \qquad s_{24} = -\frac{2872}{12079} \,, \qquad s_{34} = 4 \,, \nonumber 
\end{align}
where $\delta \to 0$ and which satisfies $q^2 = \delta$. Then the form factor is evaluated at this kinematics point with $\delta = 10^{-16}$, as
\begin{align}
	{\cal R}_{{\cal F}, 4}^{(2)} = 102.1176907432574 -4.0 \times 10^{-12} i \,,
\end{align}
and the difference with respect to the lightlike form factor ${\cal R}_{{\cal F}, 4}^{(2), {\rm LL}}$ is
\begin{align}
	{\cal R}_{{\cal F}, 4}^{(2)} -{\cal R}_{{\cal F}, 4}^{(2), {\rm LL}} = -2.2\times10^{-11} +4.0\times 10^{-12} i \,.
\end{align}
In the next section, we will show that the DDC  symmetry emerges also at the \emph{integrand level}.

\paragraph{Other numerical evaluations.}
We also compute the form factor numerically using the auxiliary mass flow method \cite{Liu:2017jxz, Liu:2021wks, Liu:2022mfb} as implemented in {\bf AMFlow} \cite{Liu:2022chg}. The package is based on the differential equation method and enables high-precision evaluations for general kinematics regions. 
More details are provided in appendix~\ref{app:consistency}.

\subsection{Triple-collinear limit}

The intriguing duality between the form factor and the amplitude is revealed by the antipodal symmetry \cite{Dixon:2021tdw, Dixon:2022xqh}. The non-trivial evidence has been shown for the three-point form factor and six-point amplitude up to eight loops \cite{Dixon:2022rse}. Furthermore, the two-loop four-point form factor was found to have an antipodal self-duality at the symbol level \cite{Dixon:2022xqh}. This duality is related closely to the triple-collinear limit behavior of the four-point form factor, which was also discussed for the two-loop six-point MHV amplitude in \cite{Bern:2008ap}. The form factor will approach the two-loop six-point amplitude in this limit. In this subsection, we study the triple-collinear limit of the four-point form factor in detail at the full functional level.

The kinematic ratios of the four-point form factor can be rewritten in terms of the OPE parameters $\{T, S, T_2, S_2, f_2\}$ as \cite{Dixon:2022xqh}
\begin{align}
	\label{uvToOPE}
	& u_1 = \frac{T^2 T_2^2}{\left(T^2+1\right) \left(S^2+T^2+T_2^2+1\right)}\,, \\
	& u_2 = \left[1 +T^2 +\frac{S^2 [S_2 T_2 (1 +f_2^2) +f_2(1 +S_2^2 +T^2 +T_2^2)]}{f_2 S_2^2}\right]^{-1}\,, \notag \\
	& u_3 = \frac{S^2}{T^2 T_2^2}u_1 \,, \qquad u_4 = \frac{S^2 T^2}{S_2^2} u_2 \,, \qquad v_1 = \frac{T_2^2+1}{S^2+T^2+T_2^2+1} \,. \nonumber
\end{align}
Then the double-collinear limit $T_2 \to 0$ and triple-collinear limit $T \to 0$ can be taken separately as
\begin{align}
	{\cal R}_{{\cal F}, 4}^{(2)} \xlongrightarrow{T_2 \to 0} {\cal R}_{{\cal F}, 3}^{(2)}(T, S) \,, \qquad
	{\cal R}_{{\cal F}, 4}^{(2)} \xlongrightarrow{T \to 0} 4\hat{{\cal R}}_{{\cal A}, 6}^{(2)}(T_2, S_2, f_2) +4\zeta_4 \,,
\end{align}
where ${\cal R}_{{\cal F}, 3}^{(2)}$ denotes the three-point form factor remainder and $\hat{\cal R}_{{\cal A}, 6}^{(2)}$ denotes the six-point MHV amplitude remainder. We comment that $\zeta_4$ on the RHS is due to the convention we define the form factor remainder (see \eqref{eq:F2loopIR} and the footnote below), and the overall factor 4 is due to the different convention of the gauge coupling.

The three cross ratios of $\hat{\cal R}_{{\cal A}, 6}^{(2)}$ in terms of the OPE parameters $\{T_2, S_2, f_2\}$ are
\begin{align}
	\label{uvwToOPE}
	& \hat{u} = \frac{\hat{s}_{12} \hat{s}_{45}}{\hat{s}_{123} \hat{s}_{345}} = \frac{S_2^2}{T_2^2} \hat{v} \hat{w} \,, \qquad \hat{v} = \frac{\hat{s}_{23} \hat{s}_{56}}{\hat{s}_{234} \hat{s}_{123}} = \frac{T_2^2}{T_2^2 +1} \,, \\
	& \hat{w} = \frac{\hat{s}_{34} \hat{s}_{61}}{\hat{s}_{345} \hat{s}_{234}} = \frac{1}{1 +(T_2 +S_2 f_2)(T_2 +S_2/f_2)} \,, \nonumber
\end{align}
where $\hat{s}_{i,i+1}$ and $\hat{s}_{i,i+1,i+2}$ are the Mandelstam variables of the six-point amplitude.

Let us explain the triple-collinear limit in more detail, see also  \cite{Dixon:2022xqh}. When we take the triple-collinear limit for both the form factor and the amplitude, for example, in the triple collinear limit $p_1 \parallel p_2 \parallel p_3$, we have the factorization properties
\begin{align}
	{\cal F}_4^{(L)}(p_1, \ldots, p_4) & \xlongrightarrow{\textrm{T.C.}} \sum_{\ell = 0}^L {\cal F}_2^{(L-\ell)}(-P, p_4) \times {\rm Sp}_4^{(\ell)}(p_1, p_2, p_3;P) \,, \\ 
	{\cal A}_6^{(L)}(\hat{p}_1, \ldots, \hat{p}_6) & \xlongrightarrow{\textrm{T.C.}} \sum_{\ell = 0}^L {\cal A}_4^{(L-\ell)}(-\hat{P}, \hat{p}_4, \hat{p}_5, \hat{p}_6) \times {\rm Sp}_4^{(\ell)}(\hat{p}_1, \hat{p}_2,\hat{p}_3; \hat{P}) \,,
\end{align}
where ${\rm Sp}_4^{(\ell)}$ is the $\ell$-loop splitting function, ${\cal F}_2^{(L-\ell)}$ and ${\cal A}_4^{(L-\ell)}$ have trivial finite remainders.
Moreover, it has been argued in  \cite{Bern:2008ap} that in ${\cal N}=4$ SYM, the finite part of triple-collinear splitting amplitude can be identified with the finite remainder of the six-gluon amplitudes, since both of them depend on three conformal cross ratios and are equivalent by the conformal symmetry of the theory.
Therefore, after the BDS subtraction for the four-point form factor and the six-gluon amplitude, their finite remainders are identical to each other as
\begin{equation}
	\label{eq:R4TCasA6}
	{\cal R}_{{\cal F}, 4}^{(2)} \xlongrightarrow{\textrm{T.C.}} 4 {\cal R}_{{\cal A}, 6}^{(2)} + 4\zeta_4  \,,
\end{equation}
up to the choice of different conventions mentioned above.

The triple-collinear limit behavior at the symbol level is confirmed in \cite{Dixon:2022xqh}. Here we check this at the function level. One subtlety is that since we have the function expression related to the kinematic region $34 \to 12(-q)$, we need to compare it with the amplitude result in the correct physical region correspondingly.

\begin{figure}[tb]
	\centering
	\includegraphics[scale=0.6]{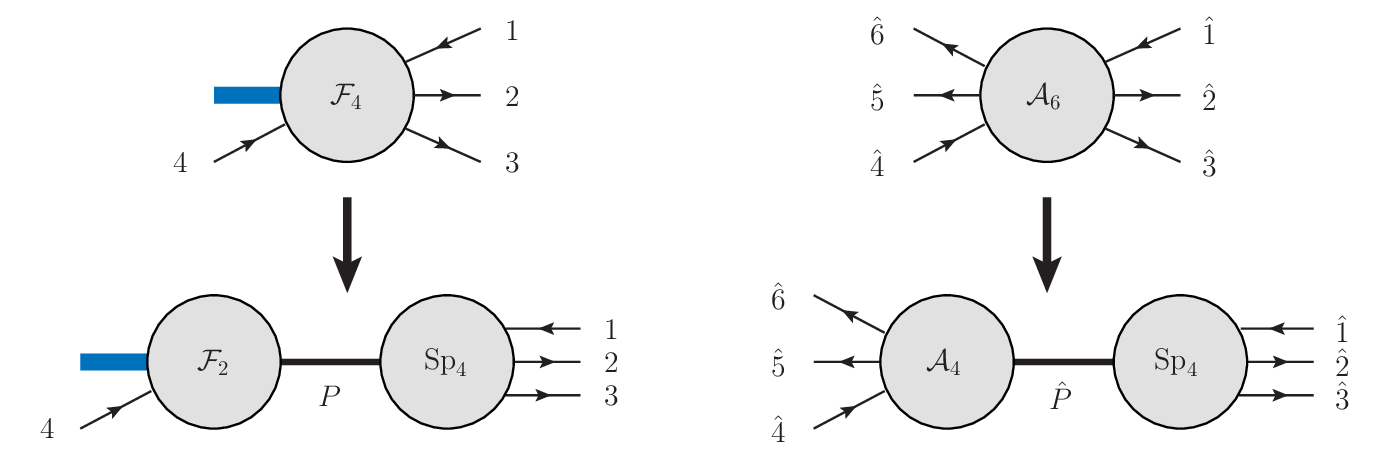}
	\caption{
	Triple-collinear limit of the form factor and the amplitude with $p_1 \parallel p_2 \parallel p_3 = P$ and $\hat{p}_1 \parallel \hat{p}_2 \parallel \hat{p}_3 = \hat{P}$, where $P^2 = \hat{P}^2 \rightarrow 0$. 
	}
	\label{fig:TripleCollinear}
\end{figure}

We note that the triple-collinear limit $T \to 0$ for parametrization \eqref{uvToOPE} corresponds to $p_4 \parallel p_1 \parallel p_2$. 
Taking triple-collinear limit $p_4 \parallel p_1 \parallel p_2$ for the form factor in the kinematic region $34 \to 12(-q)$ can be mapped to the triple-collinear limit $p_1 \parallel p_2 \parallel p_3$ for the form factor in the kinematic region $41 \to 23(-q)$.
The latter, following the above discussion, should be related to the triple-collinear limit ${\hat p}_1 \parallel {\hat p}_2 \parallel {\hat p}_3$ of six-point amplitude in the physical region ${\hat 4} {\hat 1} \to {\hat 2} {\hat 3} {\hat 5} {\hat 6}$.
This correspondence is illustrated in Figure~\ref{fig:TripleCollinear}.
In the following, we will compare the amplitude in the physical region ${\hat 4} {\hat 1} \to {\hat 2} {\hat 3} {\hat 5} {\hat 6}$ and the triple-collinear limit $p_4 \parallel p_1 \parallel p_2$ for the form factor in the kinematic region $34 \to 12(-q)$ which should be equivalent up to a cyclic shift of momenta.
%


\paragraph{Check in the self-crossing limit.}

It is notable that the full analytic remainder ${\cal R}_{{\cal A}, 6}^{(2)}$ in the physical region ${\hat 1} {\hat 4} \to {\hat 2} {\hat 3} {\hat 5} {\hat 6}$ is still not yet available in the literature.
To apply our check, we note that the amplitude has been studied in a singular configuration, which corresponds to the limit in which a hexagonal Wilson loop develops a self-crossing \cite{Dixon:2016epj}.
In this self-crossing limit, the cross ratios are highly constrained, for example, 
for the process ${\hat 1} {\hat 4} \to {\hat 2} {\hat 3} {\hat 5} {\hat 6}$ one has 
\begin{equation}
	\label{eq:defofxdelta}
	\hat{u} = \hat{w} = x \,, \qquad \hat{v} = 1-\delta \,.
\end{equation} 
The two-loop remainder function in the limit reads
\begin{align}\label{eq:RA6selfcrossing}
	{\cal R}_{{\cal A}, 6}^{(2)} \big|_\text{Self-Crossing} = & H_4^x -H_{3,1}^x +3H_{2,1,1}^x -\log(x)(H_{3}^{v}-H_{2,1}^{x})- \frac{(H_2^x)^2}{2} -\frac{5}{2} \zeta_4 \\
	& +2 \pi i \left[ \frac{\log(\delta)^3}{12} -\left( \frac{H_2^x}{2} +\frac{\log^2(x)}{4} -\frac{\zeta_2}{2} \right)\log(\delta) -H_{2,1}^x +\frac{\log^3(x)}{6} -\zeta_3 \right] \,, \nonumber
\end{align}
where $\delta \to 0$ and $H_{3,1}^x = H(0,0,1,1;1-x)$, etc; see  \cite{Dixon:2016epj}. The imaginary part of this expression is divergent when $\delta \to 0$, so we check its asymptotic behavior for small $\delta$. We take $x = 1/2$ and $\delta = 10^{-12}$ as an example.
From \eqref{eq:RA6selfcrossing}, we obtain
\begin{equation}
	{\cal R}_{{\cal A}, 6}^{(2)} \big|_\text{Self-Crossing} \approx -2.049496532670093-11125.49434747424i \,.
\end{equation}

To compute the triple-collinear limit of the form factor, we first obtain $\{T_2, S_2, f_2\}$ which can be solved by \eqref{uvwToOPE} as
\begin{align}
	T_2 = -3\sqrt{111 111 111 111} \,, \qquad S_2 = -10^6\,, \qquad f_2 = -\sqrt{1 -2(1 +i T_2) \times 10^{-12}} \,.
\end{align}
Then the Mandelstam variables for ${\cal R}_{{\cal F}, 4}^{(2)}$ are given by its OPE variables as \eqref{uvToOPE}, where $q^2$ can be specified as $1$, and $\{T, S\}$ need to be chosen as the kinematics belong to the $34 \to 12(-q)$ process. For $x=1/2$ and $0<\delta<1$, the inequalities will be
\begin{equation}
	-1 < T^2 < 0 \,, \qquad S^2 + T^2 < -\delta^{-1} \,.
\end{equation}
In addition, to achieve the triple-collinear limit, $|T^2|$ should be far less than $|\delta|$. We can specify them to be
\begin{equation}
	T^2 = -10^{-16} \,, \qquad S^2 = -\delta^{-1}-\frac{1}{2} \,,
\end{equation}
where $-\frac{1}{2}$ in $S^2$ is a random choice, since the form factor in the triple-collinear limit should be independent of $S$.

As a result, we find the two results are close enough
\begin{equation}
	\left(4{\cal R}_{{\cal A}, 6}^{(2)} \big|_\textrm{Self-Crossing} +4 \zeta_4\right) - \left( {\cal R}_{{\cal F}, 4}^{(2)} \big|_{\textrm{T.C}} \right) \approx (5.6 \times 10^{-12} +5.3 \times 10^{-9} i ) \,.
\end{equation}

\paragraph{More general numerical checks.}
To check the triple-collinear limit with more general cross ratios, we can compute the numerical result of ${\cal R}_{{\cal A}, 6}^{(2)}$ at the physical region by the package {\bf AMFlow} \cite{Liu:2022chg} based on its integrand expression provided in \cite{Bern:2008ap}.  
The strategy of determining the Mandelstam variables of the form factor is the same as before. 

For example, choosing the Mandelstam variables as
\begin{align}
	\hat{s}_{12} = \hat{s}_{34} = \hat{s}_{45} = \hat{s}_{61} = \hat{s}_{13} = \hat{s}_{15} = -1 \,, \qquad \hat{s}_{23} = \hat{s}_{56} =  \frac{4}{5} \,, \qquad \hat{s}_{35} =  \frac{1}{8} \,,
\end{align}
which correspond to 
\begin{equation}
	\hat{u}=\hat{v}=\hat{w}=\frac{4}{9} \,,
\end{equation}
the amplitude result can be computed as
\begin{equation}
	{\cal R}_{{\cal A}, 6}^{(2)} \approx 0.9348121861212953+2.813137860102412i \,.
\end{equation}
Then, the form factor result in the triple-collinear limit is obtained by solving for $\{T_2,S_2,f_2\}$ using \eqref{uvwToOPE}, with $\{T,S\}$ specified as
\begin{equation}
	T^2 = -10^{-16}\,, \qquad S^2 = -\frac{23}{10} \,.
\end{equation}
Note that the value of $S^2$ is a random choice in the allowed kinematic region, since the form factor in the triple-collinear limit should be independent of $S$.

We find the difference is
\begin{equation}
	\left( 4{\cal R}_{{\cal A}, 6}^{(2)} +4 \zeta_4 \right) - \left( {\cal R}_{{\cal F}, 4}^{(2)} \big|_{\textrm{T.C}} \right) \approx 5.5 \times 10^{-15} +2.2 \times 10^{-14}i \,.
\end{equation}

Another example is with Mandelstam variables
\begin{align}
	\hat{s}_{12} = \hat{s}_{34} = \hat{s}_{45} = \hat{s}_{61} = \hat{s}_{13} = \hat{s}_{15} = -1 \,, \qquad \hat{s}_{23} = \hat{s}_{56} = \hat{s}_{35} = \frac{1}{2} \,,
\end{align}
corresponding to 
\begin{equation}
	\hat{u} = \hat{w} = \frac{4}{9} \,, \qquad \hat{v} = \frac{1}{9} \,. 
\end{equation}
The amplitude result is evaluated as
\begin{equation}
	{\cal R}_{{\cal A}, 6}^{(2)} \approx -0.0067411176355825 -0.0287105762581458 i \,,
\end{equation}
by setting similarly 
\begin{equation}
T^2 = -10^{-16} \,, \qquad S^2 = -{49 \over40} \,,
\end{equation}
one obtains the difference as
\begin{align}
	\left( 4{\cal R}_{{\cal A}, 6}^{(2)} +4 \zeta_4 \right) - \left( {\cal R}_{{\cal F}, 4}^{(2)} \big|_{\textrm{T.C}} \right) \approx 1.2 \times 10^{-14} + 3.0 \times 10^{-15} i \,.
\end{align}

Finally, we mention that our result is also consistent with the series expansion of the analytic remainder for $2 \rightarrow 4$ Minkowski kinematics with all 3 cross ratios being equal \cite{Dixon:private}.

\section{Directional dual conformal symmetry at the integrand level}
\label{sec:DDCIintegrand}

In this section, we consider a hidden dual conformal symmetry for the form factor  in the limit where the operator momentum is taken to be lightlike, \emph{i.e.}~$q^2\rightarrow0$. We will show that the {\it integrand} of the form factor up to two loops has an exact directional dual conformal symmetry.

\subsection{Overview of the symmetry}

\begin{figure}[tb]
  \centering
  \includegraphics[scale=0.41]{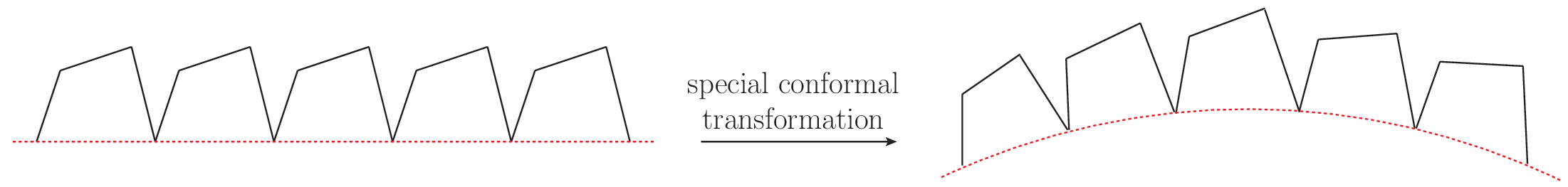}
  \caption{A special conformal transformation generically breaks the periodicity for the Wilson line configuration if $q^2 \neq 0$.}
  \label{fig:WLtransformation}
\end{figure}

The strong coupling picture for the form factor \cite{Alday:2007he, Maldacena:2010kp, Gao:2013dza} suggests a duality of the form factor and the periodic null Wilson loops. This prescription for the duality at the one-loop level was given in \cite{Brandhuber:2010ad}, and the anomalous Ward identity has been studied in \cite{Bianchi:2018rrj}. Nevertheless, the dual conformal symmetry may not be expected to be exactly true for the form factor, since a general special conformal transformation will break the periodicity of the Wilson line configuration; as shown in Figure~\ref{fig:WLtransformation}.\footnote{A ``twisted" picture for the periodic condition was considered in \cite{Ben-Israel:2018ckc}. It would be interesting to show if the symmetry can be used in a precise way along this line.} 

An exception is the case where the form factor has the lightlike period $q$. In this case, the dual Wilson line picture preserves the periodicity for the special conformal transformation made along the lightlike $q$ direction. We refer to this transformation as the directional dual conformal (DDC) transformation,\footnote{The DDC symmetry for certain non-planar amplitudes was also studied in \cite{Bern:2018oao, Chicherin:2018wes}.} 
and the form factor with $q^2=0$ will be named the lightlike form factor. This DDC symmetry is expected to be an exact symmetry for the lightlike form factor (up to a proper subtraction of the well-understood IR anomaly).

For the planar amplitudes, the dual conformal symmetry implies that there are $3n-15$ independent cross ratios for an $n$-point amplitude. A similar counting for the planar form factor, while taking into account the DDC symmetry, shows that there are $3n-9$ independent conformal ratios for an $n$-point lightlike form factor. The study of the two-loop four-point lightlike form factor has shown that its two-loop remainder depends only on three independent conformal ratios, which verifies this symmetry explicitly \cite{Guo:2022qgv}.\footnote{At the tree level, the lightlike form factor with $q^2=0$ has been considered in \cite{Bork:2015fla, Bork:2016hst}, where the Yangian symmetry and Grassmannian representation were studied.}

At the integrand level, the DDC symmetry was also shown to be true for the three-point lightlike form factor up to four loops \cite{Lin:2021lqo}.
Compared to the planar amplitudes, the planar form factor contains integrands that are of the non-planar topologies. The dual conformal coordinates involving non-planar topologies are more complicated, for which the choice of dual coordinates is not unique but depends on how to treat the operator momentum $q$. The DDC symmetry is in general not manifest for each integrand associated with a single topology but for a non-trivial combination of different topologies. 

Below we consider the DDC of the four-point form factor at the integrand level. Compared to the three-point case, some new complications are the appearance of the $\mu$ terms and parity-odd contributions. We will show that they also preserve the DDC symmetry by extending the dual conformal symmetry to be in $D$ dimensions.
A similar treatment for the five-dimensional DDC symmetry of the regularizing amplitude was discussed in \cite{Alday:2009zm}.

\begin{figure}[tb]
  \centering
  \includegraphics[scale=0.5]{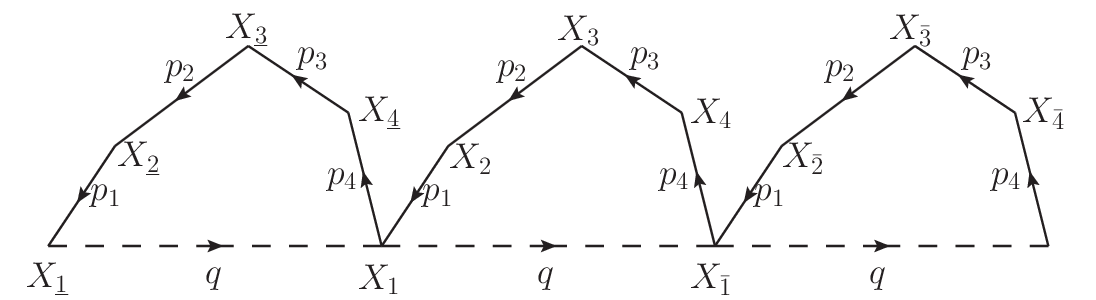}
  \caption{Dual periodic Wilson line configuration for the four-point form factor.}
  \label{fig:WL4pt}
\end{figure}

\begin{figure}[tb]
	\centering
	\includegraphics[scale=0.5]{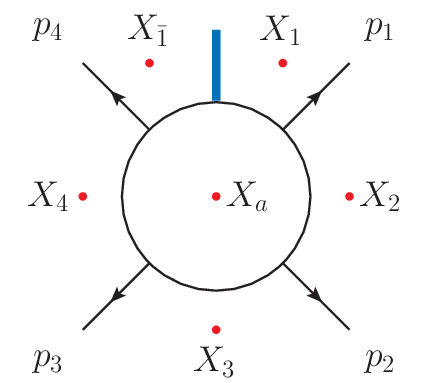}
	\caption{The dual coordinates for the one-loop four-point form factor.}
	\label{fig:oneloopdci}
  \end{figure}

\paragraph{Directional dual conformal transformation.}
We first introduce the dual coordinates (or the so-called zone variables) in the momentum space of the form factor as
 \begin{equation}
	X_i - X_{i+1} = p_i\,, \qquad X_{\underline{i}} - X_i = X_i - X_{\bar{i}} = q \,,
\end{equation}
see Figure~\ref{fig:WL4pt} and Figure~\ref{fig:oneloopdci} for the one-loop form factor.
We take the dual coordinates in $D$ dimensions as $X^M$, $M \in \{0,1,\ldots,D-1\}$, which will allow us to take the $\mu$ terms into account
\begin{equation}
\mu_{ab}  = -(X_{a}-X_i)^m (X_{b}-X_j)_m = -(X_{a})^m (X_{b})_m \,, \qquad m \in \{4,\ldots,D-1\} \,.
\end{equation}
Here $X_{i,j}$ are related to the external momenta living in the four-dimension, thus the extra-dimensional components $X_i^m$ are $0$.
The dual coordinate $X_{a,b}$ are related to the internal momenta and are fully $D$ dimensional vectors in the dimensional regularization scheme.

Next, the special dual conformal transformation with a specific conformal boost vector $B$, in the momentum space is defined as
\begin{align}
	\delta_{B} X_{a}^M \equiv \frac{1}{2} X_{a}^{2} B^{M} -\left( X_{a} \cdot B \right) X_{a}^{M} \,,
\end{align}
where the $M \in \{0,1,\ldots,D-1\}$,
the definition is the same as the equation $(28)$ in \cite{Alday:2009zm}, which specifies the dimension as $5$ and treats the extra component as a mass.

Accordingly, the transformation for a distance is
\begin{align}
	\label{formula:DDCIdelta}
	\delta_B X_{ab}^2 \equiv \delta_B \left[ (X_{a}-X_b)^M (X_{a}-X_b)_M \right] = -B \cdot(X_{a} +X_{b}) X_{ab}^2 \,,
\end{align}
where $M \in {0,1,\ldots,D-1}$.
In particular, the $\mu$-term variables satisfy
\begin{equation}
	\delta_B \mu_{ab} = \delta_B \left[ -(X_{a})^m (X_{b})_m \right] = -B \cdot(X_{a} +X_{b}) \mu_{ab} \,,
\end{equation}
where $m \in \{4,\ldots,D-1\}$.
Moreover, the transformation of the $D$-dimensional measure is
\begin{align}
	\delta_B {\rm d}^D X_a = -D (B\cdot X_a) {\rm d}^D X_a \,.
\end{align}

As we discussed before, the periodicity of the periodic Wilson line configuration in the dual conformal space will be preserved if the $B$ is parallel to the lightlike $q$. As a result, the integrand is expected to be invariant under the special dual conformal transformation $\delta_q$.

We comment that one complexity is about the special dual conformal transformation for the parity-odd part. We transform the parity-odd variables $\varepsilon(a,b,c,d)$ that involve internal momenta into Lorentz dot products using \eqref{eq:varepsilonproduct}.\footnote{One may also perform DDC transformation for $\varepsilon(a,b,c,d)$ variables directly. However, the anti-symmetry variables $\varepsilon$ satisfy subtle relation as
$\varepsilon(a,b,c,d) e^{\mu} +\text{cyclic}(a,b,c,d,e) = 0$,
where $\mu \in \{0,1,2,3\}$. In practice, we find it is more straightforward to apply  \eqref{eq:varepsilonproduct}.} 
The remaining parity-odd variable $\varepsilon(1234)$ satisfies the following transformation
\begin{equation}
	\label{eq:deltaqvarepsilon}
	\delta_q \varepsilon(1234) \equiv \delta_q \varepsilon(X_{12}, X_{23}, X_{34}, X_{4\bar{1}}) 
	= -q \cdot (X_{1} +X_{2} +X_{3} +X_{4}) \varepsilon(1234) \,, 
\end{equation}
where $X_{ij}$ in $\varepsilon(X_{12}, X_{23}, X_{34}, X_{4\bar{1}})$ is contracted only for the four-dimensional part.

\subsection{One-loop level}

We first consider the one-loop case. The one-loop integrand result can be expressed in terms of the dual coordinates as given in Figure~\ref{fig:oneloopdci} and the cycling permutations.
We use $\mathbb{I}_3^{(1)}$ as an example to explain our strategy, whose numerator and propagator are given in Table~\ref{tab:oneloopintegrand}:
\begin{align}
	{\rm d}\mathbb{I}_3^{(1)} \equiv {\rm d}^DX_a \, \mathbb{I}_3^{(1)} = {\rm d}^DX_a  \frac{{\rm tr}_{-}(X_{\underline{4}1}, X_{12}, X_{23}, X_{34}) \mu_{aa}}{X_{1a}^2 X_{2a}^2 X_{3a}^2 X_{4a}^2 X_{\bar{1}a}^2} \,,
\end{align}
We note that at the integrand level, it is safe to take the lightlike limit $q^2 = 0$ directly since the integrand is free from any pole or branch cut of $q^2$. 

The dual conformal transformation of the integrand can be obtained as
\begin{align}
	\label{eq:oneloopI3}
	\delta_q {\rm d}\mathbb{I}_3^{(1)} \big|_{q^2=0} = \left[ (4-D) q \cdot X_a +q \cdot X_{1a} \right] {\rm d}\mathbb{I}_3^{(1)} \big|_{q^2=0} \,,
\end{align}
and the trace expression ${\rm tr}_{-}(X_{\underline{4}1}, X_{12}, X_{23}, X_{34})$ can be treated similarly as \eqref{eq:deltaqvarepsilon}.
As shown in the above formula, we find that the right-hand side does not vanish when $D=4$ due to the term that is proportional to $q \cdot X_{1a}$. 
Luckily, this extra term can be reduced to sub-topology contributions, since 
\begin{equation}
q \cdot X_{1a} = (X_{1a}^2 -X_{\bar{1}a}^2)/2 \,,
\end{equation} 
which are both propagators. It turns out that other integrands can produce the same sub-topology contributions which cancel with each other after summing together. 
This property is very different from planar amplitudes where the dual conformal symmetry is manifested by each integrand separately.

To give some more details, we find that it is necessary to decompose the integrand into parity-even and parity-odd parts as for the ${\rm tr}_{-}$ factor according to \eqref{formula:traceidlength4}:
\begin{equation}
{\rm tr}_{-}(X_{\underline{4}1}, X_{12}, X_{23}, X_{34}) = {1\over2}{\rm tr}(X_{\underline{4}1}, X_{12}, X_{23}, X_{34})  +2i \varepsilon(X_{\underline{4}1}, X_{12}, X_{23}, X_{34}) \,.
\end{equation}

For the parity-even part, we find that the extra term that is proportional to $q \cdot X_{1a}$ in \eqref{eq:oneloopI3} is canceled by summing cycling the external momenta as
\begin{align}
	& \sum_{i=1}^4 {\rm d}^D X_a  \frac{q \cdot (X_i -X_a) {\rm tr}(X_{\underline{4}1}, X_{12}, X_{23}, X_{34}) \mu_{aa}}{2X_{ia}^2 X_{i+1,a}^2 X_{i+2,a}^2 X_{i+3,a}^2 X_{\bar{i}a}^2} \notag \\
	= & \sum_{i=1}^4 {\rm d}^D X_a  \frac{{\rm tr}(X_{\underline{4}1}, X_{12}, X_{23}, X_{34}) \mu_{aa}}{4X_{i+1,a}^2 X_{i+2,a}^2 X_{i+3,a}^2} \left( \frac{1}{X_{ia}^2  } -\frac{1}{ X_{\bar{i}a}^2} \right) \notag \\
	= & {\rm d}^D X_a  \frac{{\rm tr}(X_{\underline{4}1}, X_{12}, X_{23}, X_{34}) \mu_{aa}}{4} \left( \frac{1}{X_{1a}^2 X_{2a}^2 X_{3a}^2 X_{4a}^2} -\frac{1}{X_{\bar{1}a}^2 X_{\bar{2}a}^2 X_{\bar{3}a}^2 X_{\bar{4}a}^2} \right) \notag \\
	= & 0 \,,
\end{align}
where $X_5 \equiv X_{\bar{1}}$, etc. In the third line of the above formula, we shift the internal coordinate $X_i$ in the first integrand by $q$, which leads to $X_{ia}^2 \to X_{\bar{i}a}^2$. This shift can be understood as a relabelling of the propagators, which identifies the same Feynman integrands which have the different propagators labels. Since there is no need to perform any integration, we refer to the above operation as a property at the {\it integrand} level.

In other words,
\begin{align}
	& \delta_q \left( {\rm d}\mathbb{I}_{3, \text{even}}^{(1)} +\text{cyclic}(p_1, p_2, p_3, p_4)\right) \big|_{q^2=0} \notag \\
	= & (4-D) q \cdot X_a \left( {\rm d}\mathbb{I}_{3, \text{even}}^{(1)} +\text{cyclic}(p_1, p_2, p_3, p_4)\right) \big|_{q^2=0} \,.
\end{align}

As for the parity-odd part of the same extra term in \eqref{eq:oneloopI3}, they do not cancel by summing only the permutation of $\mathbb{I}_3^{(1)}$, since the parity-odd variable $\varepsilon(1234)$ will change the sign under permutation. We find that it is necessary to take the parity-odd parts of the integrands $\mathbb{I}_1^{(1)}$ and $\mathbb{I}_2^{(1)}$ into account. Here $\mu_{aa}$ in $\mathbb{I}_3^{(1)}$ can be transformed into Mandelstam variables by the formula \eqref{eq:muproduct}. As a result, we find
\begin{align}
	& \delta_q  \left( {\rm d}\mathbb{I}_{1, \text{odd}}^{(1)} +{\rm d}\mathbb{I}_{2, \text{odd}}^{(1)} +{\rm d}\mathbb{I}_{3, \text{odd}}^{(1)} +\text{cyclic}(p_1, p_2, p_3, p_4) \right) \big|_{q^2=0} \nonumber \\
	= & (4-D) q\cdot X_a \left( {\rm d}\mathbb{I}_{1, \text{odd}}^{(1)} +{\rm d}\mathbb{I}_{2, \text{odd}}^{(1)} +{\rm d}\mathbb{I}_{3, \text{odd}}^{(1)} +\text{cyclic}(p_1, p_2, p_3, p_4) \right) \big|_{q^2=0} \,.
\end{align}
We comment on the parity-odd parts of $\mathbb{I}_{1}^{(1)}$ and $\mathbb{I}_{2}^{(1)}$, which vanishes after integration but has a non-trivial contribution in verifying dual conformal symmetry.

Finally, a similar calculation can be performed for the parity-even part of  $\mathbb{I}_1^{(1)}$ and $\mathbb{I}_2^{(1)}$ as
\begin{align}
	& \delta_q \left( {\rm d}\mathbb{I}_{1, \text{even}}^{(1)} +{\rm d}\mathbb{I}_{2, \text{even}}^{(1)} +\text{cyclic}(p_1, p_2, p_3, p_4) \right) \big|_{q^2=0} \nonumber \\
	= & (4-D) q\cdot X_a \left( {\rm d}\mathbb{I}_{1, \text{even}}^{(1)} +{\rm d}\mathbb{I}_{2, \text{even}}^{(1)} +\text{cyclic}(p_1, p_2, p_3, p_4) \right) \big|_{q^2=0} \,.
\end{align}

In total, we confirm the DDC symmetry in $D=4$ for the one-loop integrand of the form factor:
\begin{align}
	\delta_q {\rm d}\mathbb{I}^{(1)} \big|_{q^2=0} = (4-D) q\cdot X_a {\rm d}\mathbb{I}^{(1)} \big|_{q^2=0} \,,
\end{align}
where $\mathbb{I}^{(1)}$ is the full one-loop integrand of the form factor as given in \eqref{eq:calI1loop}.

\subsection{Two-loop level}

To discuss the two-loop case, we will first decompose the two-loop integrand into the $\mu$-term and non-$\mu$-term contributions
\begin{equation}
	\label{eq:twopartcontribution}
	\mathbb{I}_i = \mathbb{I}_{i}^{(\text{non-}\mu)} +\mathbb{I}_{i}^{(\mu)} \,,
\end{equation}
where the $\mu$-term part $\mathbb{I}_{i}^{(\mu)}$ can be specified by expanding all higher-length traces ${\rm tr}_-$ to length-$4$ ones and collecting the terms proportional to $\mu_{ij}$.
We will ignore the two-loop superscript ``(2)" for simplicity. 
As mentioned in Section~\ref{subsec:2loopsec1}, such a decomposition is not unique. 
A specific simple choice for the numerators of $\{\mathbb{I}_{i}^{(\mu)}\}$ can be given as follows\footnote{We find that the $\mathbb{I}^{(\mu)}$ as the summation of all $\mathbb{I}_i^{(\mu)}$ is uniformly transcendental, and its parity-even part matches exactly the parity-even part of ${\cal I}_{{\cal F},4}^{(2)}\big|_{\mu\text{-term UT}}$ after the IBP reduction. Thus, it cancels out by the BDS subtraction, see also Section~\ref{subsec:remainder}.}
\begin{align}
	\label{eq:Nmu}
	& N^{(\mu)}_1 = -\frac{1}{2} q^2 \mu _{\ell_3\ell_3} {\rm tr}_-(4,1,2,3) \,, \qquad N^{(\mu)}_2 = s_{123}\mu_{\ell_1\ell_7} {\rm tr}_-(4,1,2,3) \,, \\
	& N^{(\mu)}_3 = -s_{23} \mu _{\ell\ell}  {\rm tr}_-(1,2,3,4) \,, \qquad \quad \, N^{(\mu)}_7 =- s_{12} {\rm tr}_-(4,1,2,3)\mu_{\ell_1\ell_1} \,, \nonumber \\
	& N^{(\mu)}_8 = -(s_{123}\mu_{\ell_1\ell_7}+q^2\mu_{\ell_1\ell_1}) {\rm tr}(1,2,3,4)
	\,, \nonumber \\
	& N^{(\mu)}_9 = -{\rm tr}_-(4,1,2,3) \big[ \mu_{\ell_1\ell_1} s_{34}+\mu_{\ell_4\ell_4}s_{12}+\mu_{\ell_1\ell_4}(s_{12}+s_{34}-q^2) \big] \,. \nonumber
\end{align}
For example, the numerator $N_7$ can be decomposed as
 \begin{align}
	\label{eq:N7expand}
 	N_7 = & s_{12}{\rm tr}_-(4,1,2,3,\ell_2,\ell_1) \notag \\
 	= & s_{12} \left[s_{23}\text{tr}_-(4,1,2,\ell_1)-\text{tr}_-(2,3,\ell_1,1)s_{41}-(\ell_1-p_1)^2\text{tr}_-(4,1,2,3) \right] \notag\\
	& - s_{12} \text{tr}_-(4,1,2,3)\mu_{\ell_1\ell_1} \,, 
 \end{align}
where $\ell_2 = p_1+p_2-\ell_1$, and the last term gives $N^{(\mu)}_7$ and other terms contribute to $N^{(\text{non-}\mu)}_7$.

Below we will verify the DDC symmetry for the non-$\mu$-term and $\mu$-term parts defined in \eqref{eq:twopartcontribution}, and we show that they satisfy the DDC symmetry separately. 

\begin{figure}[t]
	\centering
	\subfigure[$N_{2}$]{\includegraphics[scale=0.3]{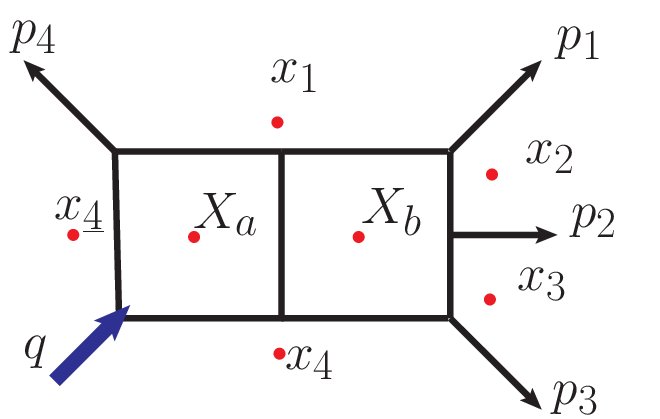}\label{Fig:Twodoubleddcitop6}}
	\subfigure[$N_{3}$]{\includegraphics[scale=0.3]{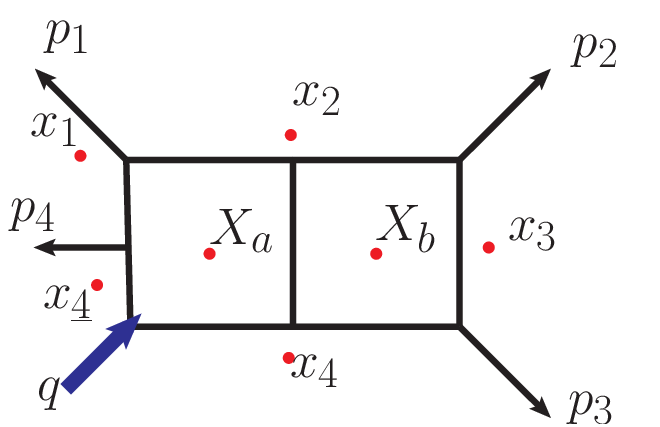}}
	\subfigure[$N_{5}$]{\includegraphics[scale=0.3]{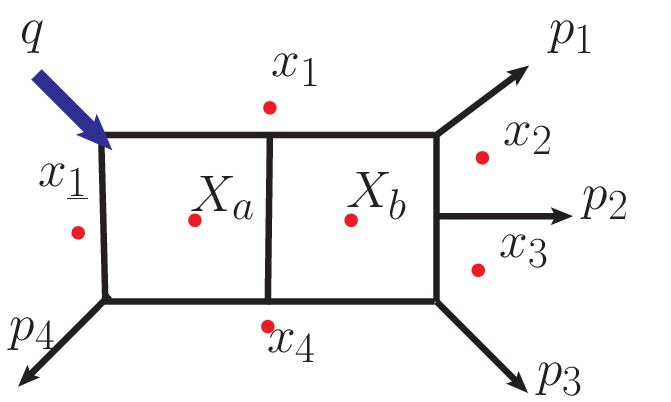}\label{Fig:Twodoubleddcitop7}}
	\subfigure[$N_{6}$]{\includegraphics[scale=0.3]{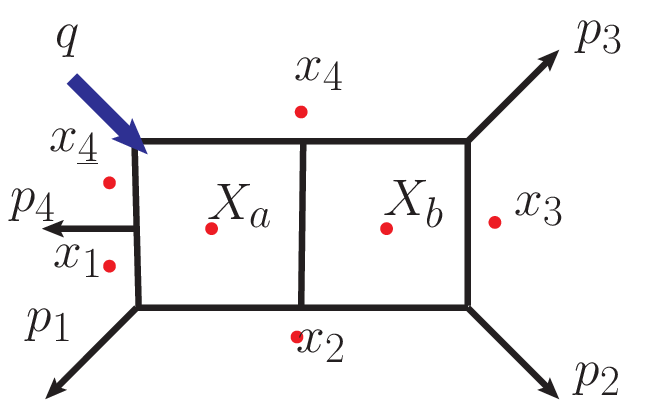}}
	\subfigure[$N_{7}$]{\includegraphics[scale=0.3]{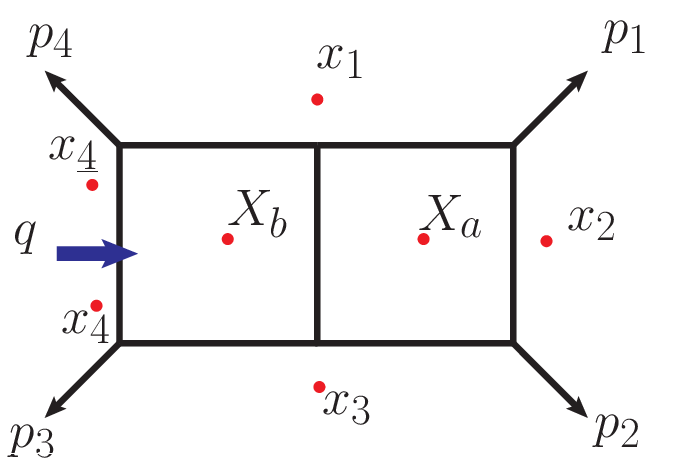}\label{Fig:Twodoubleddcitop2}}
	\subfigure[$N_{8}$]{\includegraphics[scale=0.3]{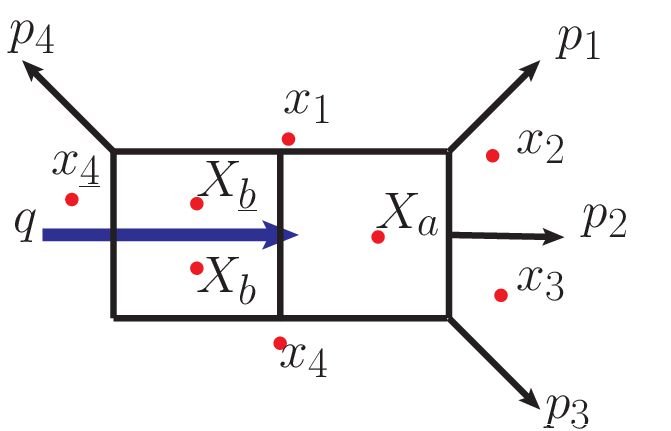}\label{Fig:Twodoubleddcitop8}}
	\subfigure[$N_{9}$]{\includegraphics[scale=0.3]{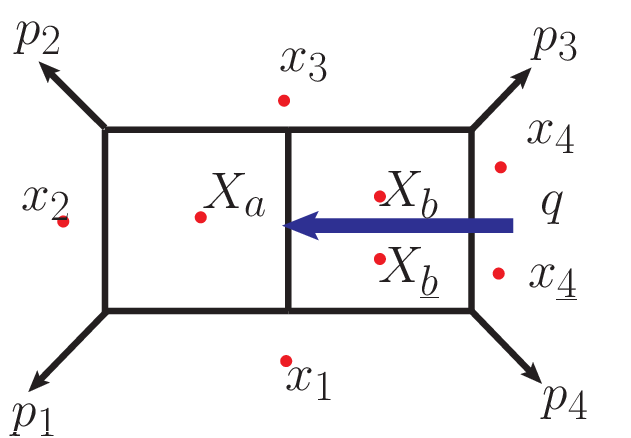}}
	\caption{The dual coordinates for the maximal topologies of the two-loop four-point form factor.}
	\label{fig:TwoLoopFourptwithLength2DCI}
\end{figure}

 When we take the lightlike limit $q^2=0$, the two numerators $N_1$ and $N_4$ simply vanish, since they are proportional to $q^2$.
Hence, we only need to consider the remaining topologies with the dual coordinates shown in Figure~\ref{fig:TwoLoopFourptwithLength2DCI},
where $X_{\underline{b}} = X_{b}+q$ is introduced for the non-planar topologies. We comment here that the labels of the dual coordinates are not unique. However, the symmetry property should be true for general assignments of dual coordinates. Different assignments can be understood as a relabelling of the propagators, while the property of DDCI is independent of the particular choice. 

The details of the computation are similar to the one-loop case, below we will mainly present the results.
We use the short notation as
\begin{equation}
{\rm d}\mathbb{I}_{i} \equiv {\rm d}^DX_a{\rm d}^DX_b \, \mathbb{I}_{i} \,.
\end{equation}

\paragraph{$\text{Non-}\mu$ part.}
 We use $\mathbb{I}_{7}^{(\text{non-}\mu)}$ as an example to explain the parametrization of the integrand in terms of dual coordinates. 
 Using \eqref{eq:N7expand}, we can decompose $N_7^{(\text{non-}\mu)}$ further into parity-even and parity-odd parts as
 \begin{align}
 	N_7^{(\text{non-}\mu)} & =	s_{12}\Big[s_{23}\text{tr}_-(4,1,2,\ell_1)-\text{tr}_-(2,3,\ell_1,1)s_{41}-(\ell_1-p_1)^2\text{tr}_-(4,1,2,3)\Big] \notag \\
 	& = \frac{s_{12}}{2}\Big[s_{23}\text{tr}(4,1,2,\ell_1)-\text{tr}(2,3,\ell_1,1)s_{41}-(\ell_1-p_1)^2\text{tr}(4,1,2,3)\Big] \notag \\
 	& \quad +2is_{12}\Big[s_{23}\varepsilon(412\ell_1)-\varepsilon(23\ell_11)s_{41}-(\ell_1-p_1)^2\varepsilon(4123)\Big]\,.
 \end{align}
The parity-even part can be rewritten as
 \begin{align}
 	N_{7, \text{even}}^{(\text{non-}\mu)} \big|_{q^2=0} 
 	= \frac{s_{12}}{2} \big[ & s_{12} s_{14} (\ell_1-p_1-p_2-p_3)^2 -\ell_1^2 s_{124} s_{23} -s_{123} s_{14} (\ell_1-p_1-p_2)^2 \notag \\
 	& +s_{123} s_{124} (\ell_1-p_1)^2 +s_{12} s_{23} (\ell_1+p_4)^2 \big] \,,
 \end{align}
and the corresponding integrand in dual coordinates will be
 \begin{align}
 	\label{eq:ddcI7even}
 	& \mathbb{I}_{7,\text{even}}^{(\text{non-}\mu)} \big|_{q^2=0} = X_{13}^2 \frac{X_{13}^2 ( X_{2\underline{4}}^2 X_{4b}^2 +X_{42}^2 X_{b\underline{4}}^2) +X_{14}^2 (X_{3\underline{4}}^2 X_{b2}^2-X_{2\underline{4}}^2 X_{b3}^2) -X_{3\underline{4}}^2 X_{42}^2 X_{b1}^2}{2X_{1a}^2 X_{2a}^2 X_{3a}^2 X_{\underline{4}b}^2 X_{1b}^2 X_{3b}^2 X_{4b}^2 X_{ab}^2} \,.
 \end{align}
 For the parity-odd part, we transform the parity-odd variables $\varepsilon(412\ell_1)$ and $\varepsilon(23\ell_11)$ into Lorentz dot products using \eqref{eq:varepsilonproduct}.
Similar parametrization applies to other topologies and it is ready to implement the dual conformal transformation.

For ${\rm d}\mathbb{I}_{7}^{(\text{non-}\mu)}$ and ${\rm d}\mathbb{I}_{9}^{(\text{non-}\mu)}$, we find that they are DDC invariant by themselves:
\begin{align}
	\delta_q \left( {\rm d}\mathbb{I}_{7}^{(\text{non-}\mu)}  \right)\big|_{q^2=0} = (4-D) q\cdot(X_a+X_b) \left( {\rm d}\mathbb{I}_{7}^{(\text{non-}\mu)} \right) \big|_{q^2=0} \,, \label{eq:I7nonmu}\\
	\delta_q \left( {\rm d}\mathbb{I}_{9}^{(\text{non-}\mu)}  \right)\big|_{q^2=0} = (4-D) q\cdot(X_a+X_b) \left( {\rm d}\mathbb{I}_{9}^{(\text{non-}\mu)} \right) \big|_{q^2=0} \,. \notag
\end{align}
For the integrands of the other three topologies, we find that they are DDC invariant by combining the integrands from the different topologies as
\begin{align}
	& \delta_q \left({\rm d}\mathbb{I}_2^{(\text{non-}\mu)} +{\rm d}\mathbb{I}_5^{(\text{non-}\mu)} +{\rm d}\mathbb{I}_8^{(\text{non-}\mu)} +\text{cyclic}(p_1, p_2, p_3, p_4)  \right) \big|_{q^2=0} \label{eq:I258A} \\
	= & (4-D) q\cdot(X_a +X_b) \left({\rm d}\mathbb{I}_2^{(\text{non-}\mu)} +{\rm d}\mathbb{I}_5^{(\text{non-}\mu)} +{\rm d}\mathbb{I}_8^{(\text{non-}\mu)} +\text{cyclic}(p_1, p_2, p_3, p_4)  \right)\big|_{q^2=0}  \,. \notag
\end{align}

\paragraph{$\mu$-term part.}
The calculation of the $\mu$-term part $\mathbb{I}_{7}^{(\mu)}$ is very similar to the one-loop case $\mathbb{I}_{3}^{(1)}$. From \eqref{eq:N7expand}, we read the numerator as
\begin{align}
	N^{(\mu)}_7 \big|_{q^2=0} = -X_{13}^2 {\rm tr}_{-}(X_{\underline{4}1}, X_{12}, X_{23}, X_{34}) \mu_{bb} \,.
\end{align}
As in the one-loop case, it is not DDC invariant by itself but satisfies
\begin{align}
	\label{eq:muN7der}
	& \delta_{q} \left( {\rm d} \mathbb{I}^{(\mu)}_7 \right)\big|_{q^2=0} = \left[ (4-D) q \cdot (X_a +X_b) +q \cdot X_{4b} \right]\left(  {\rm d} \mathbb{I}^{(\mu)}_7  \right) \big|_{q^2=0} \,,
\end{align}
where the extra factor $q \cdot X_{4b} = X_{4b}^2 -X_{\bar{4}b}^2$ will cancel the propagators.

The DDC invariance for the $\mu$-term part requires combining the integrands from the different topologies as
\begin{align}
	& \delta_q \left({\rm d}\mathbb{I}^{(\mu)}_3 +{\rm d}\mathbb{I}^{(\mu)}_6 +{\rm d}\mathbb{I}^{(\mu)}_7 +{1\over2}{\rm d}\mathbb{I}^{(\mu)}_9  +\text{cyclic}(p_1, p_2, p_3, p_4)\right) \big|_{q^2=0} \label{eq:I3679B} \\
	= & (4-D) q\cdot(X_a +X_b) \left( {\rm d}\mathbb{I}^{(\mu)}_3 +{\rm d}\mathbb{I}^{(\mu)}_6 +{\rm d}\mathbb{I}^{(\mu)}_7 +{1\over2}{\rm d}\mathbb{I}^{(\mu)}_9 +\text{cyclic}(p_1, p_2, p_3, p_4)  \right)\big|_{q^2=0} \,, \notag \\
 	& \delta_q \left( {\rm d}\mathbb{I}^{(\mu)}_2 +{\rm d}\mathbb{I}^{(\mu)}_5 +{\rm d}\mathbb{I}^{(\mu)}_8+\text{cyclic}(p_1, p_2, p_3, p_4) \right)   \big|_{q^2=0}\,,  \\=& (4-D) q\cdot(X_a +X_b) \left( {\rm d}\mathbb{I}^{(\mu)}_2 +{\rm d}\mathbb{I}^{(\mu)}_5 +{\rm d}\mathbb{I}^{(\mu)}_8+\text{cyclic}(p_1, p_2, p_3, p_4) \right)  \big|_{q^2=0} \,. \notag
\end{align}

In summary, we confirm the DDC symmetry in $D=4$ for the two-loop integrand of the form factor as
\begin{align}
	\delta_q {\rm d}\mathbb{I}^{(2)} \big|_{q^2=0} =  (4-D) q\cdot(X_a +X_b) {\rm d}\mathbb{I}^{(2)} \big|_{q^2=0} \,,
\end{align}
where $\mathbb{I}^{(2)}$ is the full integrand of the form factor as given in \eqref{eq:calI2loop}.

\section{Discussion}
\label{sec:discussion}

In this paper, we provide a first-principle computation of the two-loop four-point MHV form factor of the stress-tensor supermultiplet in ${\cal N}=4$ SYM for both the integrand and the integrated results. 
Our form factor result not only reproduces the symbol result obtained via the symbol-bootstrap method \cite{Dixon:2022xqh} but also provides the full functional results for the first time which is expected to shed new light on the study of the antipodal duality.

The form factor we consider provides a non-trivial application of the two-loop five-point one-mass integrals in the non-planar sectors that have been recently evaluated in \cite{Abreu:2023rco} together with previous integral results \cite{Abreu:2020jxa, Canko:2020ylt, Abreu:2021smk}. The consistency of our results thus serves as an important test of the new analytic expressions of the master integrals.

Our computations, which encompass both the $D$-dimensional integrand and the integral reductions, should be generalizable to four-point form factors in other theories.
In particular, the four-point form factor we consider can be viewed as an analog to the Higgs-plus-two-gluon amplitudes in QCD in the heavy top-mass limit \cite{Ellis:1975ap, Georgi:1977gs, Wilczek:1977zn, Shifman:1979eb, Dawson:1990zj, Kniehl:1995tn}.
This analogy is particularly relevant for studying the Higgs plus two jets production process at the LHC, which is a crucial aspect of current high-energy physics research. 

One interesting problem to explore is the relation between the ${\cal N}=4$ SYM and the QCD results in the context of the so-called maximal transcendentality principle, which was first observed for the anomalous-dimension-type observables \cite{Kotikov:2002ab, Kotikov:2004er}.
For the three-point form factor, it was found in \cite{Brandhuber:2012vm} that the ${\cal N}=4$ result is equal to the maximally transcendental part of the Higgs-plus-one-gluon amplitude in QCD \cite{Gehrmann:2011aa}. This correspondence was also confirmed for the three-point form factors with higher-dimensional operators \cite{Brandhuber:2014ica, Brandhuber:2017bkg, Jin:2018fak, Brandhuber:2018xzk, Brandhuber:2018kqb, Jin:2019ile, Jin:2019opr, Jin:2020pwh, Banerjee:2016kri} and with external quark states \cite{Jin:2019ile, Banerjee:2017faz}, as well as for the four-point form factor of the dimension-6 operator ${\rm tr}(F_\mu^\nu F_\nu^\rho F_\rho^\mu)$ \cite{Guo:2022pdw}.

For the four-point form factor of the stress-tensor supermultiplet, which contains the dimension-4 operator ${\rm tr}(F_{\mu\nu}F^{\mu\nu})$, 
the maximal transcendentality principle will not work directly as in the three-point case. This is because the four-point form factor should reproduce the four-gluon amplitude in the $q^2 \rightarrow 0$ limit, while it is known that the maximal transcendentality principle does not apply to the latter; see discussion in \cite{Guo:2022pdw, Dixon:2022xqh}. On the other hand, the analysis of the master-integral bootstrap method in \cite{Guo:2022pdw, Guo:2021bym} suggests that there could still be useful relations between ${\cal N}=4$ and QCD form factors since they share similar physical constraints from IR divergences and collinear limits.

Another generalization of the present work is to consider the four-point form factor in the next-to-MHV and next-to-next-to-MHV sectors, as well as beyond the planar limit. The color-kinematics duality \cite{Bern:2008qj, Bern:2010ue} is expected to be useful in studying the non-planar sectors, which have been successfully applied in the Sudakov and three-point form factors up to four or five loops \cite{Boels:2012ew, Yang:2016ear, Lin:2021kht, Lin:2021lqo}. 
We leave these for future studies.

\section*{Acknowledgements}
We would like to thank Song He, Guanda Lin, Xiao Liu, Yanqing Ma, and in particular Lance Dixon for the discussions.
This work is supported in part by the National Natural Science Foundation of China (Grants No.~12175291, 11935013, 12047503), and the Chinese Academy of Sciences (Grant No. YSBR-101). 
GY would like to thank the Simons Center for Geometry and Physics for its hospitality. 
We also acknowledge the support of the HPC Cluster of ITP-CAS.

\appendix

\section{Convention of one-loop UT master integrals}
\label{app:UT}

In this appendix, we list all one-loop UT integrals used in the paper. 
The convention is
\begin{equation}
	I^{(1)}[N(l, p_j)] = e^{\epsilon\gamma_\mathrm{E}} \int\frac{d^D l}{i\pi^{\frac{D}{2}}} 
	\frac{N(l, p_j)}{\prod_k D_k} \,.
\end{equation}
The one-loop master integrals are $I_{\text{Bub}}^{(1)}(1,\ldots,n)$, $I^{(1)}_{\text{Box}}(i,j,k)$ and $I^{(1)}_{\text{pen}}(i,j,k)$
\begin{align}
&	I^{(1)}_{\text{Bub}}(1, \ldots, n) = \frac{1-2\epsilon}{\epsilon} \times
\begin{aligned}
	\includegraphics[scale=0.35]{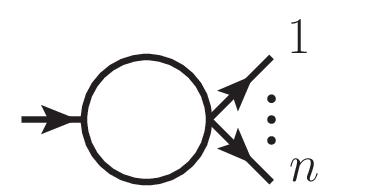}
\end{aligned} \,, \\
&	I^{(1)}_{\text{Box}}(i,j,k) = (s_{ij} s_{jk} -p_j^2 q^2) \times
\begin{aligned}
	\includegraphics[scale=0.3]{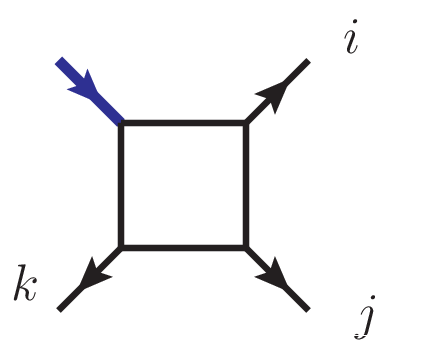}
\end{aligned} \,, \label{eq:Ibox} \\
&	I^{(1)}_{\text{Pen}}(i,j,k,l) = 4i \varepsilon(1234) \, \mu_{ll} \times
\begin{aligned}
	\includegraphics[scale=0.3]{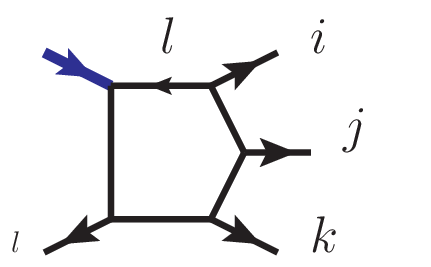}
\end{aligned} \,, \label{eq:Pen1loop}
\end{align}
where propagators are represented by the figures, $\mu_{ll}$ in \eqref{eq:Pen1loop} is understood as inside the integral,
and the momentum $p_j$ in \eqref{eq:Ibox} can be both massless or massive, the latter as the sum of two massless momenta.

\section{Further numerical evaluations}
\label{app:consistency}

In this appendix, we apply a series of consistency checks on the integrand expression without IBP reduction, such as IR divergence and collinear limit. We apply at least $16$-digits evaluation with {\bf AMFlow}.\footnote{The accuracy of the result as an option of the package is confirmed by various consistency checks of the physical requirements, such as the infrared subtraction and the collinear limit.}

\paragraph{Euclidean region}

We provide two numerical results of the two-loop four-point form factor and the remainder in Table~\ref{tab:2loopEuclidean}. The remainder function is free from the parity-odd part, which is confirmed by the result in the second column. The reason is Feynman integrals have real values in the Euclidean region. The imaginary number in the result can only originate from the odd-parity kinematics $4i\varepsilon(1234)$ (for our case, it's $12\sqrt{6}i$), and those terms are known as parity-odd part. As a result, the remainder must be a real number to be parity even.

\begin{table}[t]
	\centering
	\vskip .1 cm
	\begin{tabular}{| c | c | c | c | c |}
		\hline
		& \multicolumn{2}{c|}{${\cal I}_{{\rm tr}(\phi_{12}^2), 4}^{(2)}$} \\ \hline \hline
		$\epsilon^{-4}$ & \multicolumn{2}{c|}{$8$} \\ \hline
		$\epsilon^{-3}$ & $28.6681515076488800$ & $ 22.5391584126769984$ \\ \hline
		$\epsilon^{-2}$ & $66.6674225806393908$ & $51.1201158188825976$ \\ \hline
		$\epsilon^{-1}$ & $75.146634615188871$ & $122.302919710908763 + 0.191216954301575 i$ \\ \hline
		$\epsilon^{ 0}$ & $46.644905519809285$ & $162.415924995407944 - 0.302127906617656 i$ \\ \hline \hline
		${\cal R}_{{\rm tr}(\phi_{12}^2), 4}^{(2)}$ & $0.017537242154528$ & $0.452475160724399$ \\ \hline
	\end{tabular}
	\caption{
		The two numerical results of the two-loop four-point form factors and the finite remainders are given in the last line,
		where the kinematics for the first column are chosen as \{$s_{ij}=-1/6$, $ \varepsilon(1234) =\sqrt{3}/144$\},
		for the second column are:
		\{$s_{12} = -2$, $s_{13} = -3$, $s_{14} = -4$, $s_{23} = -5$, $s_{24} = -6$, $s_{34} = -7$, $ \varepsilon(1234) = 3\sqrt{6}$\}.
		\label{tab:2loopEuclidean}
	}
\end{table}

\paragraph{Physical region}

We provide a numerical result of the two-loop four-point form factor and the remainder in Table~\ref{tab:2loopPhysical}, which can be considered as the scattering process related to $p_1+p_2 \rightarrow p_3+p_4+q$ (compared to $p_3+p_4 \rightarrow p_1+p_2+q$ in Section~\ref{sec:subtraction}). We just point out that it requires
\begin{equation}
	\{s_{12}, s_{34}, s_{4q}, q^2\} > 0 \,, \qquad \{s_{q1}, s_{23}\} < 0 \,,
\end{equation}
where $s_{iq}=(p_i-q)^2$, and their values at the given point are $s_{12}=3$, $s_{34}=7$, $s_{4q}=4$, $q^2=9$, $s_{q1}=-2$, $s_{23}=-5$.

\begin{table}[t]
	\centering
	\vskip .1 cm
	\begin{tabular}{| c | c |}
		\hline
		& ${\cal I}_{{\rm tr}(\phi_{12}^2), 4}^{(2)}$ \\ \hline \hline
		$\epsilon^{-4}$ & $8$ \\ \hline
		$\epsilon^{-3}$ & $21.3884301228698747 - 37.69911184307751886 i$ \\ \hline
		$\epsilon^{-2}$ & $77.3973637486888853 + 107.0908232816918802 i$ \\ \hline
		$\epsilon^{-1}$ & $226.025494928740850 - 19.702307574909437 i$ \\ \hline
		$\epsilon^{ 0}$ & $245.337419852026852 - 286.267057910091597 i$ \\ \hline \hline
		${\cal R}_{{\rm tr}(\phi_{12}^2), 4}^{(2)}$ & $3.064702568868740 + 30.041245265002520 i$ \\ \hline
	\end{tabular}
	\caption{
		The numerical result of the two-loop four-point form factors and the finite remainders are given in the last line,
		where the kinematics are chosen as:
		\{$s_{12} = 3$, $s_{13} = 6$, $s_{14} = 2$, $s_{23} = -5$, $s_{24} = -4$, $s_{34} = 7$, $ \varepsilon(1234) = -{\sqrt{2065}\over4}i$\}.
		\label{tab:2loopPhysical}
	}
\end{table}

\paragraph{Collinear limit}

In Table~\ref{tab:2loopCollinear}, we consider the collinear limit as $p_3 \parallel p_4 \parallel p_3' = p_3+p_4$, where $p_3 = t p_3'$ and $p_4 = (1-t) p_3'$ but set the kinematics variable $s_{34}\sim 10^{-8}$ as a very small number instead of $0$. The kinematics of the three-point form factor are given by \{$s_{12}' = -3$, $s_{23}' = s_{23} +s_{24} = -4$, $s_{13}' = s_{13} +s_{14} = -5$\}, and we obtain the difference between the two form factors at the given kinematics point is
\begin{equation}
	{\cal R}_{{\rm tr}(\phi_{12}^2), 4}^{(2)} \big|_{p_3 \parallel p_4} - {\cal R}_{{\rm tr}(\phi_{12}^2), 3}^{(2)} = -2.29303370631 \times 10^{-7} \sim {\cal O}(s_{34} \log(s_{34})) \,,
\end{equation}
which approaches the two-loop three-point result nicely.
\begin{table}[t]
	\centering
	\vskip .1 cm
	\begin{tabular}{| c | c |}
		\hline
		& ${\cal I}_{{\rm tr}(\phi_{12}^2), 4}^{(2)}$ \\ \hline \hline
		$\epsilon^{-4}$ & $8$ \\ \hline
		$\epsilon^{-3}$ & $63.3216543140261554427469$ \\ \hline
		$\epsilon^{-2}$ & $939.045009417255828095814$ \\ \hline
		$\epsilon^{-1}$ & $8457.62209097131600061234 -0.000528002302047205608899227i$ \\ \hline
		$\epsilon^{ 0}$ & $67897.14700721552346901385 -0.00882585493368254282932391i$ \\ \hline \hline
		${\cal R}_{{\rm tr}(\phi_{12}^2), 4}^{(2)}$ & $4.1859021488936420876$ \\ \hline
	\end{tabular}
	\caption{
		The numerical result of the two-loop four-point form factors and the finite remainders are given in the last line,
		where the kinematics are chosen as:
		\{$s_{12} = -3$, $s_{13} = -5/3$, $s_{14} = -10/3$, $s_{23} = -4/3$, $s_{24} = -8/3$, $s_{34} = -10^{-8}$, $ \varepsilon(1234)=\sqrt{(48\times10^9-81)}/(12\times10^8) $\}.
		\label{tab:2loopCollinear}
	}
\end{table}

\section{Results and comments on the ${\rm tr}(\phi_{12}^3)$ form factor}
\label{app:comments}

In this appendix, we provide the integrand expressions of the two-loop four-point form factor of the length-3 half-BPS operator $\text{tr}(\phi_{12}^3)$. The integrated result has been obtained by the master-integral bootstrap method in~\cite{Guo:2021bym}. We consider the external particle configuration $\{1_\phi, 2_\phi, 3_\phi, 4_{g^+}\}$, and the tree-level form factor is
\begin{align}
	{\cal F}_{{\rm tr}(\phi_{12}^3), 4}^{(0)}(1_\phi, 2_\phi, 3_\phi, 4_{g^+})=\frac{\langle31\rangle}{\langle34\rangle \langle41\rangle}\,.
\end{align}

\begin{figure}[t]
	\begin{center}
		\includegraphics[scale=0.25]{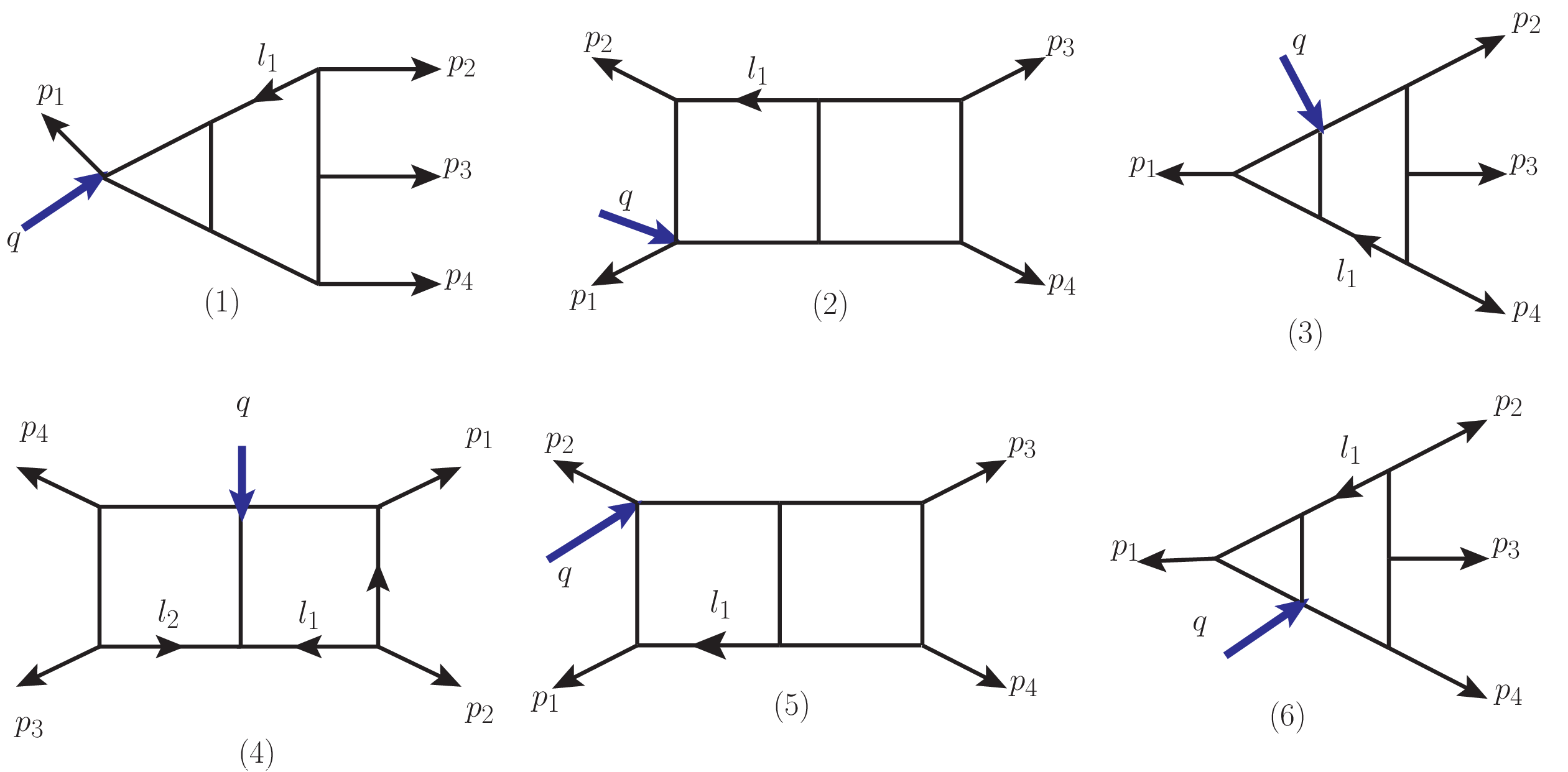}
		\caption{There are six planar topologies for the two-loop four-point form factor of $
			{\rm tr}(\phi_{12}^3)$. Figures (5) and (6) are related to (2) and (3) respectively by symmetry.}
		\label{Fig:2loopTopo}
	\end{center}
\end{figure}

\subsection*{The two-loop integrands  of $\text{tr}(\phi_{12}^3)$}
\label{sec:R2l4p}

There are in total six planar trivalent topologies to consider as given in Figure~\ref{Fig:2loopTopo}. We comment here that the topologies (5) and (6) can be seen as flipping (2) and (3).

The loop correction of the two-loop four-point form factor is
\begin{align}
	\label{eq:FexpinI}
	{\cal I}_{{\rm tr}(\phi_{12}^3), 4}^{(2)}(1_\phi, 2_\phi, 3_\phi, 4_{g^+}) =
	& \Big( {1\over2} I_{1,1}(1,2,3,4) + I_{1,2}(1,2,3,4)  + I_{2,1}(1,2,3,4)\\
	& + I_{2,2}(1,2,3,4) + I_{2,3}(1,2,3,4)+ I_{3,1}(1,2,3,4) \nonumber\\
	& + I_{3,2}(1,2,3,4) + I_{3,3}(1,2,3,4) + I_{4,1}(1,2,3,4) \nonumber\\
	& + I_{4,2}(1,2,3,4) \Big) + (p_1 \leftrightarrow p_3) \,, \nonumber
\end{align}
where $I_{i,j}(a,b,c,d)$ are the two-loop integrals of topologies $(i)$ in Figure \ref{Fig:2loopTopo} and associated with the numerator $N_{i,j}(a,b,c,d)$ given explicitly in Table~\ref{tab:integrandp}.

\begin{center}
	\begin{longtable}{|c|c|c|}
	\caption{The numerators of the topologies.\label{tab:integrandp}} \\ \hline
	\makecell*[c]{$N_{1,1}$} &\makecell*[c]{\includegraphics[scale=0.2]{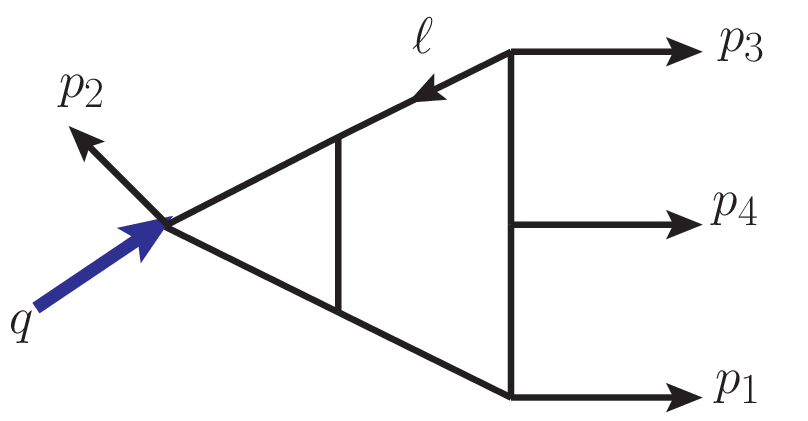}} & \makecell*[c]{$s_{134} [-\text{tr}_-\left(\ell, 3, 1, 4\right)+s_{34\ell}(s_{13} +s_{34}) +s_{14} (s_{3 \ell}+s_{34})]$} \\ \hline
	\makecell*[c]{$N_{1,2}$} &\makecell*[c]{\includegraphics[scale=0.2]{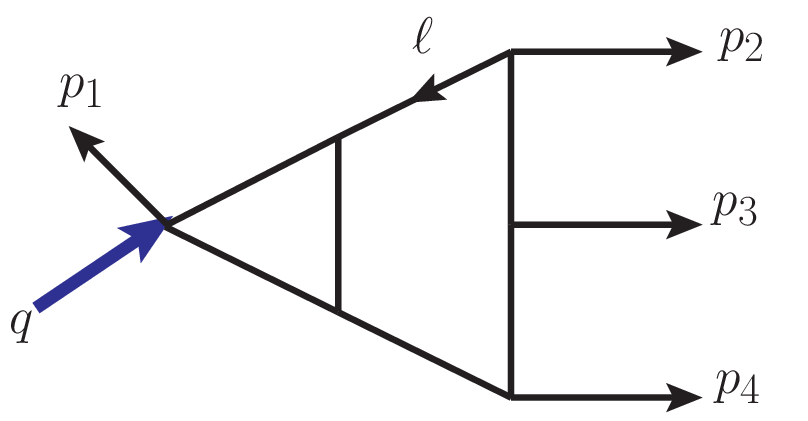}}&\makecell*[c]{$ -\frac{s_{234}}{s_{13}} [s_{23} \text{tr}_-(\ell+p_2, 3, 1, 4)-s_{23\ell} \text{tr}_-\left( 2, 3, 1, 4\right)]$}\\ \hline
	\makecell*[c]{$N_{2,1}$}	&\makecell*[c]{\includegraphics[scale=0.2]{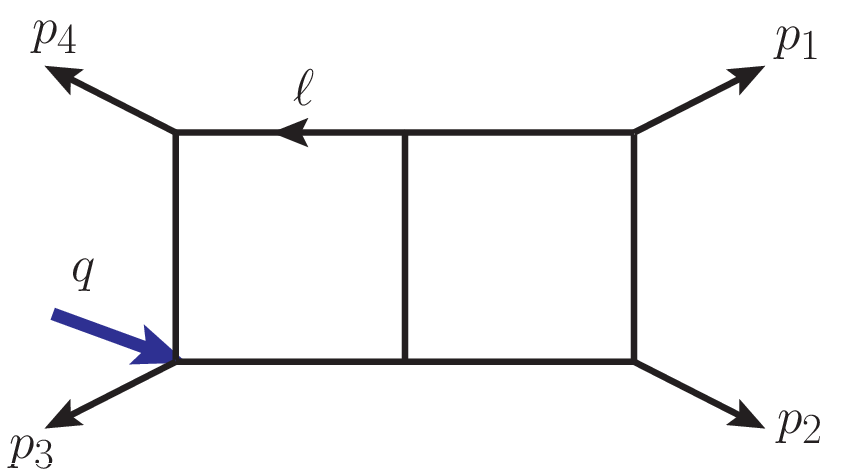}} & \makecell*[c]{ $\frac{s_{12}}{s_{13}} [\ell^2 \text{tr}_-( 3,p_1+p_4, 2, 1)-s_{12} \text{tr}_-(\ell, 3, 1, 4)]$}  \\ \hline
	\makecell*[c]{$N_{2,2}$}	&\makecell*[c]{\includegraphics[scale=0.2]{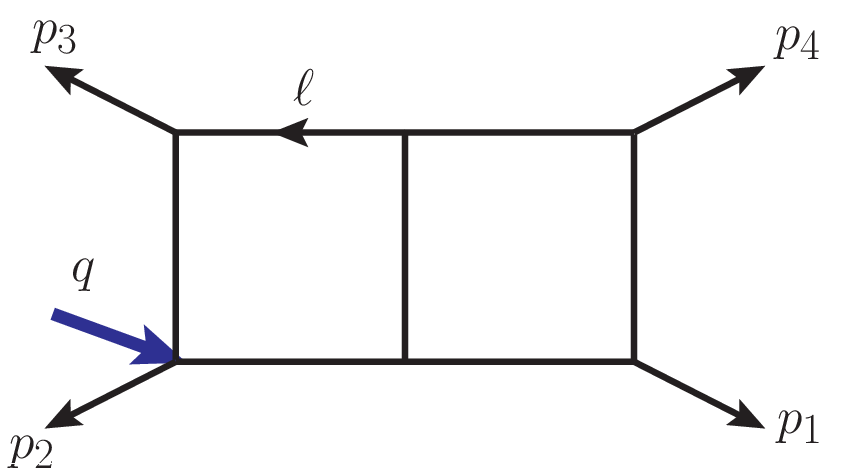}} & \makecell*[c]{ $-\ell^2 \text{tr}_-( 4, 3, 4, 1)-s_{14} \text{tr}_-( 3, 4, 1,\ell)$}  \\ \hline
	\makecell*[c]{$N_{2,3}$}	&\makecell*[c]{\includegraphics[scale=0.2]{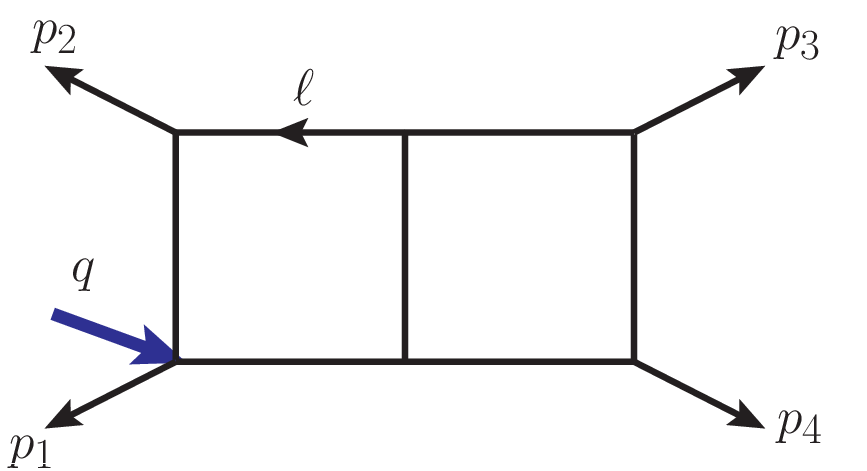}} & \makecell*[c]{ $\frac{\ell^2 s_{34} \text{tr}_-( 1, 4, 2, 3)}{s_{13}}+s_{34} \text{tr}_-\left( 2, 4, 3,\ell\right)-\frac{s_{34}^2 \text{tr}_-\left( 2, 1, 3,\ell\right)}{s_{13}}$}  \\ \hline
	\makecell*[c]{$N_{3,1}$}	&\makecell*[c]{\includegraphics[scale=0.2]{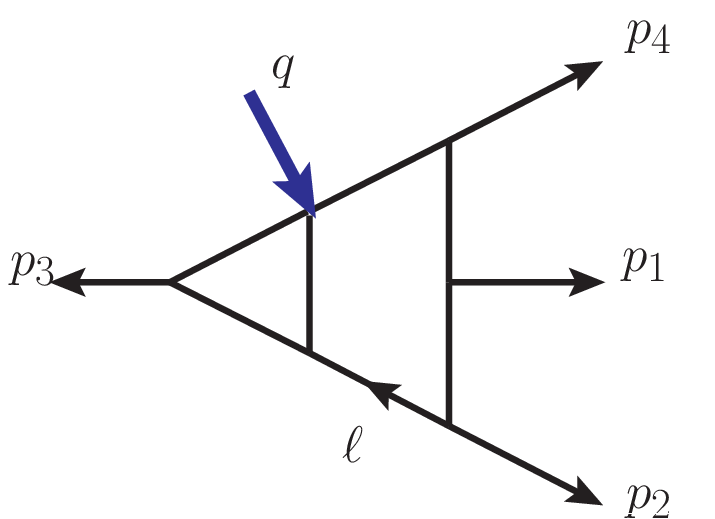}}&\makecell*[c]{$\frac{(\ell-p_ 3){}^2 }{s_{13}}[s_{12\ell} \text{tr}_-( 2, 1, 3, 4)-s_{12} \text{tr}_-\left(\ell+p_2, 1, 3, 4\right)] $}\\ \hline
	\makecell*[c]{$N_{3,2}$}	&\makecell*[c]{\includegraphics[scale=0.2]{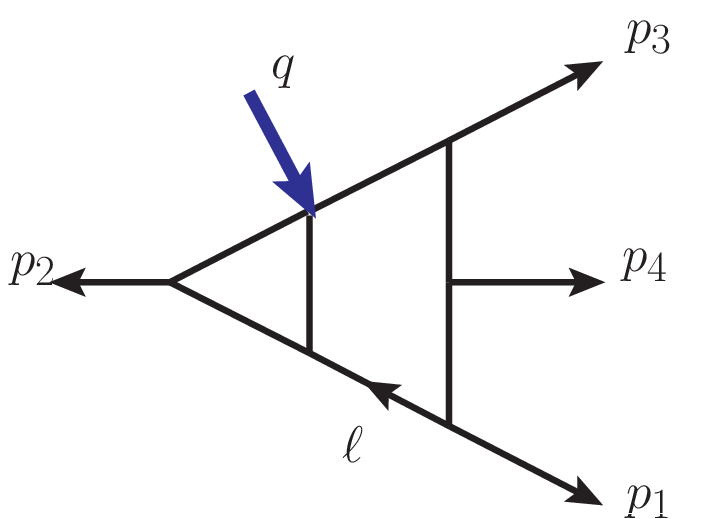}}&\makecell*[c]{$(\ell-p_2){}^2 [-\text{tr}_-(\ell, 1, 3, 4)+s_{34} s_{1\ell}+s_{13} s_{14\ell}+s_{14} \left(s_{14\ell}+s_{34}\right)]$}\\ \hline
	\makecell*[c]{$N_{3,3}$}	&\makecell*[c]{\includegraphics[scale=0.2]{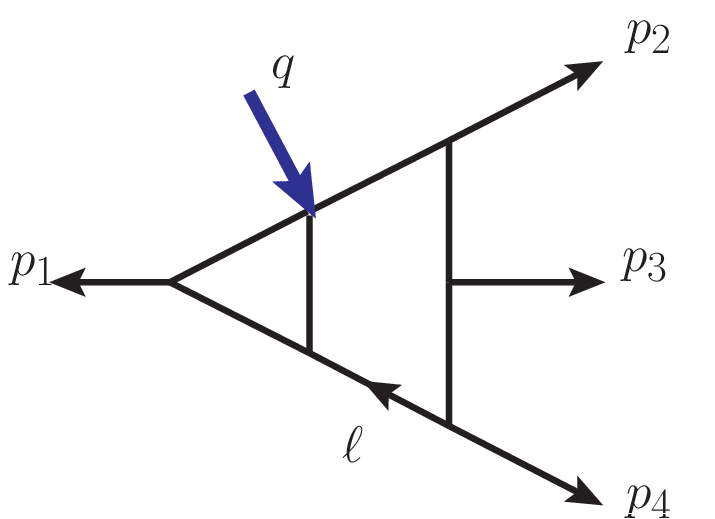}}&\makecell*[c]{$\frac{\left(\ell-p_1\right){}^2}{s_{13}}(s_{23} \text{tr}_-\left(\ell, 3, 1, 4\right)+s_{4\ell} \text{tr}_-\left( 2, 3, 1, 4\right)+\ell^2 s_{13} s_{23})$}\\ \hline
	\makecell*[c]{$N_{4,1}$}	&\makecell*[c]{\includegraphics[scale=0.2]{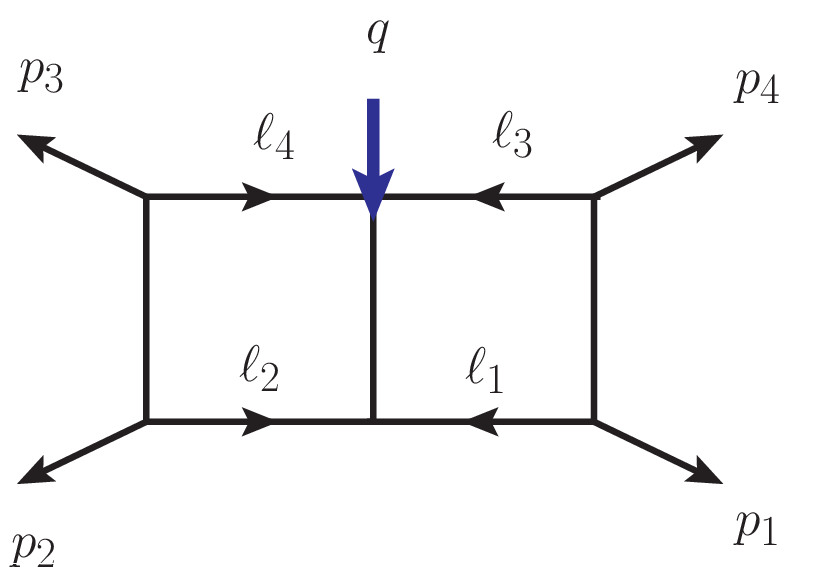}}&\makecell*[c]{$-\frac{s_{23}}{s_{13}} [s_{1\ell_1} \text{tr}_-(\ell_2, 1, 3, 4)+(p_1-\ell_2)^2 \text{tr}_-(\ell_1, 1, 3, 4)]$\\$+\ell_3^2 (s_{12} \ell_4^2+s_{23} (p_1-\ell_2)^2)+\frac{\ell_2^2 }{s_{13}}[s_{123} \text{tr}_-(\ell_1, 1, 3, 4)$\\$-s_{1\ell} (-\text{tr}_-\left( 3, 4, 1, 2\right)+s_{12} s_{34}+s_{13} s_{34})$\\$-\frac{1}{2} s_{123}s_{13} \ell_3^2]+\frac{\ell_4^2}{s_{13}} \left(s_{1\ell_1} \text{tr}_-\left( 2, 1, 3, 4\right)-s_{12} \text{tr}_-\left(\ell_1, 1, 3, 4\right)\right)$}\\ \hline
	\makecell*[c]{$N_{4,2}$}	&\makecell*[c]{\includegraphics[scale=0.2]{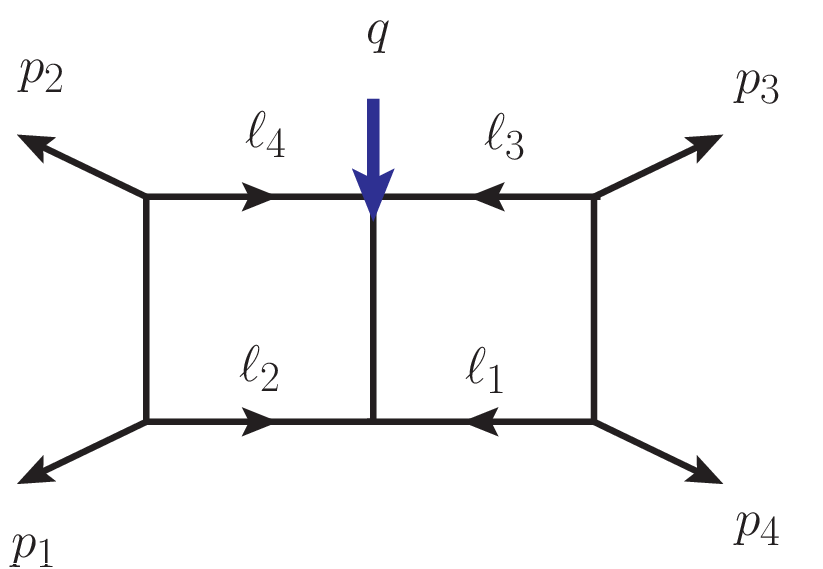}}&\makecell*[c]{$\ell_2^2 \left(\frac{\left(\ell_1+p_3\right){}^2 \text{tr}_-( 2, 3, 1, 4)}{s_{13}}+\frac{\left(-s_{13}-s_{23}\right) \text{tr}_-\left(\ell_1, 3, 1, 4\right)}{s_{13}}+s_{14} \left(\ell_1+p_3\right){}^2\right)$\\$-\ell_3^2 \ell_2^2 \left(\frac{\text{tr}_-\left( 4, 1, 3, 2\right)}{s_{13}}+s_{14}\right)-\frac{2 s_{12} (\ell_2\cdot p_3) \text{tr}_-\left(\ell_1, 3, 1, 4\right)}{s_{13}}$\\$-\frac{s_{12} (2 (\ell_1\cdot p_4)+s_{34}) \text{tr}_-\left(\ell_2, 3, 1, 4\right)}{s_{13}}$\\$-\ell_4^2 \text{tr}_-\left( 4, 1,\ell_1, 3\right)-q^2\ell_1^2 \ell_2^2 +\frac{1}{2} \ell_1^2 \ell_4^2 s_{13}+\ell_3^2 \ell_4^2 s_{14}$}\\ \hline
	\end{longtable}
\end{center}

We give some comments on the difference between the form factors of ${\rm tr}(\phi_{12}^3)$ and ${\rm tr}(\phi_{12}^2)$. 
First, the form factor of ${\rm tr}(\phi_{12}^3)$ has only the flipping symmetry of $p_1$ and $p_3$, which is less than ${\rm tr}(\phi_{12}^2)$. Second, the form factor of  ${\rm tr}(\phi_{12}^2)$ involves non-planar topologies, while ${\rm tr}(\phi_{12}^3)$ does not. Furthermore, we observe the spurious pole ${1\over s_{13}}$ appears in the integrand expression of the form factor of ${\rm tr}(\phi_{12}^3)$, while ${\rm tr}(\phi_{12}^2)$ is free of spurious poles.
Finally, there is no manifest DDC symmetry for the form factor of ${\rm tr}(\phi_{12}^3)$ in the lightlike limit.


\providecommand{\href}[2]{#2}\begingroup\raggedright\endgroup

\end{document}